\documentclass[a4paper,portrait]{article}

\usepackage[a4paper, top=0.5in, bottom=0.8in, hmargin=0.5in, footnotesep=0.55in]{geometry}
\usepackage{amsmath}
\usepackage{amssymb}
\usepackage{enumitem}
\usepackage{changepage}
\usepackage[title, toc, titletoc]{appendix}
\usepackage{listings}
\usepackage{adforn}
\usepackage{fancyvrb}
\usepackage[most]{tcolorbox}

\usepackage{tikz}
\usepackage{tikz-3dplot}
\usetikzlibrary{intersections}
\usetikzlibrary{calc}
\usetikzlibrary{positioning}
\usetikzlibrary{arrows.meta}
\usetikzlibrary{decorations.text}
\usetikzlibrary{decorations.markings}
\usetikzlibrary{backgrounds}
\usetikzlibrary{math}

\usepackage{pgfplots}
\pgfplotsset{width=12cm,compat=1.14}

\usepackage{hyperref}
\hypersetup{
	hypertexnames=false, 
	colorlinks=true,
	allcolors=black,
    urlcolor=blue!50!black,
	bookmarksnumbered=true,
	bookmarksopen=true,
	pdfauthor=Ross Ure Anderson
}

\newcounter{listCtr}

\newcommand{\te}{\ensuremath{\theta}}
\newcommand{\al}{\ensuremath{\alpha}}
\newcommand{\ao}{\ensuremath{\alpha_1}}
\newcommand\p{\ensuremath{\psi}}
\newcommand\de{\ensuremath{\delta}}
\newcommand\g{\ensuremath{\gamma}}
\newcommand{\bt}{\ensuremath{\beta}}
\newcommand{\la}{\ensuremath{\lambda}}
\newcommand{\si}{\ensuremath{\sigma}}
\newcommand{\mv}{\ensuremath{\mu}}
\newcommand{\mo}{\ensuremath{\mu_0}}
\newcommand{\no}{\ensuremath{\nu_0}}
\newcommand{\ta}{\ensuremath{\tau}}
\newcommand{\ph}{\ensuremath{\phi}}
\newcommand{\dg}{^{\circ}}
\newcommand{\bd}[1]{\mathbf{#1}}
\newcommand{\nL}{\bd{n}_{\rule{0pt}{7.5pt}L}}
\newcommand{\nH}{\bd{n}_{\rule{0pt}{7.5pt}H}}
\newcommand{\aPerp}{\bd{a}_{\rule{0pt}{7.5pt} \perp}}
\newcommand{\tf}{\:\therefore\:}
\newcommand{\AST}{\mathrm{AST}}
\newcommand{\MST}{\mathrm{MST}}
\newcommand{\CT}{\mathrm{CT}}
\newcommand{\EOT}{\mathrm{EOT}}
\newcommand{\EC}{\mathrm{EC}}
\newcommand{\OB}{\mathrm{OB}}

\def\earthAxialTilt{23.44}

\newcommand{\vc}[1]{\underline{\mathbf{#1}}}
\newcommand{\rPar}{\vc{r}_{\rule{0pt}{9pt} \hspace{2.5pt} \rule{0.5pt}{6pt} \hspace{2pt} \rule{0.5pt}{6pt}}}
\newcommand{\rpPar}{{\vc{r}'}_{\rule{0pt}{9pt} \rule{0.5pt}{6pt} \hspace{2pt} \rule{0.5pt}{6pt}}}
\newcommand{\rPerp}{\vc{r}_{\rule{0pt}{9pt} \perp}}
\newcommand{\rpPerp}{{\vc{r}'}_{\hspace{-3pt} \rule{0pt}{9pt} \perp}}

\newcommand{\rParB}{\vc{r}_{\rule{0pt}{10pt} \hspace{3pt} \rule{0.8pt}{6pt} \hspace{1.8pt} \rule{0.8pt}{6pt}}}
\newcommand{\rPerpB}{\vc{r}_{\rule{0pt}{8.5pt} \perp}}
\newcommand{\rpPerpB}{{\vc{r}'}_{\hspace{-3pt} \rule{0pt}{8.5pt} \perp}}

\title{\vspace{-2ex}Derivation of Solar Position Formulae}
\author{Ross Ure Anderson \thanks{E-mail: ruanderson100@yahoo.com; web: archive.org/details/@ross\_ure\_anderson} }
\date{31th August, 2020}

\begin{document}

\newgeometry{top=0.5in, bottom=0.8in, hmargin=0.9in, footnotesep=0.55in}

\maketitle

\vspace{2ex}

\begin{adjustwidth}{0.6in}{0.6in}
\noindent \textbf{Abstract.} Derivation of the following formulae for solar position as seen from orbiting planet based on a simplified model: sunrise direction formula, solar declination formula, sunrise equation, daylight duration formula, solar altitude formula, solar azimuth formula. Use of notion of effective axial tilt, and Rodrigues Rotation Formula, and reflections of the orbital quadrants to reduce the general case to the simpler case of the day of the winter solstice. Derivation of equations for solar time to clock time conversion, and sunrise, sunset, and solar noon times. Implementation of an analemma calculator. Comparison of the sunrise direction formula with 304 point dataset of actual sunrise data from Earth, obtaining average accuracy of $1.25\dg$, and estimate of Earth's axial tilt of $23.52\dg$~---~within $0.1\dg$ of the currently accepted value $23.44\dg$.
\end{adjustwidth}

\vspace{0.5ex}

\section{Introduction}

\label{sec:intro}

This article first derives a formula for the sunrise direction for a planet orbiting a central sun, in terms of the day of the year and the latitude, under a simplified model. Then, building on the method of proof used a number of further formulae relating to the solar position \cite{wikipedia:sun-position}, \cite{noaa:solar-calculator} are derived. The approach adopted reduces the cases of the four orbital quadrants to a single quadrant and then reduces the latter case to a single point (the winter solstice). The simplifying assumptions made throughout are : 
\begin{enumerate}
\item The planet is a sphere
\label{hypothesis:first}
\item The orbit of the planet is an ellipse with the sun at one focus\footnote{The proofs below do not specifically make use of the elliptical property --- the only orbital parameter they require is the orbital angle \p, thus they would apply to any planar orbit shape so long as the other simplifying assumptions applied. In the case of Earth the gravity of the moon and other planets perturbs its orbit away from a perfect ellipse.}
\item The planet radius is negligible compared with its minimal distance from the sun \label{hypothesis:distant-sun}
\item Whilst orbiting, the planet rotates about a fixed direction axis through its center
\item The sun can be approximated as a point source of light \label{hypothesis:point-source}
\item A day of the year is approximated as a single stationary point in the orbit at which the planet completes a full $360\dg$ revolution \label{hypothesis:last}
\end{enumerate}

\noindent The sunrise direction formula gives the sunrise direction \te\ as an angle north of east :

\begin{equation}
\te = -\arcsin \left( \frac{\sin \al \cos \p}{\cos \de} \right)
\label{eq:sunrise-formula}
\end{equation}

\noindent where : \\

{\setlength{\tabcolsep}{0.7em}
\begin{tabular}{rcl}
\te & $\in$ & $[-90\dg, 90\dg]$, \\
\al & = & planet axial tilt $\in [0\dg, 180\dg]$, \\
\de & = & latitude $\in (-90\dg, 90\dg)$, \\
\p & = & day of year angle = orbital angle swept out from winter solstice \\
\end{tabular}} \\[2ex]

In a circular orbit/constant speed model the stationary days are evenly spaced around the circle and the orbital angle \p\ can be defined as $\frac{d}{N}(360\dg)$ for day $d \in [0, N - 1]$, where $d$ is the day offset from the day of the winter solstice, and $N$ is the number of days in the year. A more accurate relation between the day of the year and \p, in the case of Earth, is described in \cite{jenkins:sun-position} with a formula expressing the angle swept out from the spring equinox as a function of time --- we can then take this angle at say 12 noon on a day and \\ 

\restoregeometry
\newpage

\noindent add $90\dg$ to obtain \p, the time in the formula being measured in units of mean solar days of 24 hours duration from UTC midnight on 1\textsuperscript{st} January 2013. In the comparison with real data from Earth in Appendices \ref{sec:actual-sunrise-data} and \ref{sec:axial-tilt-estimate} the former model is used. To use the latter model instead, the program code which computes the formula's predictions can be changed as described in Appendix~\ref{sec:perl-scripts}. The former model is a reasonable approximation for Earth whose orbit is very close to circular, but for a planet with more eccentric orbit a procedure such as in \cite{jenkins:sun-position} would be required. \\

In the above simplified model the sunset position is symmetrically opposite to the sunrise position, on the west side of the horizon, at angle \te\ north of west. A negative \te\ means south of east/west. An alternative expression is :

\begin{equation}
\te = -\arcsin \left( \frac{\sin \ao}{\cos \de} \right)
\label{eq:sunrise-formula-effective}
\end{equation}

\noindent where 
\begin{equation}
\sin \ao = \sin \al \cos \p
\label{eq:effective-axial-tilt}
\end{equation}
defines an `effective axial tilt' \ao\ at orbital angle \p\ (Figure~\ref{fig:general-day}), which makes the geometric situation identical to the winter solstice day of $\p = 0\dg$, when viewed from a different angle. The effective axial tilt equals the actual axial tilt \al\ on the winter solstice, and falls to zero at the spring equinox, so $\ao \in [0, \al]$. Any day of the first quadrant $Q1$ of the orbit is thus equivalent to the winter solstice of Figures~\ref{fig:winter-solstice} and \ref{fig:winter-solstice-towards-sun} (or Figures~\ref{fig:winter-solstice-south} and \ref{fig:winter-solstice-towards-sun-south} in the case of the southern hemisphere) with the effective axial tilt $\ao$ replacing \al. Days in $Q2$, $Q3$, $Q4$ are then obtained by simple reflections from $Q1$. Thus with the substitution of \ao\ for \al\ the simpler case of the winter solstice in Figure~\ref{fig:winter-solstice} applies to the entire orbit, with the sun being placed on the opposite side of that in Figure~\ref{fig:winter-solstice} in the cases of $Q2$ and $Q3$. The effective axial tilt is defined only for $Q1$ and is in the range $[0, \al]$ but if the definition (\ref{eq:effective-axial-tilt}) is extended throughout the whole year to a value \ao\ in the range $[-\al, \al]$ then it equals minus the solar declination $\la$ (\S\ref{sec:solar-declination}). Note the notion of effective axial tilt is only applicable to a stationary day at a fixed point in the orbit, since otherwise it would imply the wrong direction of motion for the planet in its orbit during the day.  \\

\noindent The formula (\ref{eq:sunrise-formula}) can also be written as :
\begin{equation}
\sin \te \cos \de + \sin \al \cos \p = 0
\label{eq:sunrise-formula-sum-zero}
\end{equation}

\noindent or, to estimate the axial tilt from the sunrise direction (Appendix~\ref{sec:axial-tilt-estimate}) :
\begin{equation}
\sin \al = -\frac{\sin \te \cos \de}{\cos \p}
\label{eq:sunrise-formula-axial-tilt}
\end{equation}

\noindent provided $\cos \p \neq 0$, ie we are not at an equinox, where the formula becomes $0/0$. Similarly (\ref{eq:sunrise-formula-sum-zero}) can be rearranged to express day of the year in terms of sunrise position at a known latitude (or Pole Star elevation, Appendix~\ref{sec:latitude-pole-star}) if the axial tilt is known. \\

The term `winter solstice' refers to the standard notion of `winter' of the northern hemisphere, even though towards the equator the terms winter and summer lose their normal meanings --- for example on the equator maximal solar radiation is received at the equinoxes and is minimal at both summer and winter solstices \cite{wikipedia:equatorial-seasons}. \\

Though astronomical terminology is used throughout, the problems of the solar position under the simplified model are purely geometric involving a sphere on an elliptical path intersected by parallel rays from the direction of one of the path's foci. For example the point of sunrise/sunset on a planet is simply when these rays become parallel to the tangent plane at the point. The desired sunrise/sunset direction is then the direction of the rays wrt true north \cite{wikipedia:true-north} in the local tangent (or horizon) plane. \\

In Appendix~\ref{sec:actual-sunrise-data}, the sunrise direction formula with a circular orbit/constant speed model is compared with actual data for the case of the Earth, using 304 data points at 8 latitudes for the year 2018--2019 --- the actual sunrise directions being taken from \href{https://www.timeanddate.com}{\nolinkurl{www.timeanddate.com}} \cite{timeanddate:sun-calculator} and the latitudes from  \href{https://www.google.com/maps}{\nolinkurl{Google} \nolinkurl{Maps}}. Using an Earth axial tilt of $\earthAxialTilt\dg$~\cite{wikipedia:earth-axial-tilt}, the formula then matches the actual sunrise directions to within an average error of $1.25\dg$, which is a quite good approximation for the simplified model. Using the formula's own predicted axial tilt of $23.52\dg$ (Appendix~\ref{sec:axial-tilt-estimate}) the error is $1.26\dg$. \\

A chief interest of the sunrise direction formula is how in relating the above quantities measurable by naked eye from the Earth using only primitive instruments, the hypothesis of a spherical Earth in a circular orbit and rotating about a fixed axis is arrived at using only a geometric argument --- since the close correlation with the actual data would be highly unlikely to occur unless the above simplified model was a good approximation in practice. Thus it is one method from which the ancients could have determined the Earth shape and motions from only naked eye observations and geometry, without the benefit of modern astronomy. \\

The naked eye measurements corresponding to the above parameters are :
\begin{enumerate}[label=(\Alph*)]
\item $n$, the day offset from the winter solstice, where the winter solstice is determined by the day on which the Sun is at its lowest southerly elevation at solar noon (with the convention of a northerly Sun elevation being $> 90\dg$), and $n$ is counted from that day starting at zero. For southerly latitudes such lowest southerly elevation would actually be in the summer as there the Sun reaches into the northern half of the sky and is low in the northern sky in the winter --- the day offset would thus be counted from the summer solstice when the Sun is at its highest \emph{northerly} elevation at solar noon, which is at the same time as the northern winter solstice.
\label{observable:first}
\label{observable:day-offset}
\item $\delta$, the latitude, is obtained from the fixed angle of elevation of the Pole Star at the given location. Where the Pole Star is no longer visible a corresponding fixed point in the southern sky would be used to determine latitude.
\item \te, the position of the sunrise, is measured wrt due east, where due east is determined wrt true north, and true north is determined from the position of the Pole Star projected onto the horizon orthogonally (the `azimuthal' position).
\label{observable:last}
\end{enumerate}

The latitude \de\ being measured as elevation of the Pole Star is based on the hypothesis of a spherical Earth and very distant Pole Star (see Appendix~\ref{sec:latitude-pole-star}). Measurement \ref{observable:day-offset} may be recorded by some kind of calendrical system marking mid-winter and mid-summer. \\

Originally ancient astronomers would only have these observed quantities to work with, gathering data over long periods of time, and hypothesizing what may be the larger scale structure that caused them \cite{pannekoek:history-astronomy}, \cite{lippincott:story-of-time}, \cite{bbc-in-our-time:calendar}, \cite{bbc-in-our-time:measurement-of-time}. \\

The formula would then relate the above quantities with the right choice of axial tilt, which could be hypothesized by some other method or be estimated from actual sunrise data using equation (\ref{eq:sunrise-formula-axial-tilt}) (Appendix~\ref{sec:axial-tilt-estimate}). The constancy of the rhs of equation (\ref{eq:sunrise-formula-axial-tilt}) for actual data would give evidence for the hypotheses \ref{hypothesis:first} -- \ref{hypothesis:last} above. \\

In short the above mathematical relationship between the observable quantities \ref{observable:first} -- \ref{observable:last} gives a strong evidence for the hypotheses \ref{hypothesis:first} -- \ref{hypothesis:last} with a circular orbit describing, to a good approximation, the Earth shape and motion through space --- since it would be very difficult for any other configuration to produce the same correlation. \\

The derivation below of the sunrise direction formula envisages a prograde (wrt orbital direction) rotation axis of tilt up to $90\dg$. For a retrograde rotation ($\al > 90\dg$) the axial tilt angle $(180 - \te)\dg$ could be used, and would be clockwise, so the sunrise and sunset positions would simply be swapped, and the formula would remain applicable, since $\sin \al = \sin(180 - \al)$. The `north' side of the orbital plane is defined as the side from which the orbit appears anti-clockwise. In using (\ref{eq:sunrise-formula-axial-tilt}) to estimate axial tilt, the result chosen is $\al \leq 90\dg$ or $\al > 90\dg$ according as the rotation is prograde or retrograde, the latter being when the sun rises in the west and sets in the east. From \S\ref{sec:solar-declination} onwards we will only consider the case of $\al \in [0\dg, 90\dg)$. \\

The sunrise direction formula applies to all latitudes \de\ except for the poles, at $90\dg$ and $-90\dg$, where the sunrise/sunset is caused by the orbital motion of the planet rather than the daily rotation --- eg on Earth at the North Pole the Sun rises once per year on the spring equinox and sets once per year on the autumnal equinox, the opposite being true for the South Pole, and this type of sunrise takes much longer than a diurnal sunrise, around 30 hours versus a few minutes, for the sun's disc to fully cross the horizon, and the type of discrete sunrise direction we consider here is not applicable \cite{quora:polar-sunrise}, \cite{web:sunrise-at-poles}, \cite{wikipedia:angular-diameter}, \cite{wikipedia:twilight-definitions}, \cite{astronomy-se:sunrise-duration}. \\

For non-pole latitudes beyond one of the arctic circles such non-diurnal sunrises and sunsets also occur at certain times of the year --- when periods of perpetual day or night are entered. Outside those periods the sunrise and sunset are diurnal. These changes occur as the effective axial tilt changes so the actual latitude travels above and below the effective arctic circle (the latter becoming a point at the pole on the spring equinox). In the sunrise direction formula the periods of perpetual day and night correspond with the $\arcsin$ argument going out of range, for there is no sunset nor sunrise at these times~---~and this can be used to determine the range of dates for these periods for a given polar latitude, as shown in some examples in Appendix~\ref{sec:perpetual-day-night}. \\

\newpage

\section{Sunrise Direction Formula}

The formula is first proved for the simplest case of the winter solstice (ie $\p = 0$), from which the summer solstice case then readily follows. The equinoxes are readily checked because formula (\ref{eq:sunrise-formula}) then gives $\te = 0$ at all latitudes which is clear from the symmetry --- the effective axial tilt is zero, ie. vertical, and each latitude has equal lengths of day and night, the sunset and sunrise being along the east/west line everywhere (except the poles which are seeing the non-diurnal sunset/sunrise). The general case for the first quadrant $Q1$ is obtained from the winter solstice case by calculating the effective axial tilt using the Rodrigues Rotation Formula (Appendix~\ref{sec:rodrigues-rotation-formula} and \cite{wikipedia:rodrigues}). The cases of $Q2$, $Q3$, and $Q4$ are obtained from the $Q1$ case by considering reflections into these quadrants (Figure~\ref{fig:general-day}).

\subsection{Winter Solstice}

\label{sec:winter-solstice}

We require to show :
\begin{equation}
\te = -\arcsin \left( \frac{\sin \al}{\cos \de} \right)
\label{eq:sunrise-formula-ws}
\end{equation}

\subsubsection{Northern Hemisphere}

\label{sec:Latitude-North}

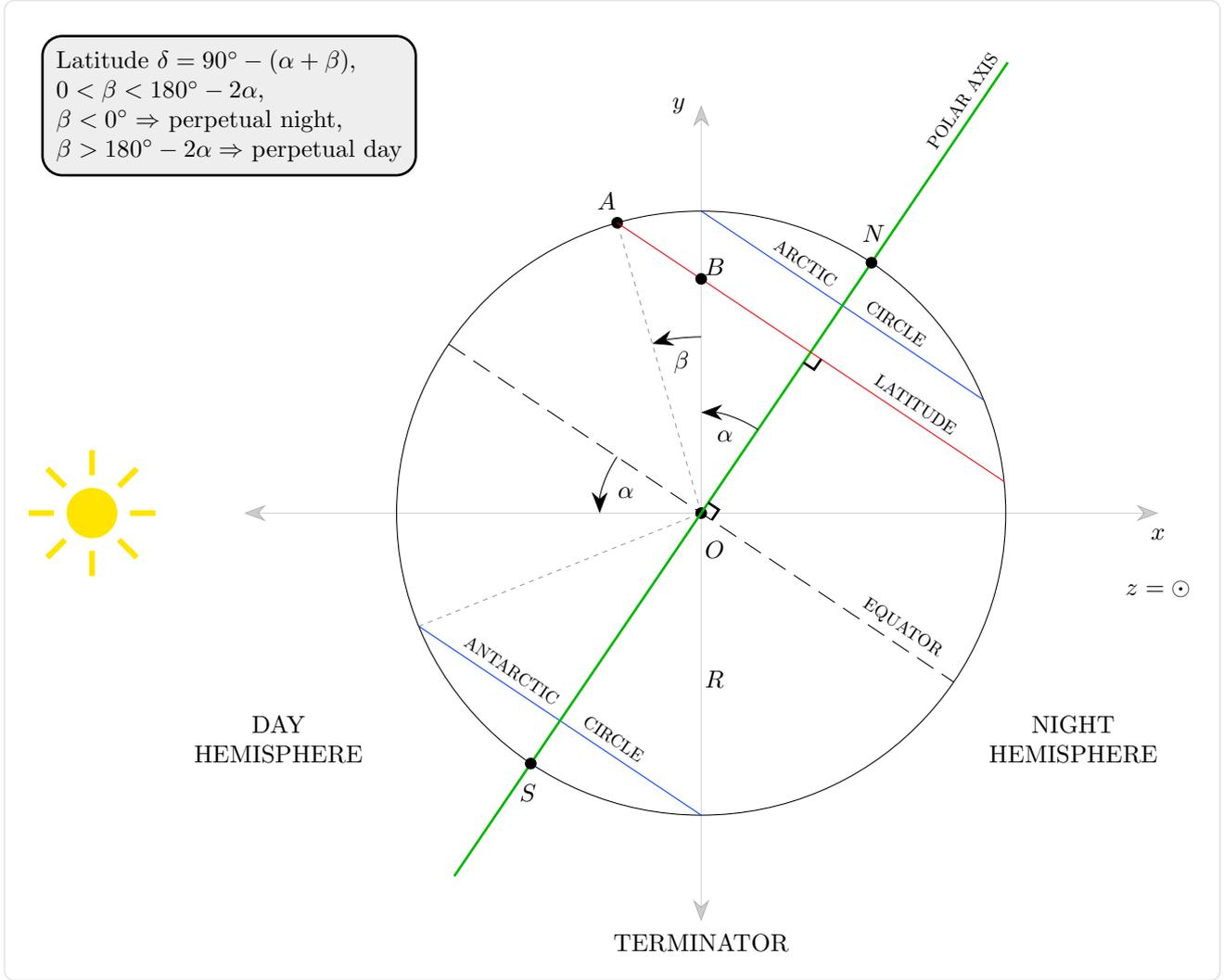
\begin{figure}[!h]
\begin{center}
\caption{Winter Solstice \label{fig:winter-solstice}}
\vspace{3ex}
\begin{tikzpicture} [
	scale=0.36,
	node font=\normalsize,
	framed,
	background rectangle/.style={draw, rounded corners, Frame Color},
	inner frame sep=2ex,
%
	coord axis/.style = {very thin, double arrow=3mm, opacity=0.2},
	polar axis/.style = {Polar Axis Color, line width=1pt},
	equator/.style = {dash pattern=on 8pt off 4pt},
	arctic circle/.style = {Arctic Circle Color},
	antarctic circle/.style = {arctic circle},
	latitude line/.style = {Latitude Color},
	dashed line/.style = {dash pattern=on 2pt off 2pt, opacity=0.4},
	right angle style/.style = {line width=1pt},
	sun/.style = {Sun Color},
	sun ray/.style = {sun, line width=2.3pt},
%
	arctic circle text/.style = {transform shape, black, scale=2},
	antarctic circle text/.style = {arctic circle text},
	equator text/.style = {arctic circle text},
	latitude text/.style = {arctic circle text},
	polar axis text/.style = {arctic circle text, sloped},
	terminator text/.style = {opacity=1},
	hemisphere text/.style={align=center},
	info box/.style={rounded corners=8pt, draw, fill=Info Box Color, inner sep=1.3ex, align=left, line width=1pt},
%
	single arrow/.style=-{Stealth[length=#1]},
	double arrow/.style={Stealth[length=#1]}-{Stealth[length=#1]},
]

\definecolor{Polar Axis Color}{HTML}{00b800}
\definecolor{Arctic Circle Color}{HTML}{003cff}
\definecolor{Latitude Color}{HTML}{ef0006}
\definecolor{Sun Color}{HTML}{ffe400}
\definecolor{Info Box Color}{HTML}{eeeeee}
\definecolor{Frame Color}{HTML}{d8d8d8}

\def\axialTilt{34}
\def\latitude{40}
\def\R{12}
\def\sunRadius{1}
\def\sunRayStart{\sunRadius*1.5}
\def\sunRayLength{\sunRadius}
\def\sunPosition{-2*\R}
\def\rightAngleSize{0.5}
\def\pointRadius{6pt}

\node at (0, 1.6*\R) {};

\filldraw (0, 0) coordinate(O) circle[radius=\pointRadius] node[below right=8pt and -2pt] {$O$};
\draw[name path=circle] (O) circle[radius=\R];
\node[right=-2pt] at (0, -\R*0.55) {$R$};

\node[left, hemisphere text] at (-13, -9) {DAY \\ HEMISPHERE};
\node[right, hemisphere text] at (11, -9) {NIGHT \\ HEMISPHERE};

\draw[coord axis] (-1.5*\R, 0) -- (1.5*\R, 0) node[below=3pt, opacity=1] {$x$};
\draw[coord axis, name path=y axis] (0, -1.35*\R) node[terminator text, below=2pt] {TERMINATOR} -- (0, 1.35*\R) node[left=3pt, opacity=1] {$y$};
\node at (18, -3) {$z = \odot$};

\begin{scope}[rotate=-\axialTilt]

\draw[equator] (-\R, 0) -- (\R, 0) node[pos=0.88, equator text, above=3pt] {EQUATOR};

\draw[arctic circle] (90 + \axialTilt:\R) -- (90 - \axialTilt:\R) node[pos=0.34, arctic circle text, above=3pt] {ARCTIC} node[pos=0.66, arctic circle text, above=3pt] {CIRCLE};

\draw[antarctic circle] (270 - \axialTilt:\R) coordinate(D) -- (270 + \axialTilt:\R) node[pos=0.3, antarctic circle text, above=3pt] {ANTARCTIC} node[pos=0.66, antarctic circle text, above=3pt] {CIRCLE};

\draw[dashed line] (D) -- (O);

\draw[polar axis, name path global=polar axis] (0, -\R*1.45) -- (0, \R*1.8) node[polar axis text, pos=1, above left=10pt and 0pt, inner sep=0ex] {POLAR AXIS};

\draw[dashed line] (O) -- (180 - \latitude:\R) coordinate(A) node[above left=2pt and -3pt, opacity=1] {$A$};
\filldraw (A) circle[radius=\pointRadius];
\draw[latitude line, name path global=latitude] (180 - \latitude:\R) -- (\latitude:\R) node[pos=0.75, latitude text, above=3pt] {LATITUDE};

\path [name intersections={of=latitude and polar axis, by={E}}];

\draw[right angle style] (E) ++(\rightAngleSize, 0) -- ++(0, -\rightAngleSize) -- ++(-\rightAngleSize, 0);
\draw[right angle style] (O) ++(\rightAngleSize, 0) -- ++(0, \rightAngleSize) -- ++(-\rightAngleSize, 0);

\draw[single arrow=0.3cm] (0, 4) arc [start angle=90, delta angle=\axialTilt, radius=4] node[pos=0.6, below=3pt] {$\alpha$};
\draw[single arrow=0.3cm] (90+\axialTilt:7) arc [start angle=90+\axialTilt, end angle=180-\latitude, radius=7] node[pos=0.4, below=2pt] {$\beta$};
\draw[single arrow=0.3cm] (-4, 0) arc [start angle=180, delta angle=\axialTilt, radius=4] node[pos=0.65, right=3pt] {$\alpha$};

\end{scope}

\path [name intersections={of=latitude and y axis, by={B}}];
\filldraw (B) circle[radius=\pointRadius] node[above right=-2pt and -2pt] {$B$};

\path [name intersections={of=polar axis and circle, sort by=polar axis, by={S, N}}];
\filldraw (S) circle[radius=\pointRadius] node[below right=5pt and -8pt] {$S$};
\filldraw (N) circle[radius=\pointRadius] node[above left=5pt and -9pt] {$N$};

\fill[sun] (\sunPosition, 0) coordinate (S) circle [radius=\sunRadius];
\foreach \a in {0, 45,...,315} \draw[sun ray, rotate around={\a:(S)}] (S) ++(\sunRayStart, 0) -- +(\sunRayLength, 0);

\node[info box, below right] at (-26, 19) {
	Latitude $\delta = 90^{\circ} - (\alpha + \beta)$, \\
	$0 < \beta < 180^{\circ} - 2\alpha$, \\
	$\beta < 0^{\circ} \Rightarrow$ perpetual night, \\
	$\beta > 180^{\circ} - 2\alpha \Rightarrow$ perpetual day
};

\end{tikzpicture}
\end{center}
\end{figure}

\begin{figure}[!h]
\begin{center}
\caption{Winter Solstice - View Towards Sun}
\label{fig:winter-solstice-towards-sun}
\vspace{3ex}
\begin{tikzpicture} [
    scale=0.32,
    node font=\normalsize,
	framed,
	background rectangle/.style={draw, rounded corners, Frame Color},
	inner frame sep=2ex,
%
	double arrow/.style={Stealth[length=#1]}-{Stealth[length=#1]},
	coord axis/.style = {very thin, double arrow=3mm, opacity=0.28},
	latitude/.style = {Latitude Color},
	horizon plane/.style = {Horizon Plane Color, line width=1pt},
	dashed line/.style = {dash pattern=on 2pt off 2pt, opacity=0.5},
	right angle style/.style = {line width=1.5pt},
	sun/.style = {Sun Color, opacity=1},
	sun ray/.style = {sun, line width=2.3pt},
%
	info box/.style={rounded corners=8pt, draw, fill=Info Box Color, inner sep=1.3ex, align=left, line width=1pt},
]

\definecolor{Latitude Color}{HTML}{ef0006}
\definecolor{Horizon Plane Color}{HTML}{00b800}
\definecolor{Sun Color}{HTML}{ffe400}
\definecolor{Info Box Color}{HTML}{eeeeee}
\definecolor{Frame Color}{HTML}{d8d8d8}

\def\R{12}
\def\sunRadius{0.8}
\def\sunRayStart{\sunRadius*1.5}
\def\sunRayLength{\sunRadius}
\def\latitudeXRadius{10.35}
\def\latitudeYRadius{2}
\def\angleB{31}
\def\pointRadius{6pt}
\def\rightAngleSize{0.6}

\draw[name path=sphere] (0, 0) coordinate(O) circle[radius=\R];
\node[right=0pt] at (0, -\R*0.55) {$R$};
\path [thick, postaction={decorate, decoration={raise=-4ex, text along path, text align=center, text={|\normalsize|TERMINATOR}}}] (0, -\R) arc[radius=\R, start angle=270, delta angle=90];

\draw[coord axis] (1.5*\R, 0) -- (-1.5*\R, 0) node[below=3pt, opacity=1] {$z$};
\draw[coord axis, name path=y axis] (0, -1.35*\R) -- (0, 1.35*\R) node[right=3pt, opacity=1] {$y$};
\node[left] at (19, -3) {$x = \odot$};

\draw[latitude, name path=latitude circle] (0, \R*0.5) circle[x radius=\latitudeXRadius, y radius=\latitudeYRadius];

\path [name intersections={of=latitude circle and y axis, sort by=y axis, by={D, A}}];
\filldraw (A) circle[radius=\pointRadius] node[above right=2pt and -1pt] {$A$};

\coordinate (B) at (180-\angleB:\R);
\coordinate (B') at (\angleB:\R);
\fill (B) circle[radius=\pointRadius] node[above left=1pt and -2pt] {$B$};
\def\angleBPrime{-38}
\fill (B') circle[radius=\pointRadius] node[above right=1pt and -2pt] {$B'$};

\path [thick, postaction={decorate, decoration={raise=1.6ex, reverse path, text along path, text align=center, text={|\normalsize|LATITUDE}}}] (D) arc[x radius=\latitudeXRadius, y radius=\latitudeYRadius, start angle=270, end angle=200];

\draw[horizon plane] (O) (B) -- ([turn]90:16);
\draw[horizon plane] (O) (B) -- ([turn]-90:16) node[sloped, above=2pt, pos=0.7, black] {HORIZON PLANE};

\draw[dashed line] (B) -- ($(O)!(B)!(-\R, 0)$) coordinate (C);
\fill (C) circle[radius=\pointRadius] node[below=6pt] {$C$};

\draw[right angle style] (C) ++(\rightAngleSize, 0) -- ++(0, \rightAngleSize) -- ++(-\rightAngleSize, 0);

\fill[white] (O) circle[radius=\sunRayStart+\sunRayLength+0.5];
\foreach \a in {0, 45,...,315} \draw[sun ray, rotate=\a] (\sunRayStart, 0) -- +(\sunRayLength, 0);
\fill[sun] (O) circle [radius=\sunRadius];
\node[below right=14pt and 12pt] at (O) {$O$};

\node[info box, below left] at (19, 20.5) {
	$BAB'$ = latitude circle, \\
	$A$ = solar noon, \\
	$B$ = sunset, $B'$ = sunrise, \\
	$x > 0 \Rightarrow$ night, $x < 0 \Rightarrow$ day
};

\end{tikzpicture}
\end{center}
\end{figure}
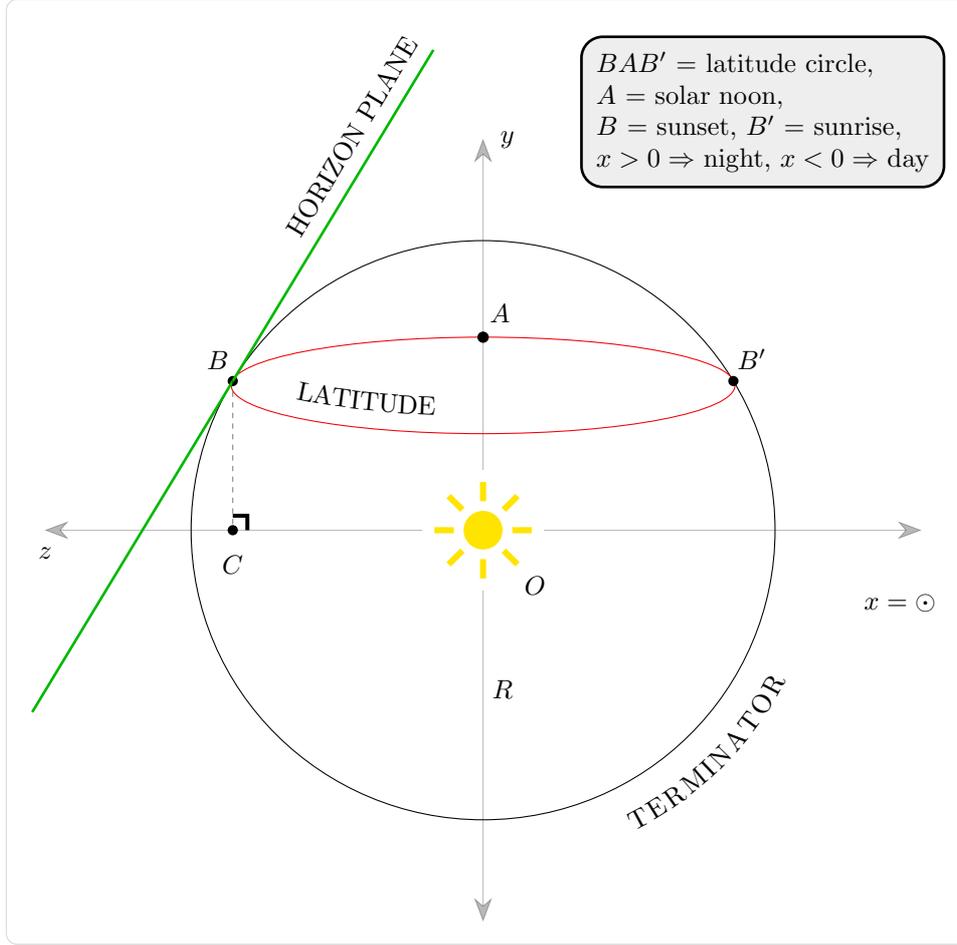

The winter solstice case ($\p = 0$) is shown in Figures~\ref{fig:winter-solstice} and \ref{fig:winter-solstice-towards-sun}, with latitude $\de \in [0, 90 - \al]$ in the northern hemisphere. The case of $\de = 90 - \al$ means points $A$ and $B$ coincide and formula (\ref{eq:sunrise-formula-ws}) reduces to $-\arcsin 1 = -90$, ie southerly, as also seen from Figure~\ref{fig:winter-solstice}. The case of $\de = 0$ means $B$ coincides with $O$ in Figure~\ref{fig:winter-solstice} and (\ref{eq:sunrise-formula-ws}) reduces to the required angle of $\te = -\al$. \\

The east-west line at any point is the intersection of the latitude plane and the horizon plane at that point. Thus to obtain the east-west line we can take the cross product of the normals $\nL$, $\nH$ to these two planes. \\

From Figure~\ref{fig:winter-solstice}, we can choose $\nL = (\sin \al, \cos \al, 0)$, and from Figure~\ref{fig:winter-solstice-towards-sun} we can choose $\nH = (0, BC, OC)$, so :
\begin{eqnarray*}
\nL \times \nH & = & 
\left|
\begin{array}{ccc}
\bd{i} & \bd{j} & \bd{k} \\
\sin \al & \cos \al & 0 \\
0 & BC & OC
\end{array}
\right| \\[1.5ex]
& = & (OC \cos \al, -OC \sin \al, BC \sin \al), \\[1ex]
\tf \mbox{parallel to} & & (\cos \al, - \sin \al, \frac{BC}{OC} \sin \al)
\end{eqnarray*}

From Figure~\ref{fig:winter-solstice} the easterly direction at point $B$ must have a positive $x$-coordinate and a negative $y$-coordinate and thus we can choose the following as east and west (non-unit) vectors at $B$ :
\[
\bd{e} = (\cos \al, - \sin \al, \frac{BC}{OC} \sin \al), \hspace{1em} \bd{w} = (-\cos \al, \sin \al, -\frac{BC}{OC} \sin \al)
\]

Then taking the mirror image of these in the $xy$-plane, which reverses the $z$-component, with east mapping onto west and vice-versa, east/west vectors at $B'$ are :
\[
\bd{e'} = (-\cos \al, \sin \al, \frac{BC}{OC} \sin \al), \hspace{1em} \bd{w'} = (\cos \al, -\sin \al, -\frac{BC}{OC} \sin \al)
\]

From the assumptions \ref{hypothesis:distant-sun} and \ref{hypothesis:point-source} the rays from the sun striking every point of the planet are parallel and come from the $\bd{s} = -\bd{i}$ direction, and it is clear geometrically that the direction of $\bd{s}$ in the horizon plane at $B'$ is south of the easterly $\bd{e'}$ at $B'$. Thus the desired magnitude of angle \te\ of sunrise satisfies :
\begin{eqnarray*}
\bd{e'} \cdot \bd{s} & = & |\bd{e'}| \: |\bd{s}| \cos \te, \\
\tf \cos \te & = & \frac{\cos \al}{|\bd{e'}|}
\end{eqnarray*}

But
$$|\bd{e'}|^2 = 1 + \frac{BC^2}{OC^2} \sin^2 \al$$

and so we need to express $BC/OC$ in terms of \al\ and $\delta$. From Figure~\ref{fig:winter-solstice-towards-sun}, $OC = \sqrt{R^2 - BC^2}$, and $BC$ equals the distance $OB$ from Figure~\ref{fig:winter-solstice}. From the triangle $\triangle AOB$ of Figure~\ref{fig:winter-solstice}, we have :
\[
\frac{\sin (90 - \al - \bt)}{OB} = \frac{\sin (90 + \al)}{R}, \hspace{1em} \mbox{ie.} \hspace{1em} \frac{\cos (\al + \bt)}{OB} = \frac{\cos \al}{R}
\]

Thus $BC$ of Figure~\ref{fig:winter-solstice-towards-sun} satisfies\footnote{The special case of $\al = 90\dg$ can be checked separately using a plan view.} :
$$BC = \frac{R \cos (\al + \bt)}{\cos \al} = \frac{R \sin \de}{\cos \al},$$

\begin{eqnarray*}
\mbox{and so : \hspace{10em}} OC & = & R \sqrt{\left(1 - \frac{\sin^2{}\de}{\cos^2{\al}}\right)} = \frac{R}{\cos \al} \sqrt{\cos^2 \al - \sin^2 \de} \\
\Rightarrow \frac{BC}{OC} & = & \frac{\sin \de}{\sqrt{\cos^{2} \al - \sin^{2} \de}} \\
\Rightarrow |\bd{e'}|^2 = 1 + \frac{BC^2}{OC^2} \cdot \sin^2 \al & = & \frac{(\cos^2 \al - \sin^2 \de) + \sin^2 \de \sin^2 \al}{\cos^2 \al - \sin^2 \de} = \frac{\cos^2 \al - \sin^2 \de \cos^2 \al}{\cos^2 \al - \sin^2 \de} = \frac{\cos^2 \al \cos^2 \de}{\cos^2 \al - \sin^2 \de}
\end{eqnarray*}

\begin{eqnarray*}
\Rightarrow \cos \te & = & \frac{\sqrt{\cos^{2} \al - \sin^{2} \de}}{\cos \de} \\
\Rightarrow \sin^2 \te & = & 1 - \left(\frac{\cos^{2} \al - \sin^{2} \de}{\cos^2 \de}\right) = \frac{1 - \cos^2 \al}{\cos^2 \de} = \frac{\sin^2 \al}{\cos^2 \de} \\
\Rightarrow \sin \te & = & \frac{\sin \al}{\cos \de}.
\end{eqnarray*}

And thus as \te\ above was an angle \emph{south} of east the required angle \emph{north} of east is given by equation~(\ref{eq:sunrise-formula-ws}).

\vspace{2em}

\subsubsection{Southern Hemisphere}

\begin{figure}[!h]
\begin{center}
\caption{Winter Solstice - Case $\al + \bt > 90\dg$}
\label{fig:winter-solstice-south}
\vspace{2ex}
\begin{tikzpicture} [
	scale=0.3,
	node font=\large,
	framed,
	background rectangle/.style={draw, rounded corners, Frame Color},
	inner frame sep=2ex,
%
	coord axis/.style = {very thin, double arrow=3mm, opacity=0.2},
	polar axis/.style = {Polar Axis Color, line width=1pt},
	latitude line/.style = {Latitude Color},
	dashed line/.style = {dash pattern=on 2pt off 2pt, opacity=0.4},
	right angle style/.style = {line width=1pt},
	sun/.style = {Sun Color},
	sun ray/.style = {sun, line width=2.3pt},
%
	latitude text/.style = {transform shape, black, scale=2},
	polar axis text/.style = {latitude text, sloped},
%
	single arrow/.style=-{Stealth[length=#1]},
	double arrow/.style={Stealth[length=#1]}-{Stealth[length=#1]},
]

\definecolor{Polar Axis Color}{HTML}{00b800}
\definecolor{Latitude Color}{HTML}{ef0006}
\definecolor{Sun Color}{HTML}{ffe400}
\definecolor{Frame Color}{HTML}{d8d8d8}

\def\axialTilt{30}
\def\latitude{45}
\def\R{12}
\def\rightAngleSize{0.5}
\def\pointRadius{6pt}
\def\sunRadius{1}
\def\sunRayStart{\sunRadius*1.5}
\def\sunRayLength{\sunRadius}
\def\sunPosition{-2*\R}

\filldraw (0, 0) coordinate(O) circle[radius=\pointRadius] node[below right=1pt and -1pt] {$O$};
\draw[name path=circle] (O) circle[radius=\R];

\draw[coord axis] (-1.5*\R, 0) -- (1.5*\R, 0) node[below=3pt, opacity=1] {$x$};
\draw[coord axis, name path=y axis] (0, -1.35*\R) -- (0, 1.35*\R) node[left=3pt, opacity=1] {$y$};
\node at (18, -3) {$z = \odot$};

\begin{scope}[rotate=-\axialTilt]

\draw[polar axis, name path global=polar axis] (0, -\R*1.45) -- (0, \R*1.7) node[polar axis text, pos=1, above left=12pt and 0pt, inner sep=0ex] {POLAR AXIS};

\draw[dashed line] (O) -- (180 + \latitude:\R) coordinate(A) node[left=2pt, opacity=1] {$A$};
\filldraw (A) circle[radius=\pointRadius];
\draw[latitude line, name path global=latitude] (180 + \latitude:\R) -- (-\latitude:\R) node[pos=0.27, latitude text, above=3pt] {LATITUDE};

\path [name intersections={of=latitude and polar axis, by={E}}];

\draw[right angle style] (E) ++(\rightAngleSize, 0) -- ++(0, -\rightAngleSize) -- ++(-\rightAngleSize, 0);

\draw[single arrow=0.3cm] (0, 4) arc [start angle=90, delta angle=\axialTilt, radius=4] node[pos=0.6, below=3pt] {$\alpha$};
\draw[single arrow=0.3cm] (90+\axialTilt:7) arc [start angle=90+\axialTilt, end angle=180+\latitude, radius=7] node[pos=0.4, below=2pt] {$\beta$};

\end{scope}

\path [name intersections={of=latitude and y axis, by={B}}];
\filldraw (B) circle[radius=\pointRadius] node[above right=-1pt and -1pt] {$B$};

\path [name intersections={of=polar axis and circle, sort by=polar axis, by={S, N}}];
\filldraw (S) circle[radius=\pointRadius] node[below right=5pt and -8pt] {$S$};
\filldraw (N) circle[radius=\pointRadius] node[above left=5pt and -7pt] {$N$};

\fill[sun] (\sunPosition, 0) coordinate (S) circle [radius=\sunRadius];
\foreach \a in {0, 45,...,315} \draw[sun ray, rotate around={\a:(S)}] (S) ++(\sunRayStart, 0) -- +(\sunRayLength, 0);

\end{tikzpicture}
\end{center}
\end{figure}
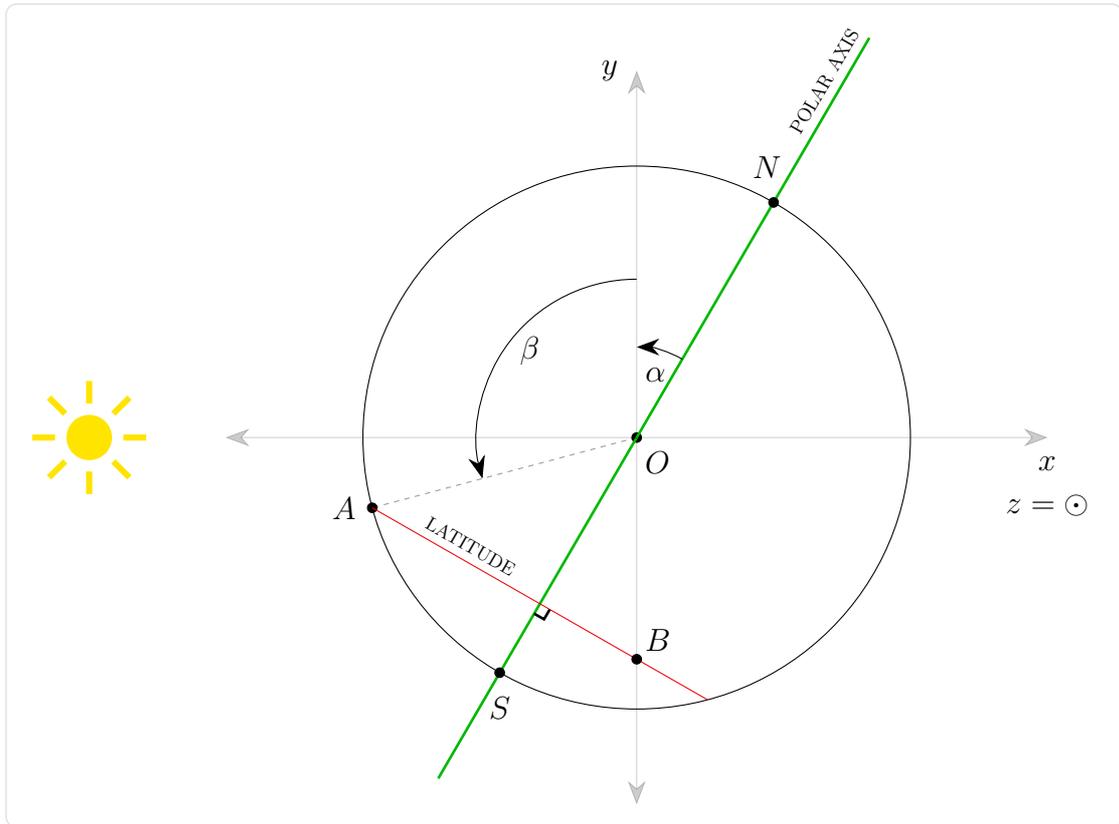

\begin{figure}[!h]
\begin{center}
\caption{Winter Solstice - View Towards Sun - Case $\al + \bt > 90\dg$}
\label{fig:winter-solstice-towards-sun-south}
\vspace{2ex}
\begin{tikzpicture} [
    scale=0.3,
    node font=\large,
	framed,
	background rectangle/.style={draw, rounded corners, Frame Color},
	inner frame sep=2ex,
%
	double arrow/.style={Stealth[length=#1]}-{Stealth[length=#1]},
	coord axis/.style = {very thin, double arrow=3mm, opacity=0.2},
	latitude/.style = {Latitude Color},
	horizon plane/.style = {Horizon Plane Color, line width=1pt},
	dashed line/.style = {dash pattern=on 3pt off 2pt, opacity=0.5},
	right angle style/.style = {line width=1.5pt},
	sun/.style = {Sun Color, opacity=1},
	sun ray/.style = {sun, line width=1.8pt},
]

\definecolor{Latitude Color}{HTML}{ef0006}
\definecolor{Horizon Plane Color}{HTML}{00b800}
\definecolor{Sun Color}{HTML}{ffe400}
\definecolor{Frame Color}{HTML}{d8d8d8}

\def\R{12}
\def\latitudeXRadius{11.1}
\def\latitudeYRadius{5.5}
\def\angleB{24}
\def\pointRadius{6pt}
\def\rightAngleSize{0.6}
\def\sunRadius{0.4}
\def\sunRayStart{\sunRadius*1.5}
\def\sunRayLength{\sunRadius}

\draw[name path=sphere] (0, 0) coordinate(O) circle[radius=\R];

\draw[coord axis] (1.5*\R, 0) -- (-1.5*\R, 0) node[below=3pt, opacity=1] {$z$};
\draw[coord axis, name path=y axis] (0, -1.35*\R) -- (0, 1.35*\R) node[right=3pt, opacity=1] {$y$};
\node[left] at (19, -3) {$x = \odot$};

\draw[latitude, name path=latitude circle] (0, -\R*0.32) circle[x radius=\latitudeXRadius, y radius=\latitudeYRadius];

\path [name intersections={of=latitude circle and y axis, sort by=y axis, by={D, A}}];
\filldraw (A) circle[radius=\pointRadius] node[above right=2pt and -1pt] {$A$};

\coordinate (B) at (180+\angleB:\R);
\coordinate (B') at (-\angleB:\R);
\fill (B) circle[radius=\pointRadius] node[below left=2pt and 0pt] {$B$};
\def\angleBPrime{-38}
\fill (B') circle[radius=\pointRadius] node[below right=-1pt and -3pt] {$B'$};

\path [thick, postaction={decorate, decoration={raise=1.2ex, text along path, text color=Latitude Color, text align=center, text={|\small|LATITUDE}}}] (D) arc[x radius=\latitudeXRadius, y radius=\latitudeYRadius, start angle=270, end angle=320];

\draw[horizon plane] (O) (B) -- ([turn]90:15);
\draw[horizon plane] (O) (B) -- ([turn]-90:18) node[sloped, above right=5pt and 0pt, pos=1, inner sep=0pt, black, node font=\small] {HORIZON PLANE};

\draw[dashed line] (B) -- ($(O)!(B)!(-\R, 0)$) coordinate (C);
\fill (C) circle[radius=\pointRadius] node[above=4pt] {$C$};

\draw[right angle style] (C) ++(-\rightAngleSize, 0) -- ++(0, -\rightAngleSize) -- ++(\rightAngleSize, 0);

\fill[white] (O) circle[radius=\sunRayStart+\sunRayLength+0.25];
\foreach \a in {0, 45,...,315} \draw[sun ray, rotate=\a] (\sunRayStart, 0) -- +(\sunRayLength, 0);
\fill[sun] (O) circle [radius=\sunRadius];
\node[below right=5pt and 4pt] at (O) {$O$};

\end{tikzpicture}
\end{center}
\end{figure}
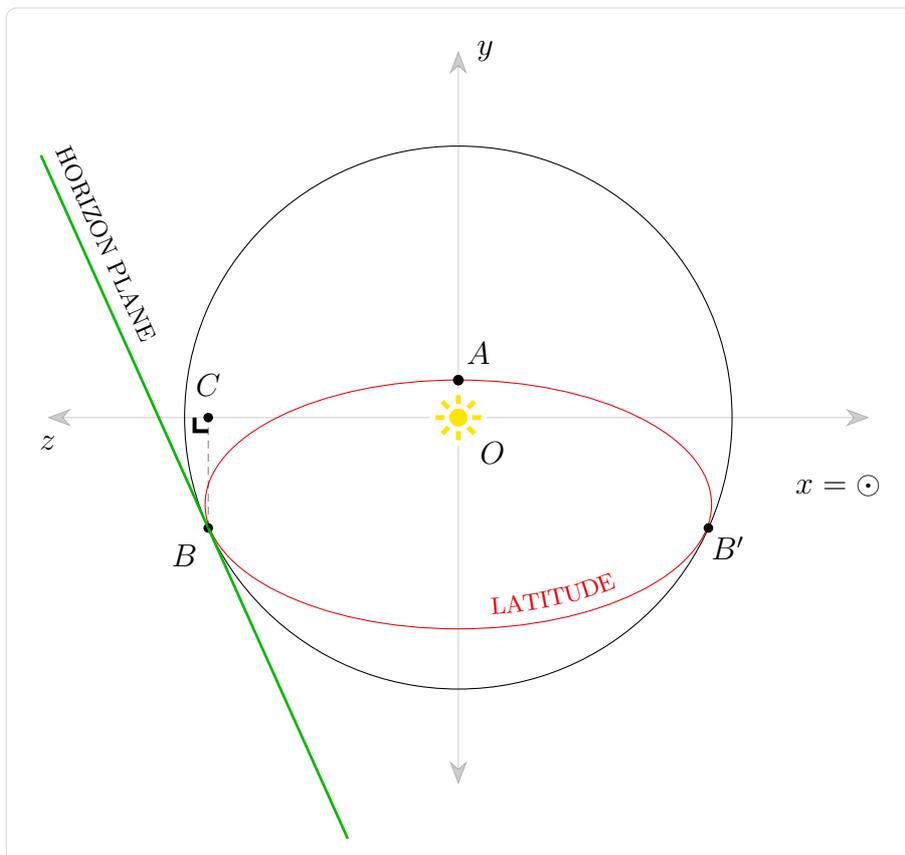

The case $\de \in [\al - 90, 0)$ in the southern hemisphere is shown in Figures~\ref{fig:winter-solstice-south} and \ref{fig:winter-solstice-towards-sun-south}. \\

From the triangle $\triangle AOB$ of Figure~\ref{fig:winter-solstice-south}, we have :
\begin{eqnarray*}
\frac{\sin (\al + \bt - 90)}{OB} & = & \frac{\sin (90 - \al)}{R} \\
\mbox{ie.} \hspace{2em} \frac{-\cos (\al + \bt)}{OB} & = & \frac{\cos \al}{R}
\end{eqnarray*}

Thus the $y$-coordinate $BC$ of Figure~\ref{fig:winter-solstice-towards-sun-south}, which is negative of $OB$ of Figure~\ref{fig:winter-solstice-south}, satisfies :
$$BC = \frac{R \cos (\al + \bt)}{\cos \al} = \frac{R \sin \de}{\cos \al},$$

as before --- thus we obtain the same east-west line as above and since the easterly direction at point $B$ still must have a positive $x$-coordinate and a negative $y$-coordinate the same expressions for $\bd{e'}$ and $\bd{w'}$, and hence \te, are obtained as in  \S\ref{sec:Latitude-North}, thus again giving equation (\ref{eq:sunrise-formula-ws}).

\newpage

\vspace{2ex}

\newpage

\subsection{Summer Solstice}

Here $\p = 180\dg$ so we require to show \te\ is minus what it is in the winter solstice case. The situation is as Figure~\ref{fig:winter-solstice} with the sun at the opposite side so $\bd{s} = \bd{i}$, and then it is clear a sunset $\te$ south of west becomes a sunrise $\te$ north of east, and a sunrise $\te$ south of east becomes a sunset $\te$ north of west, as required.

\subsection{General Day of Year}

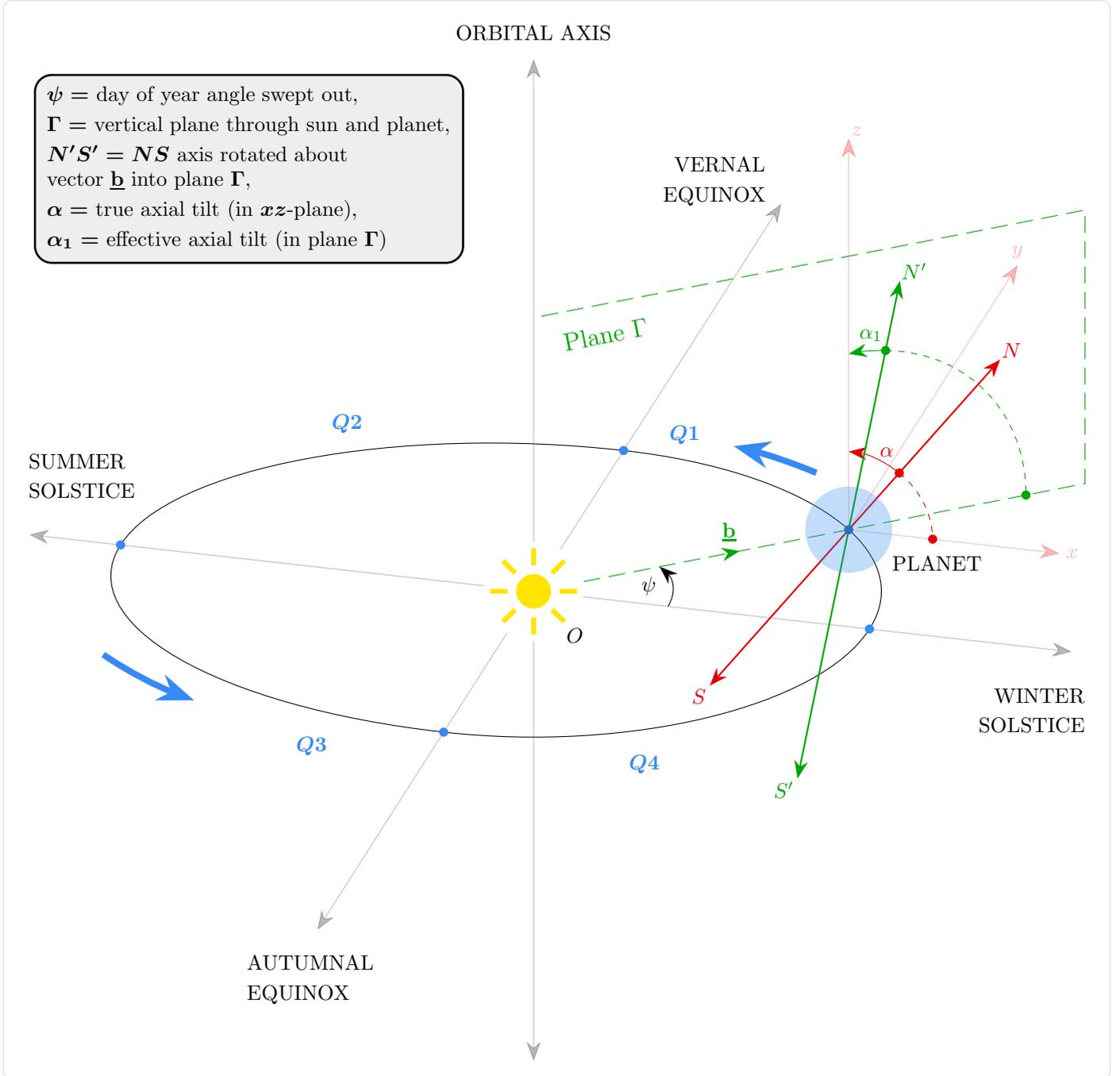
\begin{figure}[!h]
\begin{center}
\caption{General Day Of Year}
\label{fig:general-day}
\vspace{3ex}
\begin{tikzpicture} [
    scale=0.36,
    node font=\normalsize,
	framed,
	background rectangle/.style={draw, rounded corners, Frame Color},
	inner frame sep=2ex,
%
	coord axis/.style = {very thin, double arrow=3mm, opacity=0.28},
	planet coord axis/.style = {Planet Coord Axis Color, very thin, opacity=0.28, single arrow=3mm},
	gamma plane/.style = {Gamma Plane Color, dash pattern=on 8pt off 4pt},
	planet/.style = {Planet Color, opacity=0.3},
	sun/.style = {Sun Color, opacity=1},
	sun ray/.style = {sun, line width=2.3pt},
%
	single arrow/.style = -{Stealth[length=#1]},
	single arrow reverse/.style = {Stealth[length=#1]}-,
	double arrow/.style = {Stealth[length=#1]}-{Stealth[length=#1]},
	orbital arrow/.style = {single arrow=0.6cm, line width=3pt, Planet Color},
	->-/.style={
	decoration={
		markings,
		mark=at position #1 with {\arrow{Stealth[length=0.3cm]}}
	},
	postaction={decorate}
	},
	-<-/.style={
	decoration={
		markings,
		mark=at position #1 with {\arrow{Stealth[length=0.3cm,reversed]}}
	},
	postaction={decorate}
	},
%
	info box/.style={rounded corners=8pt, draw, fill=Info Box Color, inner sep=1.3ex, align=left, line width=1pt, text width=6.7cm},
]

\definecolor{Planet Color}{HTML}{358af3}
\definecolor{Planet Coord Axis Color}{HTML}{e20006}
\definecolor{Gamma Plane Color}{HTML}{00a400}
\definecolor{Sun Color}{HTML}{ffe400}
\definecolor{Info Box Color}{HTML}{eeeeee}
\definecolor{Frame Color}{HTML}{d8d8d8}

\def\axialTilt{37}
\def\effectiveAxialTilt{12}
\def\orbitalRadius{16}
\def\ellipseElongation{1.23}
\def\dayOfYearAngle{40}
\def\planetRadius{2}
\def\sunRadius{0.8}
\def\sunRayStart{\sunRadius*1.5}
\def\sunRayLength{\sunRadius}
\def\pointRadius{6pt}

\def\mainXRotation{65}
\def\mainZRotation{0}

\def\coordSystemAlpha{-15}
\def\coordSystemBeta{0}
\def\coordSystemGamma{0}

\tdplotsetmaincoords{\mainXRotation}{\mainZRotation}
\tdplotsetrotatedcoords{\coordSystemAlpha}{\coordSystemBeta}{\coordSystemGamma}

\coordinate (O) at (0, 0);

\begin{scope}[tdplot_main_coords, tdplot_rotated_coords]

\begin{scope}[every node/.style={opacity=1}]

\draw[coord axis] (-1.5*\orbitalRadius, 0, 0) node[above right=14pt and -4pt, align=left] {SUMMER \\[0.5ex] SOLSTICE} -- (1.6*\orbitalRadius, 0, 0) node[below left=14pt and -10pt, align=right] {WINTER \\[0.5ex] SOLSTICE};

\draw[coord axis] (0, -2.4*\orbitalRadius, 0) node[below left=9pt and -30pt, align=left] {AUTUMNAL \\[0.5ex]EQUINOX} -- (0, 2.75*\orbitalRadius, 0) node[above left=-4pt and 4pt, align=right] {VERNAL \\[0.5ex]EQUINOX};

\draw[coord axis] (0, 0, -1.5*\orbitalRadius) -- (0, 0, 1.7*\orbitalRadius) node[above=6pt, align=center] {ORBITAL AXIS};

\end{scope}

\draw (0, -\orbitalRadius) arc[radius=\orbitalRadius, start angle=-90, end angle=90];
\draw (0, \orbitalRadius) arc[x radius=\ellipseElongation*\orbitalRadius, y radius=\orbitalRadius, start angle=90, end angle=270];

\begin{scope}[Planet Color]
\node at (83:1.16*\orbitalRadius) {{\boldmath $Q1$}};
\node at (130:1.27*\orbitalRadius) {{\boldmath $Q2$}};
\node at (254:1.24*\orbitalRadius) {{\boldmath $Q3$}};
\node at (300:1.23*\orbitalRadius) {{\boldmath $Q4$}};
\fill (0:\orbitalRadius) circle[radius=\pointRadius] (90:\orbitalRadius) circle[radius=\pointRadius] (180:\ellipseElongation*\orbitalRadius) circle[radius=\pointRadius] (270:\orbitalRadius) circle[radius=\pointRadius];
\end{scope}

\coordinate (PLANET) at (\dayOfYearAngle:\orbitalRadius);

\draw[orbital arrow] (60:1.15*\orbitalRadius) arc [start angle=60, end angle=75, radius=1.15*\orbitalRadius];
\draw[orbital arrow] (220:{1.15*\ellipseElongation*\orbitalRadius} and {1.15*\orbitalRadius}) arc[x radius=1.15*\ellipseElongation*\orbitalRadius, y radius=1.15*\orbitalRadius, start angle=220, end angle=238];

\draw[single arrow=3mm] (0.4*\orbitalRadius, 0) arc[start angle=0, end angle=\dayOfYearAngle, radius=0.4*\orbitalRadius] node[pos=0.5, left=4pt] {$\psi$};

\draw[planet coord axis] (PLANET) -- +(10, 0, 0) node[right] {$x$};
\draw[planet coord axis] (PLANET) -- +(0, 30, 0) node[above] {$y$};
\draw[planet coord axis] (PLANET) -- +(0, 0, 20) node[above right=-3pt] {$z$};

\draw[gamma plane,->-=0.15] (O) -- ++($ 1.75*(PLANET) $) -- ++(0, 0, 14) -- (0, 0, 14) node[pos=0.88, below=6pt, sloped, node font=\large] {Plane $\Gamma$};
\path (O) -- (PLANET) node[Gamma Plane Color, above=2pt, pos=0.62] {$\vc{b}$};

\tdplotsetrotatedcoordsorigin{(PLANET)};

\tdplotsetrotatedthetaplanecoords{0};
\begin{scope}[tdplot_rotated_coords, Planet Coord Axis Color]
\draw[thick, double arrow=3mm] (\axialTilt:-11) node[below left=-2pt and -2pt] {$S$} -- (\axialTilt:12) node[above right=-3pt and -3pt] {$N$};
\draw[single arrow reverse=3mm] (4, 0) arc[start angle=0, end angle=\axialTilt, radius=4] node[pos=0.73, above] {$\alpha$} coordinate(ARC END);
\draw[dashed] (ARC END) arc[start angle=\axialTilt, end angle=90, radius=4];
\fill (ARC END) circle[radius=\pointRadius] (0, 4) circle[radius=\pointRadius];
\end{scope}

\tdplotsetrotatedthetaplanecoords{\dayOfYearAngle};
\begin{scope}[tdplot_rotated_coords, Gamma Plane Color]
\draw[thick, double arrow=3mm] (\effectiveAxialTilt:-12.5) node[below left=-2pt and -2pt] {$S'$} -- (\effectiveAxialTilt:12.5) node[above right=-3pt and -3pt] {$N'$};
\draw[single arrow reverse=3mm] (9, 0) arc[start angle=0, end angle=\effectiveAxialTilt, radius=9] node[pos=0.6, above=1pt] {$\alpha_1$} coordinate(ARC END);
\draw[dashed] (ARC END) arc[start angle=\effectiveAxialTilt, end angle=90, radius=9];
\fill (ARC END) circle[radius=\pointRadius] (0, 9) circle[radius=\pointRadius];
\end{scope}

\end{scope}

\fill[planet] (PLANET) circle[radius=\planetRadius];
\fill[Planet Color!80!black] (PLANET) circle[radius=\pointRadius];
\node[below right=9pt and 17pt] at (PLANET) {PLANET};

\fill[white] (O) circle[radius=\sunRayStart+\sunRayLength+0.3];
\foreach \a in {0, 45,...,315} \draw[sun ray, rotate=\a] (\sunRayStart, 0) -- +(\sunRayLength, 0);
\fill[sun] (O) circle [radius=\sunRadius];
\node[below right=14pt and 12pt] at (O) {$O$};

\node[info box, below right] at (-23, 24) {
	{\boldmath $\psi =$ day of year angle swept out, \\[0.5ex]
	$\Gamma =$ vertical plane through sun and planet, \\[0.5ex]
	$N'S' = NS$ axis rotated about \\ vector $\vc{b}$ into plane $\Gamma$, \\[0.5ex]
	$\alpha =$ true axial tilt (in $xz$-plane), \\[0.5ex]
	$\alpha_1 =$ effective axial tilt (in plane $\Gamma$)} \\[0.5ex]
};

\end{tikzpicture}
\end{center}
\end{figure}

Consider a day in the first quadrant $Q1$ as shown in Figure~\ref{fig:general-day}. The $xyz$-axes are fixed to the planet, but have constant direction as the planet orbits (note this is a different $xyz$-axis system than was used in Figures~\ref{fig:winter-solstice}--\ref{fig:winter-solstice-towards-sun-south} of \S\ref{sec:winter-solstice}). The vertical planes through the solstice and equinox axes intersect at right angles in the vertical line of the orbital axis passing through the sun \cite{wikipedia:ecliptic-longitude-se}. By rotating the $NS$ axis by some angle $\rho$ about vector $\bd{b}$ into plane $\Gamma$ the effective rotation axis $N'S'$ is produced, making effective axial tilt angle $\ao$ with the vertical. This is equivalent to simply viewing the sun and planet from a different angle, by rotating ourselves backwards by $\rho$ from plane $\Gamma$. After that rotation the situation is then exactly the same as the winter solstice of Figure~\ref{fig:winter-solstice}, except with the axial tilt angle $\ao$ instead of $\al$. Thus the required angle \te\ is given by equation~(\ref{eq:sunrise-formula-ws}) with \al\ replaced by $\ao$. Thus to obtain formula (\ref{eq:sunrise-formula}) we need to show $\sin \ao = \sin \al \cos \p$. \\

Let $\bd{a} = \sin \al\: \bd{i} + \cos \al\: \bd{k}$ be the $NS$ axis unit vector, and let $\bd{a'}$ be the rotated $N'S'$ axis unit vector. Then the required angle $\ao$ satisfies :
$$\cos \ao = \bd{a'} \cdot \bd{k}$$

Wrt right hand rule the rotation about $\bd{b} = \cos \p\: \bd{i} + \sin \p\: \bd{j}$ is by $-\rho$, thus from the Rodrigues Rotation Formula (Appendix~\ref{sec:rodrigues-rotation-formula}) :
$$\bd{a}' = (\cos \rho)\,\bd{a} + (\sin \rho)\,(\bd{a} \times \bd{b}) + (1 - \cos \rho)\,(\bd{b} \cdot \bd{a})\,\bd{b}$$

Viewing along the $\bd{b}$ axis the rotation angle $\rho$ satisfies :
$$\aPerp \cdot \bd{k} = |\,\aPerp| \cos \rho$$

where $\aPerp$ is the component of $\bd{a}$ perpendicular to $\bd{b}$. Since $\bd{a} \cdot \bd{b} = \sin \al \cos \p$,
\begin{eqnarray*}
\aPerp & = & \bd{a} - (\bd{a} \cdot \bd{b})\bd{b} \\
& = & \bd{a} - (\sin \al \cos \p)\bd{b} \\
& = & \sin \al \sin^2 \p\:\bd{i} - \sin \al \sin \p \cos \p\: \bd{j} + \cos \al\: \bd{k}
\end{eqnarray*}

Thus 
\begin{eqnarray*}
|\,\aPerp|^2 & = & \sin^2 \al \sin^4 \p + \sin^2 \al \sin^2 \p \cos^2 \p + \cos^2 \al \\
& = & \sin^2 \al \sin^2 \p + \cos^2 \al \\
& = & 1 - \sin^2 \al \cos^2 \p
\end{eqnarray*}

Since $\aPerp \cdot \bd{k} = \cos \al$ we then have (noting the case of the denominator zero ($\Rightarrow \p = 0$) has already been covered) :
\begin{eqnarray*}
\cos \rho & = & \frac{\cos \al}{\sqrt{1 - \sin^2 \al \cos^2 \p}} \\
\mbox{and,} \hspace{1em} \sin^2 \rho & = & 1 - \frac{\cos^2 \al}{1 - \sin^2 \al \cos^2 \p}
\end{eqnarray*}
\begin{eqnarray*}
& = & \frac{1 - \sin^2 \al \cos^2 \p - \cos^2 \al}{1 - \sin^2 \al \cos^2 \p} \\
& = & \frac{\sin^2 \al \sin^2 \p}{1 - \sin^2 \al \cos^2 \p}
\end{eqnarray*}

Using
\begin{eqnarray*}
\bd{a} \times \bd{b} & = &
\left|
\begin{array}{ccc}
\bd{i} & \bd{j} & \bd{k} \\
\sin \al & 0 & \cos \al \\
\cos \p & \sin \p & 0
\end{array}
\right| \\[1.5ex]
& = & -\cos \al \sin \p\: \bd{i} + \cos \al \cos \p\: \bd{j} + \sin \al \sin \p\: \bd{k}
\end{eqnarray*}

and the $\bd{k}$-components of the above Rodrigues formula for $\bd{a'}$ we obtain 

$$\bd{a'} \cdot \bd{k} = \cos \rho \cos \al + \sin \rho \sin \al \sin \p$$

Then, using the above expressions for $\cos \rho$ and $\sin \rho$,
\begin{eqnarray*}
\cos^2 \ao & = & (\bd{a'} \cdot \bd{k})^2 = \frac{(\cos^2 \al + \sin^2 \al \sin^2 \p)^2}{1 - \sin^2 \al \cos^2 \p} \\
& = & \frac{(1 - \sin^2 \al \cos^2 \p)^2}{1 - \sin^2 \al \cos^2 \p} = 1 - \sin^2 \al \cos^2 \p \\
\tf \sin^2 \ao & = & \sin^2 \al \cos^2 \p \\
\tf \sin \ao & = & \sin \al \cos \p
\end{eqnarray*}

as required. \\

For quadrant $Q4$, the situation at day of year angle $(360 - \p)\dg$ is exactly the same as for $Q1$ at day of year angle $\p$, with the same effective axial tilt $\ao$, just on the opposite side of the solstice axis. Thus \te\ must be the same, and then as $\cos(360\dg - \p) = \cos \p$, the formula (\ref{eq:sunrise-formula}) gives the correct result for $Q4$.

\newpage 

For quadrant $Q3$, the situation at day of year angle $(180 + \p)\dg$ is exactly the same as for $Q1$ at day of year angle $\p$ except with the sun in the exact opposite direction, and thus as with the case of deriving the summer solstice case from the winter solstice case above the sunrise direction \te\ just changes sign. Since $\cos(180\dg + \p) = -\cos \p$, formula (\ref{eq:sunrise-formula}) then gives the correct sunrise angle for $Q3$. \\

Finally for $Q2$, the situation at day of year angle $(180 - \p)\dg$ is exactly the same as for $Q3$ at day of year angle $(180 + \p)\dg$, just on the opposite side of the solstice axis. Thus \te\ must be the same, and then as $\cos(180\dg - \p) = \cos(180\dg + \p)$, formula (\ref{eq:sunrise-formula}) gives the correct sunrise angle for $Q2$. \\

\subsection{Properties of Curve}

For a given \al\ and \de\ the equation (\ref{eq:sunrise-formula}) in radians has form $f(x) = -\sin^{-1}(k \cos x)$, for $x \in [0, 2\pi]$, where $k = \sin \al / \cos \de$, and $k \in (0, 1)$ for non-polar latitudes (ie $|\de| < \frac{1}{2}\pi - \al$). Then :
\begin{eqnarray*}
f'(x) & = & \frac{k \sin x}{\surd(1 - k^2\cos^{2}x)} \\
\mbox{and} \hspace{2.5em} f''(x) & = & \frac{k (1 - k^2) \cos x}{(1 - k^2\cos^{2}x)^{3/2}}
\end{eqnarray*}

Thus the curve is `smooth' at non-polar latitudes, with the following `bell shape' gradient~: \\

{\setlength{\tabcolsep}{1em}
\renewcommand{\arraystretch}{1.5}
\begin{tabular}{c|ccccccccc}
$x$ & $0$ & & $\pi/2$ & & $\pi$ & & $3\pi/2$ & & $2\pi$ \\ \hline
$f'(x)$ & $0$ & +$\uparrow$ & $k$ & +$\downarrow$ & $0$ & --$\downarrow$ & $-k$ & --$\uparrow$ & $0$
\end{tabular}} \\[3ex]

At a polar circle (ie $\de = \pm(\frac{1}{2}\pi - \al)$), $k = 1$ and so in $Q1$ ($x \in [0, \frac{1}{2}\pi]$), $f(x) = -\sin^{-1}(\cos x) = x - \frac{1}{2}\pi$, ie a linear function. Then by reflecting into $Q2$, $Q3$, and $Q4$ as above the overall function is a triangle. In the plots in Appendix~\ref{sec:actual-sunrise-data} and in Figure~\ref{fig:sunrise-direction-curves} below it can be seen how this triangle is approached at the higher latitudes such as Reykjavik at $64.15\dg$. Beyond the polar circles, $k > 1$ and $f(x)$ is undefined at certain times of the year when there is perpetual day or night, namely where $|\cos x| > 1/k$.

\vspace{0.7ex}

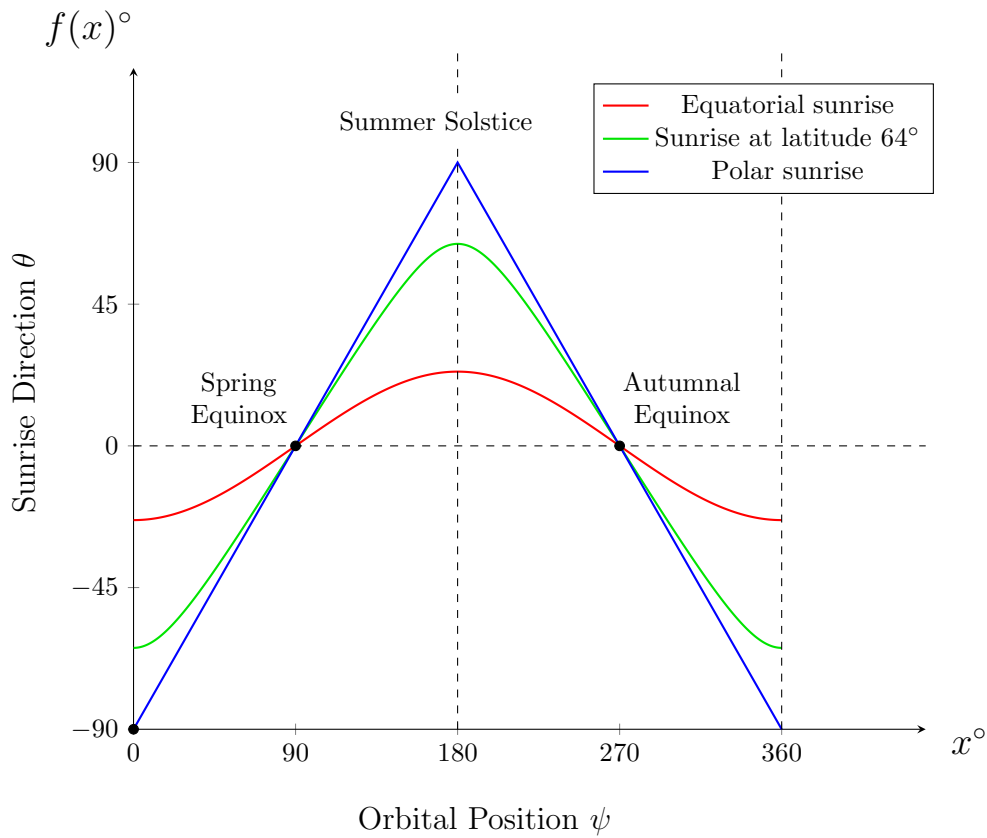
\begin{figure}[!h]
\begin{center}
\caption{Sunrise Direction Curves}
\label{fig:sunrise-direction-curves}
\vspace{3ex}
\begin{tikzpicture}[
	declare function={
		polarSunrise(\n) = {ifthenelse(\n<=180, \n - 90, 270 - \n)};
		kSunrise(\k, \n) = {-asin(\k*cos(\n))};
	},
]

\def\xtop{440}
\def\ytop{120}

\definecolor{Blue Color}{HTML}{0000ff}
\definecolor{Red Color}{HTML}{ff0000}
\definecolor{Green Color}{HTML}{00e400}

\begin{axis}[
	clip=false,
	axis x line=bottom,
	axis y line=left,	
	xmax=\xtop,
	ymax=\ytop,
	ymin=-90,
	align=center,
	xtick distance=90,
	xlabel={\Large $x\dg$},
	xlabel style={at={(1, 0)}, anchor=west, below right=-4pt and 6pt},
	ytick distance=45,
	ylabel={\Large $f(x)\dg$},
	ylabel style={at={(0, 1)}, anchor=center, above left=4pt and -3pt, rotate=-90},
	every axis legend/.append style={anchor=south, at={(axis cs:350, 80)}},
]

\addplot [
	Red Color,
	domain=0:360,
	samples=361,
	variable=n,
	line width=0.8pt
]
{kSunrise(0.4, n)};
\addlegendentry{Equatorial sunrise};

\addplot [
	Green Color,
	domain=0:360,
	samples=361,
	variable=n,
	line width=0.8pt
]
{kSunrise(0.9, n)};
\addlegendentry{Sunrise at latitude $64\dg$};

\addplot [
	Blue Color,
	domain=0:360,
	samples=361,
	variable=n,
	line width=0.8pt
]
{polarSunrise(n)};
\addlegendentry{Polar sunrise};

\filldraw (axis cs:0, -90) circle[radius=1.8pt];
\filldraw (axis cs:90, 0) circle[radius=1.8pt];
\filldraw (axis cs:270, 0) circle[radius=1.8pt];

\path (axis cs:195, -112) node[inner sep=3pt, anchor=north, node font=\large] {Orbital Position \p};
\path (axis cs:-50, 20) node[rotate=90, inner sep=3pt, anchor=south, node font=\large] {Sunrise Direction \te};

\draw[dashed] (axis cs:0, 0) -- (axis cs:\xtop, 0);
\draw[dashed] (axis cs:180, -90) -- (axis cs:180, \ytop+6);
\draw[dashed] (axis cs:360, -90) -- (axis cs:360, \ytop+6);

\path (axis cs:90, 3) node[inner sep=3pt, anchor=south east] {Spring \\ Equinox};
\path (axis cs:267, 3) node[inner sep=3pt, anchor=south west] {Autumnal \\ Equinox};
\path (axis cs:168, 95) node[fill=white, inner sep=6pt, anchor=south] {Summer Solstice};

\end{axis}

\end{tikzpicture}
\end{center}
\end{figure}

\newpage

\section{Solar Declination Formula}

\label{sec:solar-declination}

The solar declination angle \cite{wikipedia:declination} \la\ is the angle made by the direction of the sun from the planet's center with the planet's equatorial plane, one of the two coordinates of a celestial body within the equatorial coordinate system \cite{wikipedia:equatorial-cs}. From Figure~\ref{fig:winter-solstice} it is clear that $\la = -\al$ on the winter solstice. And throughout $Q1$ since the situation of the planet is as Figure~\ref{fig:winter-solstice} with $\al$ replaced with the effective axial tilt $\ao$, we must have $\la = -\ao$. Thus in $Q1$, $\la \in [-\al, 0]$ and :

\begin{equation}
\sin \la = -\sin \al \cos \p
\label{eq:solar-declination}
\end{equation}

In $Q4$, at angle $(360 - \p)\dg$, by symmetry \la\ is the same as at angle \p\ in $Q1$, and as $\cos (360 - \p) = \cos \p$, formula~(\ref{eq:solar-declination}) thus gives the correct result for $Q4$ when \p\ is substituted by $(360 - \p)\dg$. \\

In $Q3$ at $(180 + \p)\dg$ the sun is in the exact opposite direction in space wrt planet's center as it was in $Q1$ at \p, and the solar declination is $-\la \in [0, \al]$. But from (\ref{eq:solar-declination}) :
\[
\sin (-\la) = -\sin \al \, \cos(180 + \p)
\]
so that (\ref{eq:solar-declination}) gives the correct solar declination for $Q3$. \\

In $Q2$, at $(180 - \p)\dg$, by symmetry solar declination is as $Q3$ at $(180 + \p)\dg$, ie. $-\la \in [0, \al]$. But from (\ref{eq:solar-declination}) :
\[
\sin (-\la) = -\sin \al \cos(180 - \p),
\]
thus (\ref{eq:solar-declination}) gives the correct solar declination for $Q2$. \\

\noindent Thus throughout the orbit :
\begin{equation}
\la = -\arcsin (\sin \al \, \cos \p), \hspace{1em} \la \in [-\al, \al],
\label{eq:solar-declination-arcsin}
\end{equation}

\noindent and the sunrise direction formula can be written as :
\begin{equation}
\te = \arcsin \left(\frac{\sin \la}{\cos \de}\right), \hspace{1em} \te \in [-90\dg, 90\dg] \label{eq:sunrise-formula-lambda}
\end{equation}

\noindent The formulae in the sections below are expressed in terms of \la\ and equations (\ref{eq:solar-declination}) and (\ref{eq:solar-declination-arcsin}) relate \la\ to the day of the year via~\p. The relationship of \p\ to day of the year depends on the orbital model used, for example the two orbital models considered in \S\ref{sec:intro}.

\section{Sunrise Equation and Daylight Duration Formula}

\label{sec:sunrise-equation}

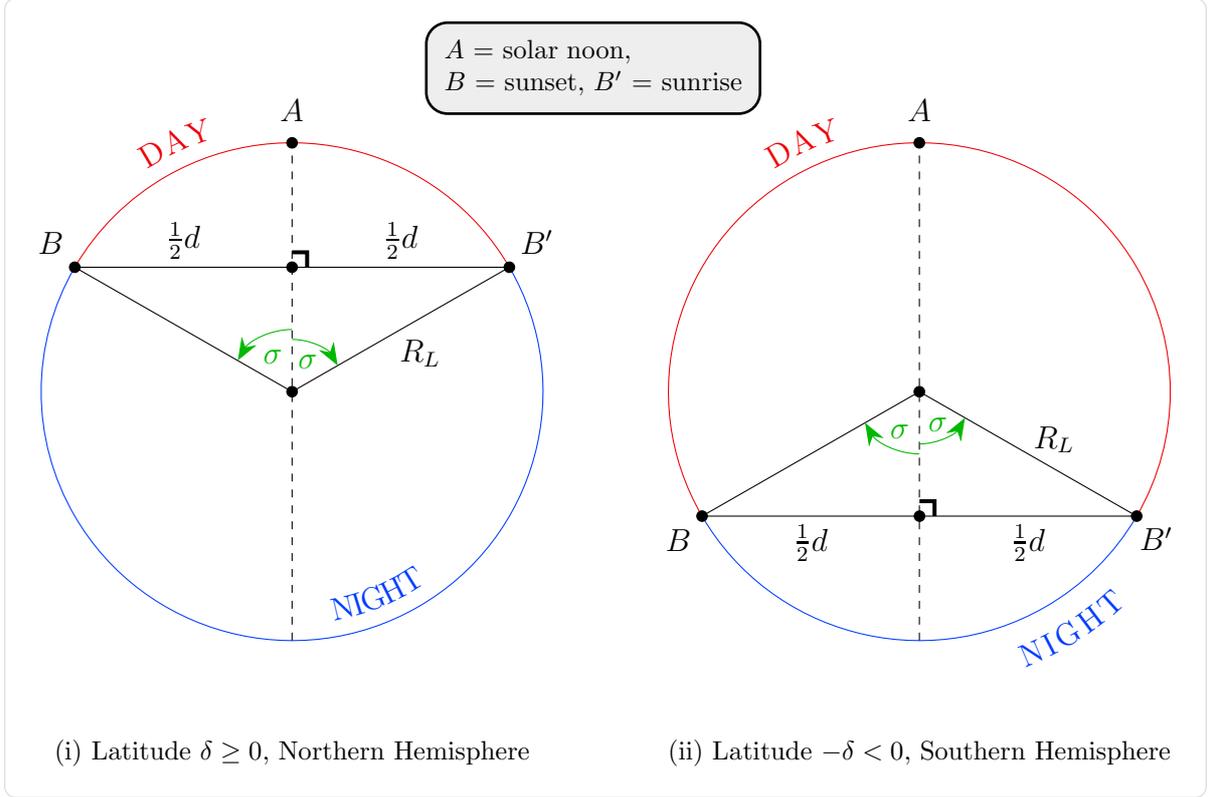
\begin{figure}[!h]
\begin{center}
\caption{Latitude Circle Hour Angles for Day/Night}
\label{fig:hour-angles}
\vspace{3ex}
\begin{tikzpicture} [
    scale=0.33,
    node font=\large,
	framed,
	background rectangle/.style={draw, rounded corners, Frame Color},
	inner frame sep=2ex,
	right angle style/.style = {line width=1.5pt},
	single arrow/.style=-{Stealth[length=#1]},
	title text/.style={node font=\normalsize},
	info box/.style={rounded corners=8pt, draw, fill=Info Box Color, inner sep=1.3ex, align=left, font=\normalsize, line width=1pt},
]

\definecolor{Angle Color}{HTML}{00b800}
\definecolor{Day Color}{HTML}{ef0006}
\definecolor{Night Color}{HTML}{003cff}
\definecolor{Info Box Color}{HTML}{eeeeee}
\definecolor{Frame Color}{HTML}{d8d8d8}

\def\RL{10}
\def\angleRadius{2.5}
\def\angleSigma{60}
\def\pointRadius{6pt}
\def\rightAngleSize{0.6}

\coordinate(O) at (0, 0);

\draw[Day Color] (90-\angleSigma:\RL) arc[radius=\RL, start angle=90-\angleSigma, end angle=90+\angleSigma];
\draw[Night Color] (90-\angleSigma:\RL) arc[radius=\RL, start angle=90-\angleSigma, delta angle=-360+2*\angleSigma];
\filldraw (O) circle[radius=\pointRadius];

\draw (O) -- (90+\angleSigma:\RL) coordinate(B) node[above left=1pt and 0pt] {$B$};
\filldraw (B) circle[radius=\pointRadius];
\draw (O) -- (90-\angleSigma:\RL) coordinate(B') node[pos=0.59, below=4pt] {$R_{L}$} node[above right=1pt and 0pt] {$B'$};
\filldraw (B') circle[radius=\pointRadius];
\draw[dashed, name path=noon radius] (-90:\RL) -- (90:\RL) coordinate(A) node[above=4pt] {$A$};
\filldraw (A) circle[radius=\pointRadius];

\draw[name path=chord, node font=\large] (B) -- (B') node[pos=0.25, above=-1pt] {$\frac{1}{2}d$} node[pos=0.75, above=-1pt] {$\frac{1}{2}d$};

\path [name intersections={of=noon radius and chord, by={C}}];
\filldraw (C) circle[radius=\pointRadius];

\draw[right angle style] (C) ++(\rightAngleSize, 0) -- ++(0, \rightAngleSize) -- ++(-\rightAngleSize, 0);

\draw[single arrow=0.3cm, Angle Color] (0, \angleRadius-0.4) arc [start angle=90, delta angle=-\angleSigma, radius=\angleRadius-0.4] node[pos=0.6, below left=-2pt and -2pt] {$\sigma$};
\draw[single arrow=0.3cm, Angle Color] (0, \angleRadius) arc [start angle=90, delta angle=\angleSigma, radius=\angleRadius] node[pos=0.6, below right=-1pt and -1pt] {$\sigma$};

\path [thick, postaction={decorate, decoration={raise=1.4ex, text along path, text color=Night Color, text align=center, text={|\large|NIGHT}}}] (0, -\RL) arc[radius=\RL, start angle=270, end angle=315];
\path [thick, postaction={decorate, decoration={raise=1.4ex, text along path, text color=Day Color, text align=center, text={|\large|DAY}}}] (140:\RL) arc[radius=\RL, start angle=140, end angle=90];

\begin{scope}[xshift=25cm]

\coordinate(O) at (0, 0);

\draw[Day Color] (270-\angleSigma:\RL) arc[radius=\RL, start angle=270-\angleSigma, delta angle=-360+2*\angleSigma];
\draw[Night Color] (270-\angleSigma:\RL) arc[radius=\RL, start angle=270-\angleSigma, delta angle=2*\angleSigma];
\filldraw (O) circle[radius=\pointRadius];

\draw (O) -- (270-\angleSigma:\RL) coordinate(B) node[below left=1pt and 0pt] {$B$};
\filldraw (B) circle[radius=\pointRadius];
\draw (O) -- (270+\angleSigma:\RL) coordinate(B') node[pos=0.62, above=2pt] {$R_{L}$} node[below right=0pt and -3pt] {$B'$};
\filldraw (B') circle[radius=\pointRadius];
\draw[dashed, name path=noon radius] (-90:\RL) -- (90:\RL) coordinate(A) node[above=4pt] {$A$};
\filldraw (A) circle[radius=\pointRadius];

\draw[name path=chord, node font=\large] (B) -- (B') node[pos=0.25, below=-1pt] {$\frac{1}{2}d$} node[pos=0.75, below=-1pt] {$\frac{1}{2}d$};

\path [name intersections={of=noon radius and chord, by={C}}];
\filldraw (C) circle[radius=\pointRadius];

\draw[right angle style] (C) ++(\rightAngleSize, 0) -- ++(0, \rightAngleSize) -- ++(-\rightAngleSize, 0);

\draw[single arrow=0.3cm, Angle Color] (0, -\angleRadius+0.4) arc [start angle=-90, delta angle=\angleSigma, radius=\angleRadius-0.4] node[pos=0.6, above left=-3pt and -3pt] {$\sigma$};
\draw[single arrow=0.3cm, Angle Color] (0, -\angleRadius) arc [start angle=-90, delta angle=-\angleSigma, radius=\angleRadius] node[pos=0.56, above right=-2pt and -2pt] {$\sigma$};

\path [thick, postaction={decorate, decoration={raise=-4ex, text along path, text color=Night Color, text align=center, text={|\large|NIGHT}}}] (0, -\RL) arc[radius=\RL, start angle=270, end angle=335];
\path [thick, postaction={decorate, decoration={raise=1.4ex, text along path, text color=Day Color, text align=center, text={|\large|DAY}}}] (140:\RL) arc[radius=\RL, start angle=140, end angle=90];
\end{scope}

\node[title text] at (0, -14.5) {(i) Latitude $\delta \geq 0$, Northern Hemisphere};
\node[title text] at (25, -14.5) {(ii) Latitude $-\delta < 0$, Southern Hemisphere};

\node[info box] at (12, 13) {
	$A$ = solar noon, \\
	$B$ = sunset, $B'$ = sunrise
};

\end{tikzpicture}
\end{center}
\end{figure}

Consider a day in $Q1$ as depicted in Figures \ref{fig:winter-solstice} and \ref{fig:winter-solstice-towards-sun} with \al\ replaced with the effective axial tilt \ao. The latitude circle is shown face on in Figure~\ref{fig:hour-angles}. $A$ is the point of solar noon on the latitude circle, $B$ is the point of sunset, and $B'$ is the point of sunrise. The angle \si\ is the sunrise (and sunset) solar hour angle \cite{wikipedia:solar-hour-angle} which is a measure of the time between the sunrise (or sunset) and solar noon \cite{wikipedia:solar-noon}. The time of solar noon is uniquely defined for any day of the year and for any arbitrary location $L$ on the planet except for the poles which are stationary throughout the day\footnote{To define solar noon at a pole a meridian longitude would have to be selected for the pole (see \S\ref{sec:solar-azimuth} re cardinal directions at the poles).} (in the simplified model). It is defined as the time when the unique half plane containing the location $L$ and hinged on the $NS$ axis intersects the sun --- solar midnight is the time when this half plane is in the exact opposite direction from that. The general angular position of the half plane through $L$ is a measure of time at $L$ throughout the day as the planet rotates at a constant angular velocity about the $NS$ axis. The solar hour angle \ta\ of $L$ at a given time is the angle of $L$'s half plane wrt the solar noon position, and is an angle in the range $[0\dg, 360\dg)$, with the +ve angle direction defined by the RH rule wrt axis $NS$. The solar hour angle is $0\dg$ at solar noon and $180\dg$ at solar midnight, and is negative at sunrise and positive at sunset. If the sun were placed in the exact opposite direction then the solar hour angle wrt the new sun position would be the previous solar hour angle plus $180\dg$ (this is used in \S\ref{sec:solar-altitude} below). \\

The latitude $-\de$ in the southern hemisphere is obtained from the latitude $+\de$ by rotating Figure~\ref{fig:winter-solstice} about $O$ by $180\dg$ so in Figure~\ref{fig:hour-angles} the day/night arc lengths in (ii) are just the reverse of those in (i). Considering case (i) of northern hemisphere the angle \si\ is in the range $[0, 90\dg]$ depending on \ao\ and \de, eg it is $90\dg$ for all \de\ at the spring equinox (12 hours of day and 12 hours of night) and is $0\dg$ at the artic circle on the winter solstice where solar noon, sunrise, and sunset all coincide. If $R_{L}$ is radius of latitude circle then from Figure~\ref{fig:winter-solstice} :
\begin{eqnarray*}
\frac{R_{L}}{R} & = & \sin (\ao + \bt) = \cos \de \\
\Rightarrow R_{L} & = & R \cos \de
\end{eqnarray*}

and from Figure~\ref{fig:hour-angles} we have $\sin \si = \frac{1}{2}d/R_{L}$. But from Figure~\ref{fig:winter-solstice-towards-sun}, $\frac{1}{2}d = OC$, and from \S\ref{sec:Latitude-North} :
\begin{eqnarray}
OC & = & \frac{R}{\cos \ao} \sqrt{\cos^2 \ao - \sin^2 \de} \nonumber \\
\tf \sin \si & = & \frac{OC}{R \cos \de} = \frac{\sqrt{\cos^2 \ao - \sin^2 \de}}{\cos \ao \cos \de} \nonumber \\
\tf \cos^2 \si & = & \frac{\cos^2 \ao \cos^2 \de - \cos^2 \ao + \sin^2 \de}{\cos^2 \ao \cos^2 \de} = \frac{\sin^2 \de \sin^2 \ao}{\cos^2 \ao \cos^2 \de} = \tan^2 \de \tan^2 \ao \nonumber \\
\tf \cos \si & = & \tan \ao \tan \de \nonumber \\
& = & -\tan \la \, \tan \de \nonumber \\
\tf \si & = & \arccos (-\tan \la \, \tan \de) \label{eq:sunrise-hour-angle}
\end{eqnarray}

The identity (\ref{eq:sunrise-hour-angle}) also applies in case (ii) of the southern hemisphere, at latitude $-\de$, since from Figure~\ref{fig:hour-angles}(ii) the required sunrise solar hour angle is $180\dg - \si$, and $\arccos$ as a function from $[-1, 1]$ to $[0, 180\dg]$ satisfies the condition  $\arccos(-x) = 180\dg - \arccos(x)$. The equation (\ref{eq:sunrise-hour-angle}) is called the `Sunrise Equation' \cite{wikipedia:sunrise-equation}, \cite{web:solar-geometry}. The argument to the $\arccos$ will go out of range for latitudes beyond the effective polar circles (ie the polar circles corresponding to the effective axial tilt), for at these latitudes and times there is no sunrise nor sunset. \\

We have shown (\ref{eq:sunrise-hour-angle}) holds for any day in $Q1$, using Figures \ref{fig:winter-solstice} and \ref{fig:winter-solstice-towards-sun} with an effective axial tilt $\ao \in [0, \al]$. For a day in $Q4$ at angle $(360 - \p)$, by symmetry the situation of the planet is the same as for a day in $Q1$ at angle \p, thus \si\ is the same. Then since $\cos (360 - \p)\dg = \cos \p$ and from (\ref{eq:solar-declination-arcsin}) the solar declination \la\ is the same, equation~(\ref{eq:sunrise-hour-angle}) produces the correct sunrise solar hour angle \si\ for $Q4$. \\

For a day in $Q3$ at angle $(180 + \p)\dg$ and with solar declination $-\la$, the geometric situation of the planet is identical to a day in $Q1$ at angle \p\ with solar declination $\la$ ($< 0$), except with the sun's light coming from exactly the opposite direction from that in Figure~\ref{fig:winter-solstice}. For the day in $Q1$, equation (\ref{eq:sunrise-hour-angle}) gives the value of \si\ with the sun at its \emph{normal side} of Figure~\ref{fig:winter-solstice}, and then the required sunrise solar hour angle for the sun at the opposite side is $(180 - \si)\dg$, since the day and night arcs on the latitude circles will be reversed. Thus the required sunrise solar hour angle for the day in $Q3$ is $(180 - \si)\dg$. But from (\ref{eq:sunrise-hour-angle}), for any latitude \de, we have :
\[
180\dg - \si = \arccos( \tan (-\la) \, \tan \de)
\]

and so (\ref{eq:sunrise-hour-angle}) gives the correct sunrise solar hour angle for the day in $Q3$ of solar declination -\la. \\

For a day in $Q2$ at angle $(180 - \p)\dg$, by symmetry the situation of the planet is the same as for a day in $Q3$ at angle $(180 + \p)\dg$, and so \si\ is the same. Then since $\cos (180 - \p)\dg = \cos (180 + \p)\dg$ and the solar declination \la\ from (\ref{eq:solar-declination-arcsin}) is the same, the equation (\ref{eq:sunrise-hour-angle}) produces the correct sunrise solar hour angle \si\ for $Q2$. \\

Thus (\ref{eq:sunrise-hour-angle}) at any point in the orbit gives the required sunrise solar hour angle $\si \in [0, 180\dg]$. Writing $\tan \la$ in terms of the orbital angle \p, (\ref{eq:solar-declination-arcsin}) and (\ref{eq:sunrise-hour-angle}) give : \\
\begin{eqnarray}
\tan \la & = & \pm \frac{\sin \la}{\surd (1 - \sin^2 \la)} = -\frac{\sin \al \, \cos \p}{\surd{(1 - \sin^2\al \, \cos^2 \p)}} \hspace{0.5em} \mbox{(noting declination is -ve in $Q1$ and $Q4$)} \nonumber \\
\mbox{and} \hspace{0.5em} \si & = & \arccos \left(\tan \de \cdot \frac{\sin \al \, \cos \p}{\surd{(1 - \sin^2\al \, \cos^2 \p)}}\right) \label{eq:sunrise-hour-angle-expanded}
\end{eqnarray}

To convert the total solar hour angle $2\si$ of daylight to the daylight duration $D$ we simply multiply the rotational period $T$ of the planet by the fraction $2\si / 360\dg = \si / 180\dg$ :
\begin{equation}
D = T \, \frac{\si}{180\dg}
\label{eq:daylight-duration}
\end{equation}

Although we are using the model of a stationary day, we could use as $T$ the mean solar day \cite{wikipedia:solar-time}, \cite{wikipedia:day-types}, of 24 hours duration in the case of Earth, rather than the sidereal day \cite{wikipedia:sidereal-time} length of 23.93447 hours, to compensate for the prograde motion of the Earth in its orbit during the day, which means it takes slightly longer than a sidereal day for the sun to return to the local meridian position at solar hour angle $0\dg$. A planet with a retrograde rotation would have a mean solar day shorter than the sidereal day. \\

In the case of Earth two sources of inaccuracy in equation~(\ref{eq:daylight-duration}) cause underestimation of the daylight duration $D$ :
\begin{itemize}
\item The non-zero angular diameter \cite{wikipedia:angular-diameter} of the solar disc, approximately $0.5\dg$, and the convention of measuring sunrise and sunset wrt the upper edge of the solar disc crossing the horizon means official sunrise time comes a few minutes before and official sunset time a few minutes after the times predicted by the simple geometric model.
\item The effect of atmospheric refraction \cite{wikipedia:atmos-refraction} of light from an object in space arriving at an observer on Earth causes the altitude of the object appear slightly higher than it really is. This is because entering the slower medium of the atmosphere from the vacuum causes the light ray to bend towards the normal between the two media, ie the angle of incidence reduces, and this implies a downwards bending, thus the eye sees the light source as higher up, similarly to a stick under water appearing higher up. This means the Sun can be seen shortly before sunrise and shortly after sunset, whilst its upper edge is still geometrically below the horizon, again lengthening the day.
\end{itemize}

\cite{jenkins:sun-position} provides details on adjustments which can be applied to (\ref{eq:daylight-duration}) to compensate for these two factors. \\

\noindent Note: the two sunrise angles \si\ and \te\ should not be confused --- \si\ is the angle of the sunrise wrt solar noon, whilst \te\ is the angle of the sunrise wrt due east on the horizon --- the former is an angle in the latitude plane and the latter is an angle in the horizon plane. Both are functions of time of year and latitude and from the equations (\ref{eq:sunrise-formula-lambda}) and (\ref{eq:sunrise-hour-angle}) $\cos \si / \sin \te = -\sin \de / \cos \la$, which varies throughout the year.

\section{Clock Time Versus Solar Time}

\label{sec:clock-time}

In this section we relate the clock time CT (or civil time) to the solar time for the Earth\footnote{We shall assume a 24 hour clock.}. The solar time is based on the daily passage of the Sun, with a day defined as the time between successive solar noons (\S\ref{sec:sunrise-equation}). This is refered to as the apparent solar time (AST) or true solar time \cite{wikipedia:AST}, and its day length varies with the time of the year, by about -20/+30 seconds about a mean value of 24 hours. AST is the type of time measured by a sundial, as it is purely dependent on the position of the Sun which is recorded by the sundial's shadow. The mean value of 24 hours forms the basis of the mean solar time (MST), and a mean solar day is of fixed duration 24 hours. As the successive daily differences accumulate the time difference $\AST - \MST$ varies within a range of about -14/+16 minutes according to the `equation of time' \cite{wikipedia:EOT}. \\

Clock time in the various time zones is based on the UTC time standard, which is within 0.9 seconds of another standard called UT1 \cite{usno:UT} which measures MST at longitude $0\dg$ (UT1 replaces the older GMT standard). Thus unless greater than this level of accuracy is required UTC can be taken as the MST at longitude $0\dg$, \cite{usno:EOT}. Each time zone is then defined as UTC plus or minus an integral number of hours\footnote{In a few time zones, eg in India, a non-integral number is used.}, and has a defining meridian \cite{usno:EOT} located at longitude the corresponding integral multiple of $15\dg$. The time zone time, ie the clock time CT, or civil time, shared by each location in the time zone, is defined as the MST at the longitude of the defining meridian for the zone. To determine the MST for any other location in the time zone we need to adjust by 4 minutes for every degree of longitude difference $\Delta L\dg$ from the longitude of the defining meridian for the zone, with the easterly direction being positive \cite{usno:EOT}, ie.
\begin{eqnarray}
\MST & = & \CT + 4\, \mbox{minutes} \times \Delta L\dg \nonumber \\
\tf \MST & = & \CT + \frac{\Delta L\dg}{15} \hspace{0.8em} \mbox{(in decimal hours)} \label{eq:mst-ct}
\end{eqnarray}

Daylight Saving Time (DST) is an adjustment to civil clock time of +1 hour for the summer months (eg from March to October in UK), but here we will define the civil clock time CT to be exclusive of any DST changes. \\

When requiring the time of day in formulae (eg the solar hour angle parameters \si\ in the sunrise equation in \S\ref{sec:sunrise-equation} and \ta\ in the altitude and azimuth formulae in \S\ref{sec:solar-altitude} and \S\ref{sec:solar-azimuth}) the AST is convenient, since geometrically it corresponds straightforwardly with the solar hour angle parameter, with 24 hours of AST being equivalent to a solar hour angle $360\dg$. AST of 12 noon is solar noon, and AST of 12 midnight is solar midnight. An `hour' of AST would be a 24\textsuperscript{th} part of the solar day and would vary from day to day, within a range of approximately 3599 to 3601 atomic seconds. To obtain the time of day solar hour angle \ta\ from AST :
\begin{eqnarray}
(\AST - 12) & = & \frac{\ta}{360} \cdot 24 = \frac{\ta}{15} \nonumber \\
\tf \ta & = & 15\,(\AST - 12), \hspace{0.8em} \mbox{AST in decimal hours, \ta\ in degrees} \label{eq:tau-ast}
\end{eqnarray}

To obtain \ta\ from the clock time CT we can use (\ref{eq:mst-ct}) and (\ref{eq:tau-ast}) together with the equation of time (EOT) which expresses AST in terms of MST, \cite{wikipedia:EOT}, \cite{usno:EOT}, \cite{hughes:EOT}, and which is a continuous function of time throughout the year, repeating itself over a yearly cycle. We can approximate EOT as a constant for each day of the year $d$ so that $\EOT(d)$ equals a constant difference $\AST - \MST$ throughout the day $d$. $\EOT(d)$ can thus produce AST from a known MST, and vice-versa. The difference $\EOT(d)$ has two components : $\EC(d)$ the difference due to the eccentricity of the Earth's orbit, and $\OB(d)$ the difference due to the obliquity of the ecliptic plane. $\EC(d)$ is an approximate sine wave of period one year with zeros at perihelion and aphelion, which are respectively about 2 weeks after the winter and summer solstices (the exact day varying with the year), \cite{timeanddate:perihelion}. $\OB(d)$ is an approximate sine wave of period a half year with zeros at the equinoxes and solstices. Using respective amplitudes of 7.66 mins and 9.87 mins for these from \cite{wikipedia:EOT}, and taking a perihelion/aphelion offset from the winter/summer solstice of 14 days (the 2020 figure --- other years will require a different value \cite{timeanddate:perihelion}, \cite{timeanddate:SE}), and refering to the graphs in \cite{wikipedia:EOT}, \cite{hughes:EOT}, approximate formulae for the discretized $\EC(d)$ and $\OB(d)$ in decimal hours for the day which is at offset $d$ from the winter solstice are :
\begin{eqnarray}
\EC(d) & = & -0.1277 \, \sin \left( \frac{d - 14}{365} \cdot 360\dg \right) \nonumber \\
\OB(d) & = & -0.1645 \, \sin \left( \frac{d}{365} \cdot 720\dg \right) \nonumber \\
\mbox{and} \hspace{0.8em} \EOT(d) & = & \EC(d) + \OB(d) \hspace{0.8em} \mbox{(in decimal hours)} \label{eq:EOT-formula}
\end{eqnarray}

Using the relation $\EOT(d) = \AST - \MST$ and equations~(\ref{eq:mst-ct}) and (\ref{eq:tau-ast}) we then have :
\begin{eqnarray}
\ta & = & 15\,(\MST + \EOT(d) - 12) \nonumber \\
& = & 15\,(\CT + \frac{\Delta L\dg}{15} + \EOT(d) - 12) \nonumber \\
\Rightarrow \ta & = & 15\,\CT - 180\dg + \Delta L\dg + 15\,\EOT(d) \hspace{0.8em} \mbox{(CT, EOT in decimal hours, \ta\ in degrees)} \label{eq:tau-ct}
\end{eqnarray}

This is an approximate relation from clock time CT (ex-DST) to solar hour angle \ta\ for the day $d$, and is applicable to all time zones. Values for the approximate $\EOT(d)$ can be calculated using the above sine wave formulae or obtained from online calculators such as \cite{minasi:EOT-calc} and \cite{mbsoft:EOT-calc}. Note the approximation is based on a discretized form of the EOT --- however it is sufficiently accurate to demonstrate the effect of the analemma and to generate approximate analemma graphs (\S\ref{sec:analemma}).

\newpage

\section{Sunrise, Sunset, and Solar Noon Times}

\label{sec:sunrise-time}

The solar noon local clock time CT\textsubscript{0} for Earth is derived from equation~(\ref{eq:tau-ct}) by putting $\ta = 0$ :
\begin{equation}
\CT_{0} = 12 - \frac{\Delta L\dg}{15} - \EOT(d) \label{eq:solar-noon-time}
\end{equation}

If $\si \in [0, 180\dg]$ is the sunrise/sunset solar hour angle of \S\ref{sec:sunrise-equation} then in equation~(\ref{eq:tau-ct}) $\ta = -\si$ at sunrise clock time CT\textsubscript{R}, and $\ta = \si$ at sunset clock time CT\textsubscript{S}. Then from (\ref{eq:tau-ct}) :
\begin{eqnarray}
\CT_{R} & = & \CT_{0} - \frac{\si}{15} \label{eq:sunrise-time} \\
\CT_{S} & = & \CT_{0} + \frac{\si}{15} \label{eq:sunset-time}
\end{eqnarray}

with \si\ given by equation~(\ref{eq:sunrise-hour-angle}) or (\ref{eq:sunrise-hour-angle-expanded}). \\

\noindent \textbf{Example} Consider the location Madrid on the date 15\textsuperscript{th} May 2019. Longitude is $-3.72\dg$ and latitude $\de = 40.42\dg$. The time zone is UTC+1 (ex-DST) with defining meridian longitude at $+15\dg$. Thus $\Delta L\dg = -18.72\dg$. Day offset from winter solstice 21\textsuperscript{st} Dec 2018 is $d = 10 + 31 + 28 + 31 + 30 + 15 = 145$. Assuming a circular orbit/constant speed model to calculate an approximation to \p, we have $\p =(d / 365) \cdot 360\dg = 143.0\dg$. Using EOT formula (\ref{eq:EOT-formula}), $\EOT(d) = 0.05917$ (the above EOT calculators \cite{minasi:EOT-calc} and \cite{mbsoft:EOT-calc} give values within about 3\% of this). Then from (\ref{eq:solar-noon-time}) :
\begin{eqnarray*}
\CT_{0} & = & 12 + \frac{18.72}{15} - 0.05917 = \mbox{13.19 hours = 13:11:24 (ex-DST)} \\
\mbox{Actual } \CT_{0} & = & \mbox{14:11:00 (DST) = 13:11:00 (ex-DST)}
\end{eqnarray*}

and from equation~(\ref{eq:sunrise-hour-angle-expanded}) :
\begin{eqnarray*}
\si & = & \arccos \left( \frac{\tan 40.42 \cdot \sin 23.44 \cdot \cos 143.0}{\surd(1 - \sin^2 23.44 \cdot \cos^2 143.0)} \right) \\
& = & 106.6\dg \\
\Rightarrow \CT_{R} & = & 13.19 - 106.6/15 = \mbox{ 6.083 hours = 06:04:59 (ex-DST)} \\
\mbox{and} \hspace{0.8em} \CT_{S} & = & 13.19 + 106.6/15 = \mbox{ 20.30 hours = 20:18:00 (ex-DST)} \\
\mbox{Actual } \CT_{R} & = & \mbox{06:58:00 (DST) = 05:58:00 (ex-DST)} \\
\mbox{Actual } \CT_{S} & = & \mbox{21:23:00 (DST) = 20:23:00 (ex-DST)}
\end{eqnarray*}

\noindent where the actual sunrise, sunset, and solar noon times are obtained from \cite{timeanddate:sun-calc-spain}. The actual sunrise comes about 7 minutes before the time predicted by the model, and the actual sunset comes about 5 minutes after the time predicted by the model --- this correlates with the two sources of discrepancy described in \S\ref{sec:sunrise-equation}. The sunrise/sunset times given by \cite{timeanddate:sun-calculator} take into account the effect of refraction \cite{timeanddate:refraction}. The solar noon prediction by contrast was more accurate and is not impacted by the above two factors.

\section{Solar Altitude Formula}

\label{sec:solar-altitude}

The solar altitude angle \mv\ \cite{wikipedia:solar-zenith-angle} is the angle the line from planet to sun makes with an observer's local horizon plane. It is well-defined at all points on the planet at all times and is in the range $[-90\dg, 90\dg]$, with the sun visible for $\mv \geq 0$ and invisible for $\mv < 0$, and with $\mv = 0$ at sunrise and sunset. The formula derived here gives \mv\ as a function of latitude \de, the day of the year (via \la, as described in \S\ref{sec:solar-declination}), and \ta\ the time of day expressed as the solar hour angle, \S\ref{sec:sunrise-equation} and \cite{wikipedia:solar-hour-angle}, wrt solar noon \cite{wikipedia:solar-noon}. Solar noon time can be looked up for any location and day in \cite{timeanddate:sun-calculator} or calculated using (\ref{eq:solar-noon-time}). \ta\ is negative before solar noon and positive after. \\

Assume we are at a day in $Q1$ with effective axial tilt $\ao = -\la$, where \la\ is the solar declination (\S\ref{sec:solar-declination}) --- ie. in Figure~\ref{fig:winter-solstice}, \al\ is replaced by \ao\ --- and consider a local horizontal coordinate system \cite{wikipedia:horizon-cs} at the point $A$ in Figure~\ref{fig:winter-solstice} at latitude \de\, with orthogonal right-handed basis vectors $\bd{u}$ (east), $\bd{v}$ (north), $\bd{w}$ (upwardly vertical) given (in terms of the $xyz$-coordinate system of Figure~\ref{fig:winter-solstice}) by $\bd{u} = (0, 0, 1)$, $\bd{v} = (\cos \bt, \sin \bt, 0)$, $\bd{w} = (-\sin \bt, \cos \bt, 0)$. Note $A$ could be any point on the half circumference of the planet including the poles $N$, $S$ and $\bt \in [-\al, 180\dg - \al]$. The other half circumference can be obtained by rotating about $NS$ by $180\dg$ --- this ensures $\bd{v}$ always points north (see \S\ref{sec:solar-azimuth} re cardinal directions at the poles). \\

The corresponding horizontal coordinate system basis vectors $\bd{u'}$ (east), $\bd{v'}$ (north), $\bd{w'}$ (upwardly vertical) for the point on latitude \de\ of Figure~\ref{fig:winter-solstice} which is at solar hour angle \ta\ wrt solar noon can be obtained by rotating $\bd{u}$, $\bd{v}$, $\bd{w}$ by angle \ta\ about the polar axis $NS$ according to the RH rule. Then at this angle \ta, if $\bd{s} = -\bd{i}$ is the direction from the planet to the Sun, making altitude angle $\mv \in [-90\dg, 90\dg]$ with the $\bd{u'}\bd{v'}$-plane, then :
\begin{eqnarray}
\mv \geq 0 & \Rightarrow & \bd{s} \cdot \bd{w'} = \cos(90\dg - \mv) = \sin \mv \nonumber \\
\mbox{and } \mv < 0 & \Rightarrow & \bd{s} \cdot \bd{w'} = \cos(90 + |\mv|) = -\sin |\mv| = \sin \mv \nonumber \\
\tf \forall \mv \in [-90\dg, 90\dg],\; \sin \mv & = & \bd{s} \cdot \bd{w'} = -\bd{i} \cdot \bd{w'} \label{eq:sine-mu-initial}
\end{eqnarray}

Since in Figure~\ref{fig:winter-solstice} the $NS$ rotation axis $\bd{a} = (\sin \ao, \cos \ao, 0)$, the Rodrigues Rotation Formula (Appendix~\ref{sec:rodrigues-rotation-formula}) gives~:
\begin{eqnarray*}
\bd{w'} & = & (\cos \ta)\bd{w} + (\sin \ta)(\bd{a} \times \bd{w}) + (1 - \cos \ta)(\bd{a} \cdot \bd{w})\bd{a} \\
\bd{a} \times \bd{w} & = &
\left|
\begin{array}{ccc}
\bd{i} & \bd{j} & \bd{k} \\
\sin \ao & \cos \ao & 0 \\
-\sin \bt & \cos \bt & 0
\end{array}
\right| 
= (0, 0, \sin \ao \, \cos \bt + \cos \ao \, \sin \bt) \\[1.5ex]
& = & (0, 0, \sin(\ao + \bt)) = (0, 0, \cos \de), \hspace{0.8em} \mbox{and } \bd{a} \cdot \bd{w} = \cos(\ao + \bt) = \sin \de \\
\tf \bd{w'} & = & \cos \ta(-\sin \bt, \cos \bt, 0) + \sin \ta(0, 0, \cos \de) + (1 - \cos \ta)\sin \de(\sin \ao, \cos \ao, 0) \\
\tf \sin \mv & = & -\bd{w'} \cdot \bd{i} = \cos \ta \, \sin \bt - (1 - \cos \ta) \sin \de \, \sin \ao
\end{eqnarray*}

Since $\de = 90 - (\ao + \bt)$, $\sin \bt = \cos(\ao + \de) \tf$
\begin{eqnarray}
\sin \mv & = & \cos \ta(\cos \ao \, \cos \de - \sin \ao \, \sin\ \de) - (1 - \cos \ta)\sin \de \sin \ao \nonumber \\
\tf \sin \mv & = & \cos \ta \, \cos \ao \, \cos \de - \sin \de \, \sin \ao \nonumber \\
\tf \mbox{at any point in $Q1$}, \hspace{0.8em} \sin \mv & = & \cos \ta \, \cos \de \, \cos \la + \sin \de \, \sin \la, \hspace{0.8em} \label{eq:solar-altitude} \mv \in [-90\dg, 90\dg]
\end{eqnarray}

Note the above derivation is valid for any latitude on the planet, including the poles which are stationary throughout the day with $\de = \pm90\dg$ and the formula (\ref{eq:solar-altitude}) reducing to $\mv = \la$ (for N) and $\mv = -\la$ (for S) which is clear intuitively as the horizon plane at a pole is parallel to the equatorial plane. \\

The above derivation was based on Figure~\ref{fig:winter-solstice} which covers case $Q1$ using the effective axial tilt \ao\ in place of \al\ in the diagram. In $Q4$ at $(360 - \p)\dg$, the situation of the planet wrt the sun is as $Q1$ at angle \p\, so \mv\ must be the same. Since \la\ given by (\ref{eq:solar-declination-arcsin}) is also the same, the angle $\mv \in [-90\dg, 90\dg]$ given by (\ref{eq:solar-altitude}) is then correct for $Q4$, for any latitude \de\ and time \ta. \\

In $Q3$ the situation of the planet at angle $(180 + \p)\dg$ is as $Q1$ at angle \p, except with the sun at the exact opposite side. Considering this symmetric $Q1$ position, with the sun at its \emph{usual side}, as depicted in Figure~\ref{fig:general-day}, the solar hour angle is $\ta + 180$, where \ta\ is the solar hour angle for the $Q3$ situation (consider the half planes which define the solar hour angles and that the solar hour angle +ve direction is always given from the $NS$ axis via the RH rule --- thus to go from a sun at one side to the complete opposite side add $180\dg$ to the solar hour angle), and the solar declination is $-\la$ where \la\ is the solar declination for the $Q3$ position. Then, writing equation~(\ref{eq:solar-altitude}) for this $Q1$ point we have :
\begin{eqnarray}
\sin \mv & = & \cos(\ta + 180) \, \cos \de \, \cos(-\la) + \sin \de \, \sin(-\la), \\
\mbox{and hence} \hspace{0.8em} \sin(-\mu) & = & \cos \ta \, \cos \de \, \cos \la + \sin \de \, \sin \la
\end{eqnarray}
where \mv\ is the solar altitude for the $Q1$ point, and $-\mu$ is the required solar altitude for the point in $Q3$. Thus equation~(\ref{eq:solar-altitude}) gives the correct solar altitude for the $Q3$ point. \\

In $Q2$ the situation of the planet at angle $(180 - \p)\dg$ is as $Q3$ at angle $(180 + \p)\dg$ so \mv\ must be the same. Since $\cos(180 - \p)\dg = \cos(180 + \p)\dg$ and \la\ given by (\ref{eq:solar-declination-arcsin}) is the same, the angle $\mv \in [-90\dg, 90\dg]$ given by (\ref{eq:solar-altitude}) is then correct for $Q2$, for any latitude \de\ and time \ta. \\

Thus equation~(\ref{eq:solar-altitude}) gives the solar altitude $\mv \in [-90\dg, 90\dg]$ at all latitudes \de\ (including the poles), at all times of the year, and all times \ta\ of the solar day. The solar declination parameter \la\ depends on the day of the year as described in \S\ref{sec:solar-declination}. Equation~(\ref{eq:solar-altitude}) also gives an alternate route to the Sunrise Equation~(\ref{eq:sunrise-hour-angle}) by setting altitude $\mv = 0$ in (\ref{eq:solar-altitude}) and then solving for solar hour angle \ta. \\

\subsection{Sun Directly Overhead}

\label{sec:sun-overhead}

From equation~(\ref{eq:solar-altitude}), since the rhs is maximal at $\ta = 0$ the sun can only be directly overhead when $\ta = 0$, ie at solar noon. The condition for that to happen for a day is then :
\begin{eqnarray}
\sin 90\dg & = & \cos \de \, \cos \la + \sin \de \, \sin \la \nonumber \\
\Rightarrow \hspace{3em} 1 & = & \cos(\de - \la) \nonumber \\
\tf \mbox{as } \de - \la \in (-180\dg, 180\dg), \hspace{0.8em} \de - \la & = & 0 \nonumber \\
\mbox{or} \hspace{0.8em} \sin \de & = & \sin \la \hspace{2ex} \mbox{(as $\de, \la \in [-90\dg, 90\dg]$)} \nonumber \\
\tf \mbox{from (\ref{eq:solar-declination})} \hspace{0.8em} \sin \de & = & -\sin \al \, \cos \p \label{eq:sun-overhead}
\end{eqnarray}

Outside the tropics, where $|\de| > \al$, (\ref{eq:sun-overhead}) has no solution in \p. On the Tropic of Capricorn at $\de = -\al$, there is a single solution $\p = 0$, ie on the winter solstice. On the Tropic of Cancer at $\de = \al$, there is a single solution $\p = 180\dg$, ie the summer solstice. Inside the tropical region :
\[
|\de| < \al \Rightarrow 0 \leq \left| \frac{\sin \de}{\sin \al} \right| < 1
\]

$\tf$ there are two distinct solutions to (\ref{eq:sun-overhead}) in $\p \in [0, 360\dg)$, ie on two days of the year the sun is directly overhead at solar noon. In the case of the equator, these solutions are $\p = 90\dg$ and $\p = 270\dg$, ie the two equinoxes.

\section{Solar Azimuth Formula}

\label{sec:solar-azimuth}

Let the azimuthal angle in the horizon plane be $\ph \in [0\dg, 360\dg)$ using the convention of measuring \ph\ clockwise from due north \cite{wikipedia:solar-azimuth-angle}. Continuing from the main $Q1$ case in \S\ref{sec:solar-altitude} above, let $\bd{p}$ be the component of $\bd{s}$ parallel to the horizon plane, ie parallel to the plane of $\bd{u'}$ and $\bd{v'}$. Assume $\bd{p} \neq \bd{0}$, ie $|\mv| < 90\dg$, otherwise \ph\ is undefined. Then as $(\bd{s} \cdot \bd{w'})\bd{w'}$ is the perpendicular component of $\bd{s}$ :
\begin{eqnarray}
\bd{p} & = & \bd{s} - (\bd{s} \cdot \bd{w'})\bd{w'} \nonumber \\
& = & -\bd{i} + (\bd{i} \cdot \bd{w'})\bd{w'} \nonumber \\
& = & -\bd{i} - \bd{w'} \sin \mv \hspace{0.8em}, \mbox{(from (\ref{eq:sine-mu-initial}))} \nonumber \\
\tf |\bd{p}|^2 = \bd{p} \cdot \bd{p} & = & 1 + \sin^2 \mv + 2 (\sin \mv) \bd{w'} \cdot \bd{i} \nonumber \\
\tf |\bd{p}| & = & \cos \mv \hspace{0.5em} (\mbox{since from (\ref{eq:sine-mu-initial}) } \sin \mv = -\bd{i} \cdot \bd{w'}) \label{eq:mod-p} 
\end{eqnarray}

Now as \ph\ is defined clockwise from north (ie from $\bd{v'}$) in the $\bd{u'}\bd{v'}$-plane, \ph\ satisfies :
\begin{eqnarray}
|\bd{p}| \sin \ph & = & \bd{p} \cdot \bd{u'}, \nonumber \\
\mbox{and} \hspace{0.5em} |\bd{p}| \cos \ph & = & \bd{p} \cdot \bd{v'}, \nonumber \\
\tf |\bd{p}| \sin \ph & = & (-\bd{i} - \bd{w'} \sin \mv) \cdot \bd{u'} = -\bd{u'} \cdot \bd{i} \hspace{0.5em} \mbox{(since basis vectors $\bd{u'}$, $\bd{v'}$, $\bd{w'}$ are orthogonal)}, \label{eq:p-sin-phi} \\
\mbox{and} \hspace{0.5em} |\bd{p}| \cos \ph & = & (-\bd{i} - \bd{w'} \sin \mv) \cdot \bd{v'} = -\bd{v'} \cdot \bd{i} \label{eq:p-cos-phi} 
\end{eqnarray}

But from the Rodrigues Rotation Formula :
\begin{eqnarray}
\bd{u'} & = & (\cos \ta)\bd{u} + (\sin \ta)(\bd{a} \times \bd{u}) + (1 - \cos \ta)(\bd{a} \cdot \bd{u})\bd{a} \nonumber \\
\bd{a} \times \bd{u} & = &
\left|
\begin{array}{ccc}
\bd{i} & \bd{j} & \bd{k} \\
\sin \ao & \cos \ao & 0 \\
0 & 0 & 1
\end{array}
\right| 
= (\cos \ao, -\sin \ao, 0), \hspace{0.5em} \mbox{and } \bd{a} \cdot \bd{u} = 0 \nonumber \\[1.5ex]
\tf \bd{u'} & = & \cos \ta(0, 0, 1) + \sin \ta (\cos \ao, -\sin \ao, 0) \nonumber \\
\tf \bd{u'} \cdot \bd{i} & = & \sin \ta \, \cos \ao = \sin \ta \, \cos \la \label{eq:u-dot-i}
\end{eqnarray}
and
\begin{eqnarray*}
\bd{v'} & = & (\cos \ta)\bd{v} + (\sin \ta)(\bd{a} \times \bd{v}) + (1 - \cos \ta)(\bd{a} \cdot \bd{v})\bd{a} \\
\bd{a} \times \bd{v} & = &
\left|
\begin{array}{ccc}
\bd{i} & \bd{j} & \bd{k} \\
\sin \ao & \cos \ao & 0 \\
\cos \bt & \sin \bt & 0
\end{array}
\right| 
= (0, 0, \sin \ao \, \sin \bt - \cos \ao \, \cos \bt) \\[1.5ex]
& = & (0, 0, -\cos(\ao + \bt)) = (0, 0, -\sin \de), \hspace{0.5em} \mbox{and } \bd{a} \cdot \bd{v} = \sin(\ao + \bt) = \cos \de \\[1.5ex]
\tf \bd{v'} & = & \cos \ta(\cos \bt, \sin \bt, 0) + \sin \ta(0, 0, -\sin \de) + (1 - \cos \ta)\cos \de(\sin \ao, \cos \ao, 0) \\
\tf \bd{v'} \cdot \bd{i} & = & \cos \ta \, \cos \bt + (1 - \cos \ta) \cos \de \, \sin \ao
\end{eqnarray*}

Since $\de = 90 - (\ao + \bt)$, $\cos \bt = \sin(\ao + \de) \tf$
\begin{eqnarray}
\bd{v'} \cdot \bd{i} & = & \cos \ta \, \sin(\ao + \de) + (1 - \cos \ta) \cos \de \, \sin \ao \nonumber \\
& = & \cos \ta (\sin \ao \, \cos \de + \cos \ao \, \sin \de) + (\cos \ta - 1) \cos \de \, \sin \ao \nonumber \\
& = & \cos \ta \, \sin \de \, \cos \ao + \cos \de \, \sin \ao = \cos \ta \, \sin \de \, \cos \la - \cos \de \, \sin \la \label{eq:v-dot-i}
\end{eqnarray}

Thus from (\ref{eq:mod-p}), (\ref{eq:p-sin-phi}),  (\ref{eq:p-cos-phi}), (\ref{eq:u-dot-i}), and (\ref{eq:v-dot-i})
\begin{eqnarray}
\sin \ph & = & \frac{-\sin \ta \, \cos \la}{\cos \mv} \label{eq:solar-azimuth-sin} \\
\mbox{and} \hspace{0.8em} \cos \ph & = & \frac{\cos \de \, \sin \la - \cos \ta \, \sin \de \, \cos \la}{\cos \mv} \label{eq:solar-azimuth-cos}
\end{eqnarray}

Equation~(\ref{eq:solar-azimuth-cos}) can also be written without a \ta\ term :
\begin{eqnarray}
\cos \ph & = & \frac{(1 - \sin^2 \de) \sin \la - \cos \ta \, \sin \de \, \cos \la \, \cos \de}{\cos \mv \, \cos \de} \nonumber \\
& = & \frac{\sin \la - \sin \de(\cos \ta \, \cos \de \, \cos \la + \sin \de \, \sin \la)}{\cos \mv \, \cos \de} \nonumber \\
& = & \frac{\sin \la - \sin \de \, \sin \mv}{\cos \mv \, \cos \de}, \hspace{1.5em} \mbox{(from (\ref{eq:solar-altitude}))} \label{eq:solar-azimuth-cos-no-tau}
\end{eqnarray}

Then equations (\ref{eq:solar-azimuth-sin}) and (\ref{eq:solar-azimuth-cos}) (or equations (\ref{eq:solar-azimuth-sin}) and (\ref{eq:solar-azimuth-cos-no-tau})) uniquely define $\ph \in [0\dg, 360\dg)$, for the case of $Q1$. For the case of $Q4$, at an angle $(360 - \p)\dg$ the situation of the planet is as at angle \p\ in $Q1$ and thus the same \ph\ applies. Since \la\ given by (\ref{eq:solar-declination-arcsin}) is the same at these two angles it follows that the rhs's of (\ref{eq:solar-azimuth-sin}) and (\ref{eq:solar-azimuth-cos}) produce the same results, and hence give the correct \ph\ for $Q4$. \\

For $Q3$ at angle $(180 + \p)\dg$, the same process of reflection as in \S\ref{sec:solar-altitude} into $Q1$ at angle \p\ can be performed, with the result that (\ref{eq:solar-azimuth-sin}) and (\ref{eq:solar-azimuth-cos}) produce negative of the $Q1$ values for the sine and cosine of \ph. Thus the required angle for $Q3$, which is $(\ph + 180\dg) \bmod 360\dg$, is produced by (\ref{eq:solar-azimuth-sin}) and (\ref{eq:solar-azimuth-cos}).  (Note the $\cos \mv$ term does not change sign in the reflection). \\

In $Q2$ the situation of the planet at angle $(180 - \p)\dg$ is as $Q3$ at angle $(180 + \p)\dg$ so \ph\ must be the same. Since $\cos(180 - \p)\dg = \cos(180 + \p)\dg$ and \la\ given by (\ref{eq:solar-declination-arcsin}) is the same, the angle $\ph \in [0\dg, 360\dg)$ given by (\ref{eq:solar-azimuth-sin}) and (\ref{eq:solar-azimuth-cos}) is then correct for $Q2$, for any latitude \de\ and time \ta. \\

Thus equations (\ref{eq:solar-azimuth-sin}) and (\ref{eq:solar-azimuth-cos}) (or equations (\ref{eq:solar-azimuth-sin}) and (\ref{eq:solar-azimuth-cos-no-tau})) uniquely define the solar azimuth angle $\ph \in [0\dg, 360\dg)$ at all latitudes \de\, at all times of the year, and all times \ta\ of the solar day, provided an azimuth is defined ie the sun is not at the zenith or nadir position. Since at the poles cardinal directions of north, south, east, and west are not well-defined\footnote{At the north pole every direction points to the south and at the south pole every direction points to the north.} a meridian longitude (a half great circle) has to be selected, wrt which the cardinal directions can be defined. Then solar noon time is well-defined as the time when the sun crosses the selected meridian, even though on a stationary day in the orbit (in the simplified model) the sun's altitude is constant (possibly negative) all throughout the day. North at a pole would then be defined as the continuation of northwards on the selected longitude line, which would be towards the north pole and away from the south pole. With this convention equations~(\ref{eq:solar-azimuth-sin}) and (\ref{eq:solar-azimuth-cos}) produce the correct azimuth angle using the north clockwise convention, with \ta\ measured wrt to the selected meridian longitude :
\begin{eqnarray*}
\mbox{North pole } \Rightarrow \mv = \la \Rightarrow \sin \ph & = & -\sin \ta \\
\mbox{and } \cos \ph & = & -\cos \ta \\
\tf \ph & = & \ta + 180\dg \pmod{360\dg}
\end{eqnarray*}

\noindent which is the correct angle wrt due north as \ta\ is wrt due south, and 
\begin{eqnarray*}
\mbox{South pole } \Rightarrow \mv = -\la \Rightarrow \sin \ph & = & -\sin \ta \\
\mbox{and } \cos \ph & = & \cos \ta \\
\tf \ph & = & 360\dg - \ta \pmod{360\dg} \\
\mbox{ie } \ph & = & -\ta \pmod{360\dg}
\end{eqnarray*}

\noindent which is the correct angle wrt due north since at the south pole north points into the selected meridian half plane, and a positive rotation \ta\ from solar noon will produce a negative solar azimuth \ph\ according to the north clockwise convention. The above process will work for any choice of meridian longitude for a pole. \\

The azimuth cosine formula (\ref{eq:solar-azimuth-cos-no-tau}) provides an alternate route to the sunrise direction formula (\ref{eq:sunrise-formula-lambda}) since putting $\mv = 0$ in (\ref{eq:solar-azimuth-cos-no-tau}) and remembering \te\ is measured anti-clockwise from due east we obtain :
\[
\te + \ph = 90\dg \Rightarrow \sin \te = \cos \ph = \frac{\sin \la}{\cos \de}
\]
This completely characterizes \te\ since from \S\ref{sec:intro} we know in all cases $\te \in [-90\dg, 90\dg]$. \\

\newpage

\noindent \textbf{Example} Continuing with the example of Madrid on 15\textsuperscript{th} May 2019 in \S\ref{sec:sunrise-time}, compare the above altitude and azimuth formulae with the actual solar altitude of $50\dg$ and azimuth of $249\dg$ at DST 16:47 from \cite{timeanddate:sun-calc-spain}. We have :
\begin{eqnarray*}
& & \mbox{CT = 15:47 = 15.78 hours} \\
& & \ta = 15 \cdot 15.78 - 180 - 18.72 + 15 \cdot 0.05917 = 38.87\dg, \tf \cos \ta = 0.7786,\; \sin \ta = 0.6275 \\
\mbox{and} & & \sin \la = -\sin 23.44 \cdot \cos 143.0 = 0.3177,\; \cos \la = 0.9482 \\
\tf & & \sin \mv = 0.7786 \cdot (\cos 40.42) \cdot 0.9482 + (\sin 40.42) \cdot 0.3177 = 0.7681 \\
\tf & & \underline{\mv = 50.2\dg}, \cos \mv = 0.6403. \\
\mbox{And} & & \sin \ph = -0.6275 \cdot 0.9482 / 0.6403 = -0.9297 \\
& & \cos \ph = (0.3178 - (\sin 40.42) \cdot 0.7681) / (0.6403 \cdot \cos 40.42) = -0.3697, \hspace{0.8em} \mbox{from (\ref{eq:solar-azimuth-cos-no-tau})} \\
\Rightarrow & & \underline{\ph = 248.4\dg}
\end{eqnarray*}

Thus the formulae predictions are accurate to within $0.6\dg$ when compared with \cite{timeanddate:sun-calc-spain}.

\section{Solar Noon Altitude}

The solar noon altitude \cite{web:solar-geometry} \mo\ for a location is by definition the solar altitude when the solar hour angle \ta\ for the location is $0\dg$. From (\ref{eq:solar-altitude}) this is the maximum value of \mv\ throughout the day, ie the highest position of the sun. On a  day of perpetual darkness this will be negative, eg as in Figure~\ref{fig:winter-solstice} at a point above the arctic circle on the winter solstice (where case \ref{sna-case-south} below applies). From Figure~\ref{fig:winter-solstice} with the effective axial tilt \ao, covering cases $Q1$ and $Q4$, it is clear the azimuth \ph\ of the sun at solar noon for any day and latitude is either :
\begin{enumerate}
\item southerly, ie $\ph = 180\dg$, for latitudes above the effective Tropic of Capricorn (at latitude $-\ao$) \label{sna-case-south}
\item northerly, ie $\ph = 0\dg$, for latitudes below the effective Tropic of Capricorn \label{sna-case-north}
\item undefined on the effective Tropic of Capricorn, as the sun is then directly overhead at solar noon (ie $\mo = 90\dg$) \label{sna-case-undefined}
\end{enumerate}

The effective Tropic of Capricorn is the southern tropic corresponding with the effective axial tilt \ao. The effective tropics converge to the equator at the spring equinox where \ao\ becomes zero. Note cases (\ref{sna-case-south}) and (\ref{sna-case-north}) do not correspond with the northern and southern hemispheres, but with the two sides of the effective tropic which changes throughout the year. For example in Figure~\ref{fig:winter-solstice} a southern latitude between $-\ao$ and $0\dg$ will see the noon sun in the southern half of the sky, as a northern latitude will. The dividing line is the effective tropic, which at the spring equinox becomes the equator. For $Q2$ and $Q3$ a similar set of three cases applies involving the effective Tropic of Cancer in the north at latitude $\ao$. For practical purposes which of the above three cases applies would be determinable whenever the sun is visible at solar noon. \\

Considering case~\ref{sna-case-south}, entering $\ta = 0, \ph = 180, \mv = \mo$ in equations (\ref{eq:solar-altitude}) and (\ref{eq:solar-azimuth-cos}) we obtain :
\begin{eqnarray*}
\sin \mo & = & \cos \de \, \cos \la + \sin \de \, \sin \la \\
\cos \mo & = & \sin \de \, \cos \la - \cos \de \, \sin \la \\
\mbox{ie. } \cos(90 - \mo) & = & \cos(\de - \la) \\
\mbox{and} \hspace{0.7em} \sin(90 - \mo) & = & \sin(\de - \la) \\
\tf (90 - \mo) - (\de - \la) & = & \mbox{integral multiple of } 360
\end{eqnarray*}

But then, defining the solar zenith angle \cite{wikipedia:solar-zenith-angle} \no\ at solar noon to be the complement of angle \mo\ : 
\begin{eqnarray}
& & \mo \in [-90, 90], \de \in (-90, 90), \la \in [-\al, \al], \al \in [0, 90) \label{eq:sna-angles-ranges} \\
& \Rightarrow & (90 - \mo) - (\de - \la) \in (-180, 360) \nonumber \\
& \Rightarrow & (90 - \mo) - (\de - \la) = 0 \nonumber \\
& \Rightarrow & \no = \de - \la \label{eq:solar-noon-south}
\end{eqnarray}

Considering case~\ref{sna-case-north}, entering $\ta = 0, \ph = 0, \mv = \mo$ in equations (\ref{eq:solar-altitude}) and (\ref{eq:solar-azimuth-cos}) we obtain :
\begin{eqnarray*}
\sin \mo & = & \cos \de \, \cos \la + \sin \de \, \sin \la \\
\cos \mo & = & \cos \de \, \sin \la - \sin \de \, \cos \la \\
\mbox{ie. } \cos(90 - \mo) & = & \cos(\de - \la) \\
\mbox{and} \hspace{0.7em} \sin(90 - \mo) & = & -\sin(\de - \la)
\end{eqnarray*}

Thus in the $xy$-plane on the unit circle the angles $90 - \mo$ and $\de - \la$ are reflections of one another in the $x$-axis.

\newpage

\begin{eqnarray}
\tf 90 - \mo & = & 360 - (\de - \la) \pmod{360} \nonumber \\
\tf 270 + \mo - \de + \la & = & \mbox{integral multiple of } 360 \nonumber \\
\mbox{but from (\ref{eq:sna-angles-ranges}),} \hspace{0.7em} \mo - \de + \la & \in & (-270, 270) \nonumber \\
\tf 270 + \mo - \de + \la & \in & (0, 540) \nonumber \\
\tf 270 + \mo - \de + \la & = & 360 \nonumber \\
\tf \no & = & \la - \de \label{eq:solar-noon-north}
\end{eqnarray}

In case~\ref{sna-case-undefined}, entering $\ta = 0, \mo = 90$ in equation (\ref{eq:solar-altitude}) we obtain :
\begin{eqnarray}
\sin 90 & = & \cos \de \, \cos \la + \sin \de \, \sin \la \nonumber \\
\mbox{ie. } 1 & = & \cos(\de - \la) \nonumber \\
\tf \de - \la & = & \mbox{integral multiple of } 360 \nonumber \\
\Rightarrow \mbox{from (\ref{eq:sna-angles-ranges}), } \de - \la & = & 0 \nonumber \\ 
\Rightarrow \la & = & \de \label{eq:solar-noon-zenith}
\end{eqnarray}

The three solar noon equations (\ref{eq:solar-noon-south}), (\ref{eq:solar-noon-north}), and (\ref{eq:solar-noon-zenith}) give a relationship between latitude \la\ and solar declination \de\ depending on which of the above cases~\ref{sna-case-south}--\ref{sna-case-undefined} applies. These equations will still apply even if $\mo < 0$, ie a day of perpetual darkness, however in case~\ref{sna-case-undefined} the one possibility not considered, ie $\mo = -90$, with the sun at the nadir position at solar noon, is not possible, as we would expect intuitively, but formally by (\ref{eq:solar-altitude}) such a position would imply~: $\sin (-90) = -1 = \cos(\de - \la) \Rightarrow \de - \la - 180 = \mbox{multiple of } 360$. But since from (\ref{eq:sna-angles-ranges}), $\de - \la - 180 \in (-360, 0)$, this is impossible. The solar noon equations can allow the latitude in daytime to be determined from a table of solar declinations (which can be computed from (\ref{eq:solar-declination-arcsin})), by measuring the zenith angle of the sun at noon.

\section{Analemma}

\label{sec:analemma}

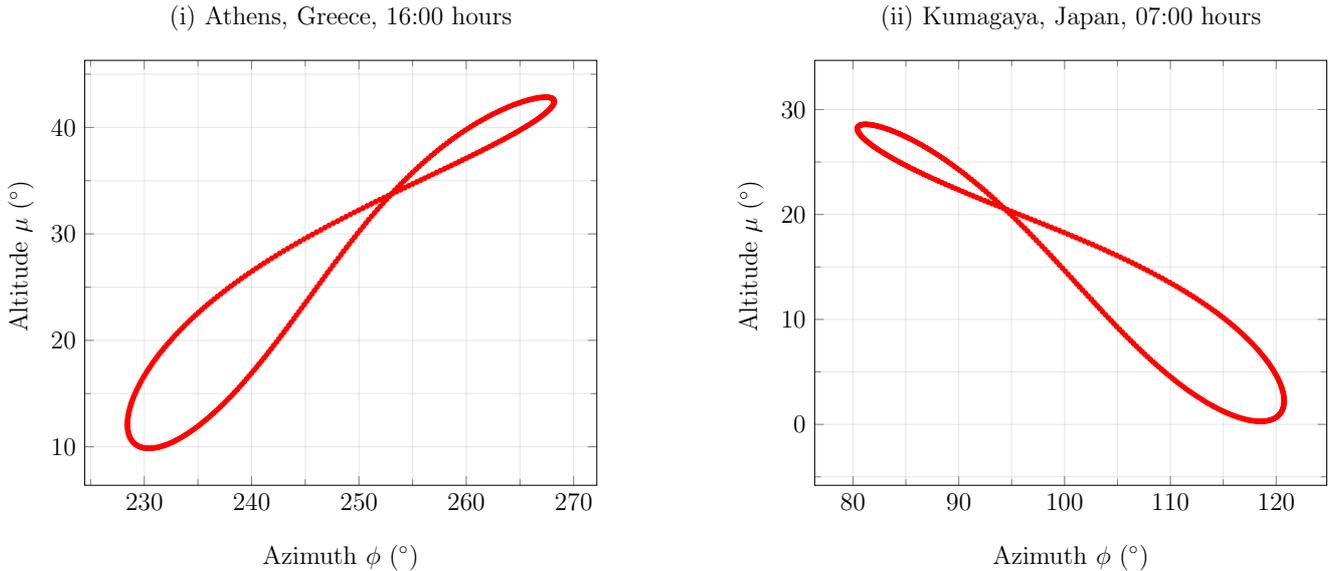
\begin{figure}[!h]
\begin{center}
\begin{tikzpicture} [
	node font=\large,
	scale=0.8
]

\pgfplotsset{width=10cm,compat=1.14}
\definecolor{Grid Color}{HTML}{aaaaaa}
\definecolor{Analemma Color}{HTML}{ff0000}

\begin{axis}[
	clip=false,
	axis equal=true,
	xlabel={Azimuth \ph\ ($\dg$)},
	xlabel style={anchor=north, below=8pt},
	ylabel={Altitude \mv\ ($\dg$)},
	ylabel style={anchor=south, above=8pt},
	minor tick num=1,
	grid=both,
	grid style={Grid Color, line width=0.3pt, opacity=0.3},
]
\addplot [
	Analemma Color,
	only marks,
	mark size=1.1pt,
] table[x index=2, y index=1, skip first n=0]  {analemma-athens.dat} -- cycle;

\path (axis description cs:0.5, 1.1) node[] {(i) Athens, Greece, 16:00 hours};

\end{axis}

\begin{axis}[
	clip=false,
	xshift=12cm,
	axis equal=true,
	xlabel={Azimuth \ph\ ($\dg$)},
	xlabel style={anchor=north, below=8pt},
	ylabel={Altitude \mv\ ($\dg$)},
	ylabel style={anchor=south, above=8pt},
	minor tick num=1,
	grid=both,
	grid style={Grid Color, line width=0.3pt, opacity=0.3},
]
\addplot [
	Analemma Color,
	only marks,
	mark size=1.1pt,
] table[x index=2, y index=1, skip first n=0]  {analemma-kumagaya.dat} -- cycle;

\path (axis description cs:0.5, 1.1) node[] {(ii) Kumagaya, Japan, 07:00 hours};

\end{axis}

\end{tikzpicture}
\vspace{3ex}
\caption{Analemmas Generated by Formulae (\ref{eq:tau-ct}), (\ref{eq:solar-altitude}), (\ref{eq:solar-azimuth-sin}), and (\ref{eq:solar-azimuth-cos})}
\label{fig:analemmas}
\vspace{4ex}
\end{center}
\end{figure}

One form of analemma \cite{wikipedia:analemma} is a graph of the Sun's position in the sky as seen from a fixed location on Earth, as altitude plotted against azimuth, at the same local clock time CT (ex-DST) each day for a period of a year. With equal horizontal and vertical scales this will have the same shape as a trace of the Sun's actual positions over these times, for example as captured in analemma photographs. The irregular figure 8 shape of the analemma (Figure~\ref{fig:analemmas}) is a result of the deviation throughout the year of the AST from the MST, as given by the equation of time (\S\ref{sec:clock-time}). In the present simplified model the analemma is given by equations~(\ref{eq:solar-altitude}), (\ref{eq:solar-azimuth-sin}), and (\ref{eq:solar-azimuth-cos}) together with equation~(\ref{eq:tau-ct}) which for a fixed daily local clock time CT, produces a varying solar hour angle \ta\ for the day $d$, which is then entered in equations~(\ref{eq:solar-altitude}), (\ref{eq:solar-azimuth-sin}), and (\ref{eq:solar-azimuth-cos}) to produce the altitude \mv\ and azimuth \ph. The parameter \la\ also varies daily, according to equation~(\ref{eq:solar-declination-arcsin}). If the solar day length were constant throughout the year so that MST = AST, and $EOT(d) = 0\;\forall d$, then \ta\ would be constant for clock time CT each day, and the analemma would be a simple curve without any loops because the solar declination parameter \la\ retraces its values in $[0, 180\dg]$ in the return journey from $180\dg$ to $360\dg$ due to the $\cos \p$ term in equation~(\ref{eq:solar-declination-arcsin}). \\

The Perl script \verb+calc-analemma.pl+ in Appendix~\ref{sec:script-analemma} calculates the analemma for a given location and time using the above formulae, producing a table of altitude and azimuth values over a year that can be directly input into the TIKZ/pgfplots package to produce a graph (eg by compiling the file \verb+analemma-graph.tex+). The script uses the circular orbit/constant speed model to calculate an approximate orbital angle \p\ for each day of the year. To use the script specify the desired local clock time CT (ex-DST), the latitude \de, and the difference $\Delta L\dg$ in degrees between the longitude of the location and the longitude of the defining meridian for the time zone\footnote{Note the longitude itself is not sufficient information as a single longitude can span more than one time zone, eg the UK and Spain span time zones UTC and UTC+1.}. \\

Uncommenting a line in the script shows the effect of a constant solar day length, and other lines can be uncommented to show the separate effects of the eccentricity and obliquity components of the EOT. \\

Note that an analemma in the tropics will become ill-defined towards solar noon because at such locations at certain times of the year the Sun is directly overhead at solar noon (\S\ref{sec:sun-overhead}) and the azimuth is then no longer well-defined. As we approach solar noon at such locations the azimuth equations (\ref{eq:solar-azimuth-sin}) and (\ref{eq:solar-azimuth-cos}) approach $0/0$. Note because of equation~(\ref{eq:solar-altitude}), which implies \mv\ is maximal at $\ta = 0$, the Sun directly overhead can only occur at solar noon (and likewise the nadir position only at solar midnight). \\

Since analemma photographs are rare the accuracy of the analemmas produced by the script are best checked by comparing the overall accuracy of the altitude and azimuth formulae (\ref{eq:solar-altitude}), (\ref{eq:solar-azimuth-sin}), and (\ref{eq:solar-azimuth-cos}) against real data. In the example of \S\ref{sec:solar-azimuth} the accuracy was $0.6\dg$. \\

In Figure~\ref{fig:analemmas} below two analemmas generated by the script are shown :
\begin{list}{(\roman{listCtr})}{\usecounter{listCtr}}
\item Athens, Greece : latitude $37.98\dg$ N, longitude = $23.73\dg$ E, UTC+2, $\Delta L\dg$ = $-6.27\dg$, CT = 16:00 \\
Script command line : \verb+perl calc-analemma.pl 37.98 16 -6.27 > analemma-athens.dat+ \\
Photo : \url{http://www.perseus.gr/Astro-Solar-Analemma-140000.htm}
\item Kumagaya, Japan : latitude $36.15\dg$ N, longitude = $139.38\dg$ E, UTC+9, $\Delta L\dg$ = $4.38\dg$, CT = 7:00 \\
Script command line : \verb+perl calc-analemma.pl 36.15 7 4.38 > analemma-kumagaya.dat+ \\
Photo : \url{https://earthsky.org/todays-image/todays-image-analemma-2013}
\end{list}

\vspace{5ex}

\vspace{4ex}

For a planet other than the Earth, with an axial tilt $\al \in [0, 90\dg)$, provided the original assumptions \ref{hypothesis:first}--\ref{hypothesis:last} in \S\ref{sec:intro} apply, the equations~(\ref{eq:solar-altitude}), (\ref{eq:solar-azimuth-sin}), and (\ref{eq:solar-azimuth-cos}) would still apply as they are in terms of solar time, but equation~(\ref{eq:tau-ct}) for clock time, and the EOT, would be different and the mean solar day length different from 24 hours. Also if the orbit were not well approximated by a circle then a more complicated mapping from day of the year $d$ to orbital angle \p\ would be required. With the required changes the script should then be able to generate analemmas for the planet.

\vspace{4ex}

\section{Acknowledgements}

Dedicated to my nephew Rory, a great sportsman, who passed away all too soon in 2015, who loved astronomy and who always encouraged my studies. God bless Rory, your light shines each day. And with much appreciation also to my parents, my teachers, and my brother and sister and their families. \\

\vspace{5ex}

\begin{center}
\adforn{45} \hspace{0.8em} Whatsoever thy hand findeth to do, do it with thy might \hspace{0.8em} \adforn{46}
\end{center}


\newpage
\begin{appendices}


\section{\textcolor{white}{Comparison With Actual Sunrise Data}}

\vspace{3ex}
{\Large \bfseries Comparison With Actual Sunrise Data} \\

\subsection{Sunrise Predictions}

\label{sec:actual-sunrise-data}

\setcounter{table}{0}
\renewcommand\thetable{\thesection.\arabic{table}}

The actual sunrise directions \te\ for locations at 8 different latitudes (4 northern hemisphere, 3 southern hemisphere, 1 equatorial) at 10 day intervals over the year 2018-2019 are shown in Table~\ref{table:actual-sunrise} below (source : 
\url{https://www.timeanddate.com}, \cite{timeanddate:sun-calculator}). \href{https://www.timeanddate.com}{\nolinkurl{www.timeanddate.com}} gives the sunrise and sunset directions to the nearest degree. Where in the \href{https://www.timeanddate.com}{\nolinkurl{www.timeanddate.com}} figures the sunrise angle north of east differs from the sunset angle north of west on a day the average of these two is quoted in the table below. The day offset $n$ is from the 2018 winter solstice date of 21/12/2018. \\

\begin{table}[!h]
\begin{center}
\begin{tabular}{l|l|l|l|l|l|l|l|l}
Day Offset&		Abu Dhabi&		Edinburgh&		Melbourne&		Milan&		Quito&		Reykjavik&		Rio&		Stanley		\\
$n$&				$24.45\dg$&		$55.95\dg$&		$-37.81\dg$&	$45.47\dg$&	$-0.17\dg$&	$64.15\dg$&		$-22.88\dg$& $-51.69\dg$	\\ \hline
0&				-25	&			-44  &			-31  &			-34&		-23&		-62  &			-26  &		-41		\\
10&				-25 &			-43  &			-31  &			-33&		-23&		-60  &			-26  &		-41		\\
20&				-24 &			-40  &			-29  &			-31&		-22&		-56  &			-24  &		-38		\\
30&				-22 &			-36  &			-27  &			-28&		-20&		-49  &			-22  &		-35		\\
40&				-19 &			-31  &			-23.5&			-25&		-18&		-42  &			-19.5&		-30.5		\\
50&				-16 &			-25.5&			-19.5&			-20&		-15&		-33.5&			-16  &		-25		\\
60&				-12 &			-19  &			-15  &			-15&		-11&		-25  &			-12.5&		-19.5		\\
70&				-8  &			-12.5&			-10.5&			-10&		 -7.5&		-16  &			-8.5 &		-13		\\
80&				-4  &			-5.5 &			-5.5 &			 -4.5&		 -4&		-7   &			-4   &		 -7		\\
90&				0.5 &			1.5  &			0.5  &			  1&		  0&		2.5  &			0    &		 -0.5		\\
100&			5   &			8.5  &			4.5  &			  7&		  4&		11.5 &			4    &		  6		\\
110&			9   &			15.5 &			9    &			 12&		  8&		20.5 &			8    &		 12		\\
120&			13  &			22   &			14   &			 17.5&		 12&		29.5 &			12   &		 18		\\
130&			17  &			28.5 &			18   &			 22&		 15&		38   &			16   &		 23		\\
140&			20  &			34   &			22   &			 26.5&		 18&		46.5 &			19   &		 28		\\
150&			22  &			39   &			25   &			 30&		 20&		54.5 &			21   &		 32		\\
160&			24  &			43   &			27   &			 33&		 22&		61.5 &			23   &		 35.5		\\
170&			26  &			46   &			29   &			 35&		 23&		67.5 &			25   &		 38		\\
180&			26  &			47   &			29   &			 36&		 23&		70   &			25   &		 39		\\
190&			26  &			46.5 &			29   &			 35&		 23&		69   &			25   &		 38		\\
200&			25  &			44.5 &			28   &			 34&		 22&		64.5 &			24   &		 36.5		\\
210&			23.5&			41   &			26   &			 31.5&		 21&		57.5 &			22   &		 34		\\
220&			21  &			36.5 &			23   &			 28&		 19&		50   &			20   &		 30		\\
230&			18  &			31   &			20   &			 24&		 16&		41.5 &			17   &		 25		\\
240&			15  &			25   &			16   &			 20&		 13&		33.5 &			14   &		 20		\\
250&			11  &			19   &			12   &			 15&		 10&		24.5 &			10   &		 14.5		\\
260&			7   &			12   &			7    &			  9.5&		  6&		15.5 &			6    &		  9		\\
270&			3   &			5    &			2.5  &			  4&		  2&		7    &			2    &		  2.5		\\
280&			-1  &			-2   &			-2.5 &			 -1.5&		 -2&		-1   &			-2   &		 -4		\\
290&			-5.5&			-8.5 &			-7.5 &			 -7&		 -5.5&		-11  &			-6   &		-10		\\
300&			-10 &			-15.5&			-12  &			-12&		 -9&		-20  &			-10.5&		-16		\\
310&			-14 &			-22  &			-17  &			-17.5&		-13&		-28.5&			-14  &		-22		\\
320&			-17 &			-28  &			-21  &			-22&		-16&		-37  &			-18  &		-27.5		\\
330&			-20 &			-33.5&			-24.5&			-26&		-19&		-45  &			-21  &		-32.5		\\
340&			-23 &			-38  &			-27.5&			-30&		-21&		-52  &			-23  &		-36.5		\\
350&			-24 &			-41  &			-30  &			-32&		-22.5&		-58  &			-25  &		-39.5		\\
360&			-25 &			-43  &			-31  &			-33&		-23&		-61  &			-26  &		-41		\\
365&			-25 &			-44  &			-31  &			-34&		-23&		-62  &			-26  &		-41
\end{tabular}
\vspace{3ex}
\caption{Actual Sunrise Directions \te\ for 2018-2019 \label{table:actual-sunrise}}
\end{center}
\end{table}

The charts below compare the actual sunrise directions shown in red with the curve of sunrise directions predicted by the sunrise direction formula (with the currently accepted Earth axial tilt $\al = \earthAxialTilt\dg$) shown in blue. The circular orbit/constant speed model is used with the orbital angle \p\ in the formula calculated as $\frac{d}{N}(360\dg)$ for day offset $d \in [0, N - 1]$ from the day of the winter solstice, where $N = 365$ is the number of days in the year. The average error over the 8 charts is $1.25\dg$. When the formula's own axial tilt estimate of $23.52\dg$ (Appendix~\ref{sec:axial-tilt-estimate}) is used the average error is $1.26\dg$.

\vspace{5ex}

\begin{figure}[!ph]
\begin{center}
\begin{tikzpicture}[
	scale=0.9,
	declare function={
		sunrise(\d,\x) = {-asin( sin(\earthAxialTilt)*cos(360*\x/365) / cos(\d) )};
	}
]

\def\latitude{55.95} 
\def\avgerror{2.14} 
\def\ytop{70} 

\begin{axis}[
	axis lines=left,
	xmax=400,
	ymax=\ytop,
	align=center,
	title={\textbf{ \large Actual/Computed Sunrise in} \\ \textbf{ \large Edinburgh, 2018-19} },
	xlabel={\large Day Offset $(n)$ from Winter Solstice},
	ylabel={\large Degrees $(\theta)$ North of East},
	title style={outer ysep=1.3em},
	xlabel style={at={(ticklabel cs:0.5)},anchor=near ticklabel,outer ysep=1.5em},
	ylabel style={at={(ticklabel cs:0.5)},anchor=near ticklabel,outer ysep=0.9em}
]

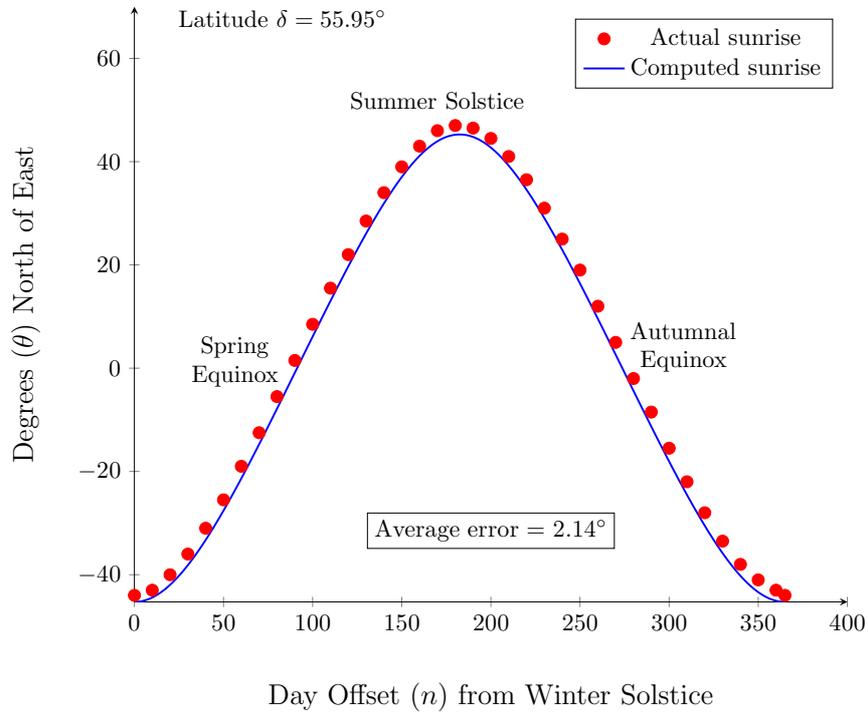
\addplot [
	red,
	only marks,
	mark size=2.5pt
] table {Graph-edinburgh.dat};
\addlegendentry{Actual sunrise};

\addplot [
	blue,
	domain=0:364,
	samples=365,
	variable=n,
	line width=0.8pt
]
{sunrise(\latitude, n)} 
[xshift=-4pt,yshift=3pt] node [pos=0.25,black,anchor=east] {Spring \\ Equinox}
[xshift=4pt,yshift=5pt] node [pos=0.75,black,anchor=west] {Autumnal \\ Equinox}
[xshift=-9pt,yshift=-1pt] node [pos=0.5,black,anchor=south] {Summer Solstice}
;
\addlegendentry{Computed sunrise};

\path (axis cs:200,-35) node[draw,inner sep=3pt,anchor=south] {Average error = $\avgerror^\circ$};

\path (axis cs:20, \ytop) node[below right=-2pt and 0pt] {Latitude $\delta = \latitude^\circ$};

\end{axis}

\end{tikzpicture}
\end{center}
\end{figure}

\vspace{4ex}

\begin{figure}[!ph]
\begin{center}
\begin{tikzpicture}[
	scale=0.9,
	declare function={
		sunrise(\d,\x) = {-asin( sin(\earthAxialTilt)*cos(360*\x/365) / cos(\d) )};
	}
]

\def\latitude{45.47} 
\def\avgerror{1.51} 
\def\ytop{55} 

\begin{axis}[
	axis lines=left,
	xmax=400,
	ymax=\ytop,
	align=center,
	title={\textbf{ \large Actual/Computed Sunrise in} \\ \textbf{ \large Milan, 2018-19} },
	xlabel={\large Day Offset $(n)$ from Winter Solstice},
	ylabel={\large Degrees $(\theta)$ North of East},
	title style={outer ysep=1.3em},
	xlabel style={at={(ticklabel cs:0.5)},anchor=near ticklabel,outer ysep=1.5em},
	ylabel style={at={(ticklabel cs:0.5)},anchor=near ticklabel,outer ysep=0.9em}
]

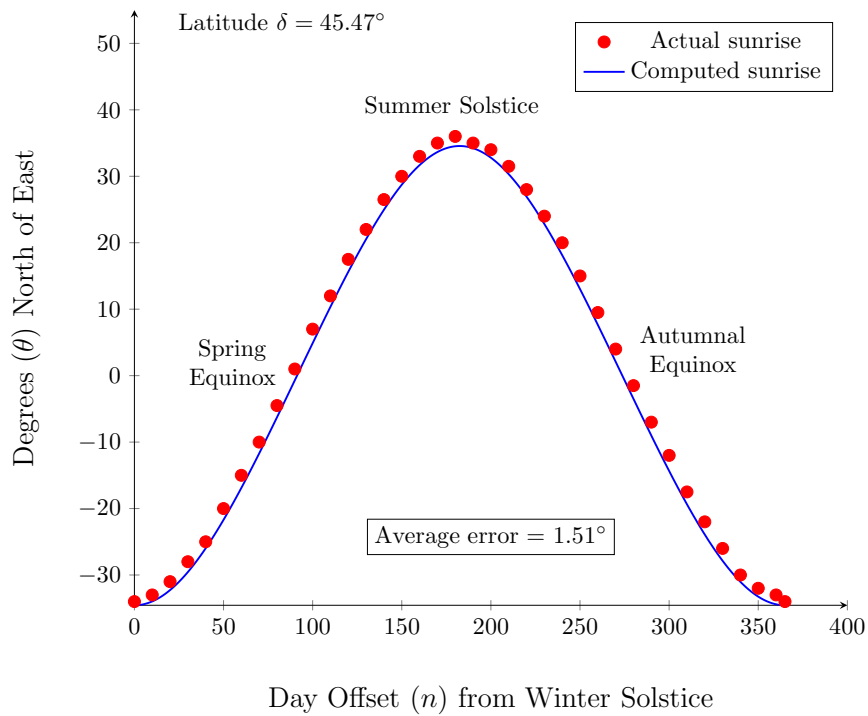
\addplot [
	red,
	only marks,
	mark size=2.5pt
] table {Graph-milan.dat};
\addlegendentry{Actual sunrise};

\addplot [
	blue,
	domain=0:364,
	samples=365,
	variable=n,
	line width=0.8pt
]
{sunrise(\latitude, n)} 
[xshift=-5pt,yshift=5pt] node [pos=0.25,black,anchor=east] {Spring \\ Equinox}
[xshift=9pt,yshift=4pt] node [pos=0.75,black,anchor=west] {Autumnal \\ Equinox}
[xshift=-7pt,yshift=1pt] node [pos=0.5,black,anchor=south] {Summer Solstice}
;
\addlegendentry{Computed sunrise};

\path (axis cs:200,-27) node[draw,inner sep=3pt,anchor=south] {Average error = $\avgerror^\circ$};

\path (axis cs:20, \ytop) node[below right=-2pt and 0pt] {Latitude $\delta = \latitude^\circ$};

\end{axis}

\end{tikzpicture}
\end{center}
\end{figure}

\begin{figure}[!ph]
\begin{center}
\begin{tikzpicture}[
	scale=0.9,
	declare function={
		sunrise(\d,\x) = {-asin( sin(\earthAxialTilt)*cos(360*\x/365) / cos(\d) )};
	}
]

\def\latitude{64.15} 
\def\avgerror{3.62} 
\def\ytop{103} 

\begin{axis}[
	axis lines=left,
	xmax=400,
	ymax=\ytop,
	align=center,
	title={\textbf{ \large Actual/Computed Sunrise in} \\ \textbf{ \large Reykjavik, 2018-19} },
	xlabel={\large Day Offset $(n)$ from Winter Solstice},
	ylabel={\large Degrees $(\theta)$ North of East},
	title style={outer ysep=1.3em},
	xlabel style={at={(ticklabel cs:0.5)},anchor=near ticklabel,outer ysep=1.5em},
	ylabel style={at={(ticklabel cs:0.5)},anchor=near ticklabel,outer ysep=0.9em}
]

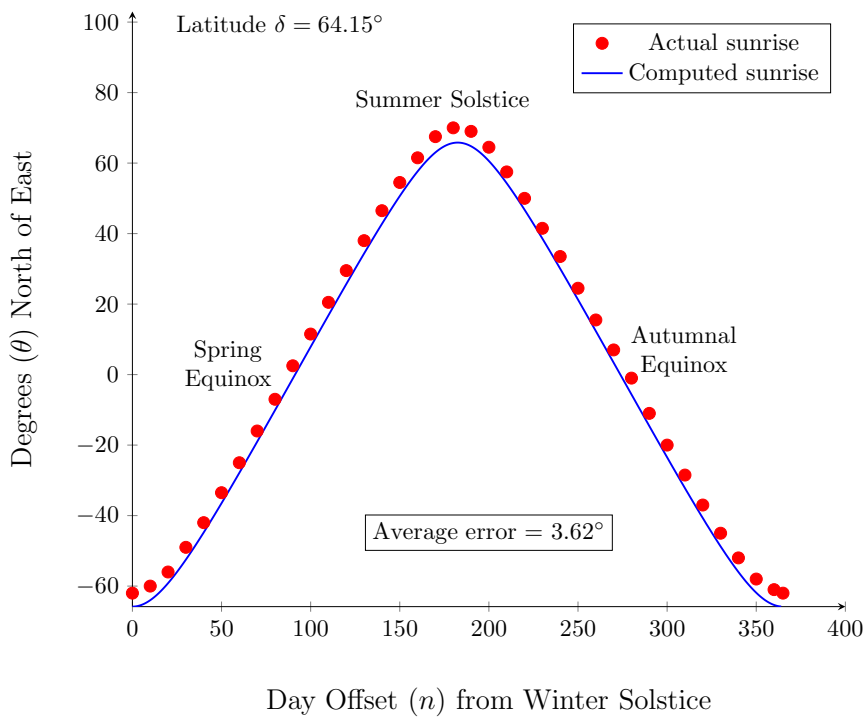
\addplot [
	red,
	only marks,
	mark size=2.5pt
] table {Graph-reykjavik.dat};
\addlegendentry{Actual sunrise};

\addplot [
	blue,
	domain=0:364,
	samples=365,
	variable=n,
	line width=0.8pt
]
{sunrise(\latitude, n)} 
[xshift=-6pt,yshift=4pt] node [pos=0.25,black,anchor=east] {Spring \\ Equinox}
[xshift=7pt,yshift=5pt] node [pos=0.75,black,anchor=west] {Autumnal \\ Equinox}
[xshift=-7pt,yshift=2pt] node [pos=0.5,black,anchor=south] {Summer Solstice}
;
\addlegendentry{Computed sunrise};

\path (axis cs:200,-50) node[draw,inner sep=3pt,anchor=south] {Average error = $\avgerror^\circ$};

\path (axis cs:20, \ytop) node[below right=-2pt and 0pt] {Latitude $\delta = \latitude^\circ$};

\end{axis}

\end{tikzpicture}
\end{center}
\end{figure}

\begin{figure}[!ph]
\begin{center}
\begin{tikzpicture}[
	scale=0.9,
	declare function={
		sunrise(\d,\x) = {-asin( sin(\earthAxialTilt)*cos(360*\x/365) / cos(\d) )};
	}
]

\def\latitude{24.45} 
\def\avgerror{0.83} 
\def\ytop{40} 

\begin{axis}[
	axis lines=left,
	xmax=400,
	ymax=\ytop,
	align=center,
	title={\textbf{ \large Actual/Computed Sunrise in} \\ \textbf{ \large Abu Dhabi, 2018-19} },
	xlabel={\large Day Offset $(n)$ from Winter Solstice},
	ylabel={\large Degrees $(\theta)$ North of East},
	title style={outer ysep=1.3em},
	xlabel style={at={(ticklabel cs:0.5)},anchor=near ticklabel,outer ysep=1.5em},
	ylabel style={at={(ticklabel cs:0.5)},anchor=near ticklabel,outer ysep=0.9em}
]

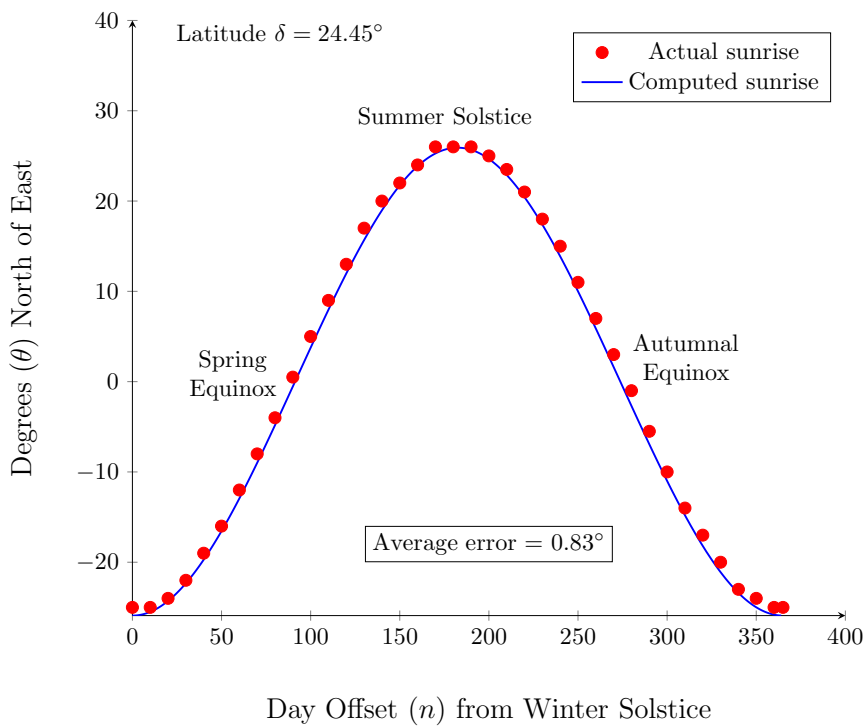
\addplot [
	red,
	only marks,
	mark size=2.5pt
] table {Graph-abudhabi.dat};
\addlegendentry{Actual sunrise};

\addplot [
	blue,
	domain=0:364,
	samples=365,
	variable=n,
	line width=0.8pt
]
{sunrise(\latitude, n)} 
[xshift=-4pt,yshift=3pt] node [pos=0.25,black,anchor=east] {Spring \\ Equinox}
[xshift=6pt,yshift=5pt] node [pos=0.75,black,anchor=west] {Autumnal \\ Equinox}
[xshift=-7pt,yshift=-2pt] node [pos=0.5,black,anchor=south] {Summer Solstice}
;
\addlegendentry{Computed sunrise};

\path (axis cs:200, -20) node[draw,inner sep=3pt,anchor=south] {Average error = $\avgerror^\circ$};

\path (axis cs:20, \ytop) node[below right=-2pt and 0pt] {Latitude $\delta = \latitude^\circ$};

\end{axis}

\end{tikzpicture}
\end{center}
\end{figure}

\begin{figure}[!ph]
\begin{center}
\begin{tikzpicture}[
	scale=0.9,
	declare function={
		sunrise(\d,\x) = {-asin( sin(\earthAxialTilt)*cos(360*\x/365) / cos(\d) )};
	}
]

\def\latitude{-37.81} 
\def\avgerror{0.49} 
\def\ytop{45} 

\begin{axis}[
	axis lines=left,
	xmax=400,
	ymax=\ytop,
	align=center,
	title={\textbf{ \large Actual/Computed Sunrise in} \\ \textbf{ \large Melbourne, 2018-19} },
	xlabel={\large Day Offset $(n)$ from Winter Solstice},
	ylabel={\large Degrees $(\theta)$ North of East},
	title style={outer ysep=1.3em},
	xlabel style={at={(ticklabel cs:0.5)},anchor=near ticklabel,outer ysep=1.5em},
	ylabel style={at={(ticklabel cs:0.5)},anchor=near ticklabel,outer ysep=0.9em}
]

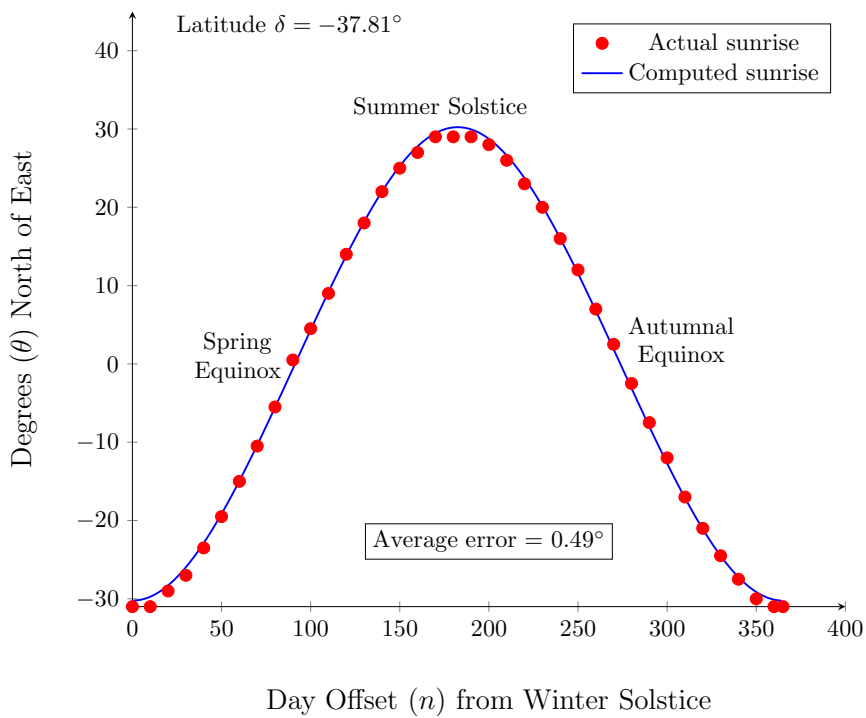
\addplot [
	red,
	only marks,
	mark size=2.5pt
] table {Graph-melbourne.dat};
\addlegendentry{Actual sunrise};

\addplot [
	blue,
	domain=0:364,
	samples=365,
	variable=n,
	line width=0.8pt
]
{sunrise(\latitude, n)} 
[xshift=-2pt,yshift=3pt] node [pos=0.25,black,anchor=east] {Spring \\ Equinox}
[xshift=2pt,yshift=5pt] node [pos=0.75,black,anchor=west] {Autumnal \\ Equinox}
[xshift=-7pt,yshift=-6pt] node [pos=0.5,black,anchor=south] {Summer Solstice}
;
\addlegendentry{Computed sunrise};

\path (axis cs:200,-25) node[draw,inner sep=3pt,anchor=south] {Average error = $\avgerror^\circ$};

\path (axis cs:20, \ytop) node[below right=-2pt and 0pt] {Latitude $\delta = \latitude^\circ$};

\end{axis}

\end{tikzpicture}
\end{center}
\end{figure}

\begin{figure}[!ph]
\begin{center}
\begin{tikzpicture}[
	scale=0.9,
	declare function={
		sunrise(\d,\x) = {-asin( sin(\earthAxialTilt)*cos(360*\x/365) / cos(\d) )};
	}
]

\def\latitude{-51.69} 
\def\avgerror{0.65} 
\def\ytop{58} 

\begin{axis}[
	axis lines=left,
	xmax=400,
	ymax=\ytop,
	align=center,
	title={\textbf{ \large Actual/Computed Sunrise in} \\ \textbf{ \large Stanley, Falkland Islands, 2018-19} },
	xlabel={\large Day Offset $(n)$ from Winter Solstice},
	ylabel={\large Degrees $(\theta)$ North of East},
	title style={outer ysep=1.3em},
	xlabel style={at={(ticklabel cs:0.5)},anchor=near ticklabel,outer ysep=1.5em},
	ylabel style={at={(ticklabel cs:0.5)},anchor=near ticklabel,outer ysep=0.9em}
]

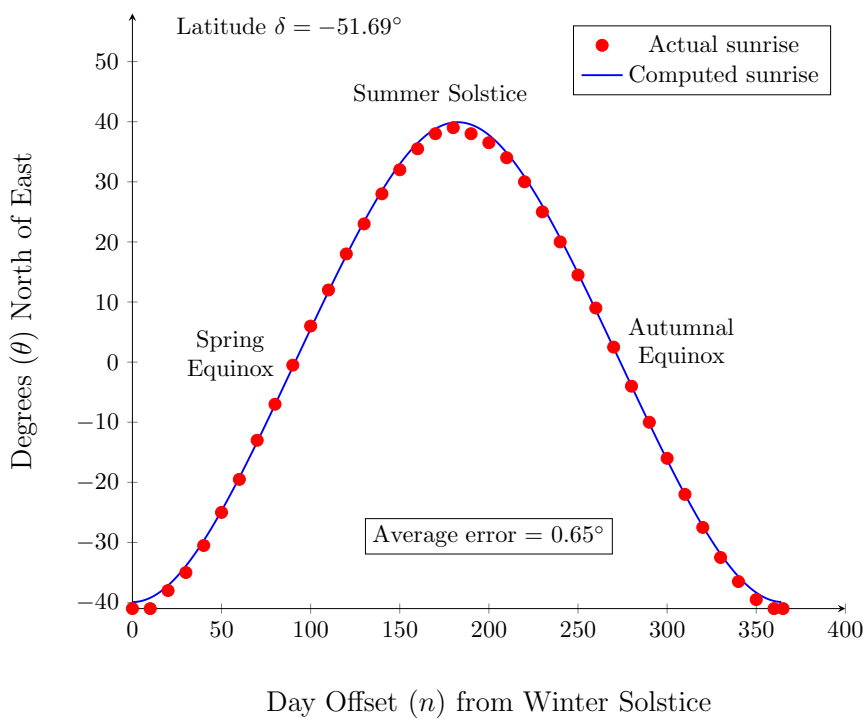
\addplot [
	red,
	only marks,
	mark size=2.5pt
] table {Graph-stanley.dat};
\addlegendentry{Actual sunrise};

\addplot [
	blue,
	domain=0:364,
	samples=365,
	variable=n,
	line width=0.8pt
]
{sunrise(\latitude, n)} 
[xshift=-5pt,yshift=3pt] node [pos=0.25,black,anchor=east] {Spring \\ Equinox}
[xshift=5pt,yshift=4pt] node [pos=0.75,black,anchor=west] {Autumnal \\ Equinox}
[xshift=-7pt,yshift=-2pt] node [pos=0.5,black,anchor=south] {Summer Solstice}
;
\addlegendentry{Computed sunrise};

\path (axis cs:200,-32) node[draw,inner sep=3pt,anchor=south] {Average error = $\avgerror^\circ$};

\path (axis cs:20, \ytop) node[below right=-2pt and 0pt] {Latitude $\delta = \latitude^\circ$};

\end{axis}

\end{tikzpicture}
\end{center}
\end{figure}

\begin{figure}[!ph]
\begin{center}
\begin{tikzpicture}[
	scale=0.9,
	declare function={
		sunrise(\d,\x) = {-asin( sin(\earthAxialTilt)*cos(360*\x/365) / cos(\d) )};
	}
]

\def\latitude{-22.88} 
\def\avgerror{0.35} 
\def\ytop{40} 

\begin{axis}[
	axis lines=left,
	xmax=400,
	ymax=\ytop,
	align=center,
	title={\textbf{ \large Actual/Computed Sunrise in} \\ \textbf{ \large Rio De Janeiro, 2018-19} },
	xlabel={\large Day Offset $(n)$ from Winter Solstice},
	ylabel={\large Degrees $(\theta)$ North of East},
	title style={outer ysep=1.3em},
	xlabel style={at={(ticklabel cs:0.5)},anchor=near ticklabel,outer ysep=1.5em},
	ylabel style={at={(ticklabel cs:0.5)},anchor=near ticklabel,outer ysep=0.9em}
]

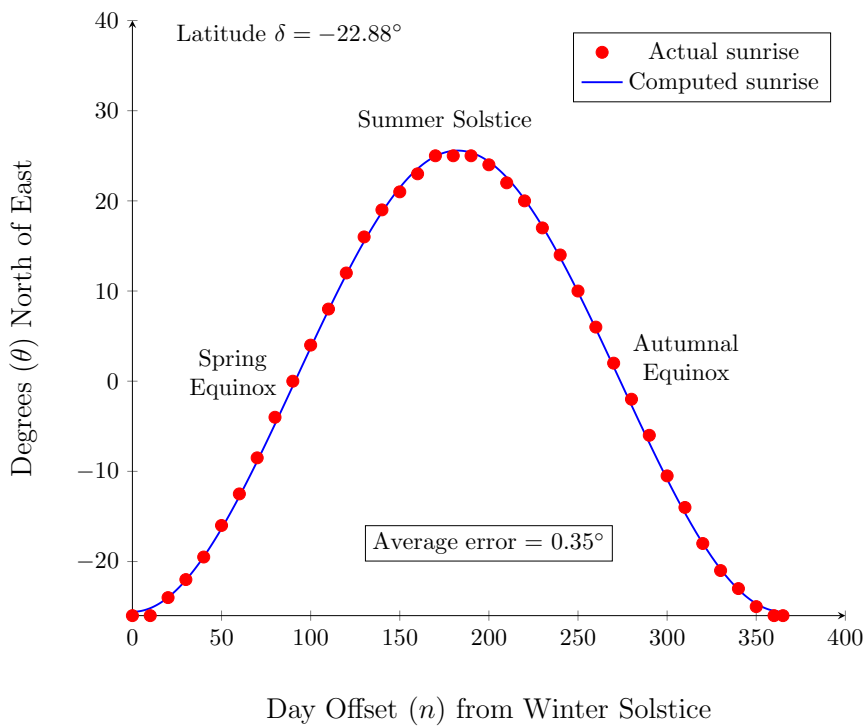
\addplot [
	red,
	only marks,
	mark size=2.5pt
] table {Graph-rio.dat};
\addlegendentry{Actual sunrise};

\addplot [
	blue,
	domain=0:364,
	samples=365,
	variable=n,
	line width=0.8pt
]
{sunrise(\latitude, n)} 
[xshift=-4pt,yshift=3pt] node [pos=0.25,black,anchor=east] {Spring \\ Equinox}
[xshift=6pt,yshift=5pt] node [pos=0.75,black,anchor=west] {Autumnal \\ Equinox}
[xshift=-7pt,yshift=-2pt] node [pos=0.5,black,anchor=south] {Summer Solstice}
;
\addlegendentry{Computed sunrise};

\path (axis cs:200,-20) node[draw,inner sep=3pt,anchor=south] {Average error = $\avgerror^\circ$};

\path (axis cs:20, \ytop) node[below right=-2pt and 0pt] {Latitude $\delta = \latitude^\circ$};

\end{axis}

\end{tikzpicture}
\end{center}
\end{figure}

\begin{figure}[!ph]
\begin{center}
\begin{tikzpicture}[
	scale=0.9,
	declare function={
		sunrise(\d,\x) = {-asin( sin(\earthAxialTilt)*cos(360*\x/365) / cos(\d) )};
	}
]

\def\latitude{-0.17} 
\def\avgerror{0.44} 
\def\ytop{35} 

\begin{axis}[
	axis lines=left,
	xmax=400,
	ymax=\ytop,
	align=center,
	title={\textbf{ \large Actual/Computed Sunrise in} \\ \textbf{ \large Quito, Ecuador, 2018-19} },
	xlabel={\large Day Offset $(n)$ from Winter Solstice},
	ylabel={\large Degrees $(\theta)$ North of East},
	title style={outer ysep=1.3em},
	xlabel style={at={(ticklabel cs:0.5)},anchor=near ticklabel,outer ysep=1.5em},
	ylabel style={at={(ticklabel cs:0.5)},anchor=near ticklabel,outer ysep=0.9em}
]

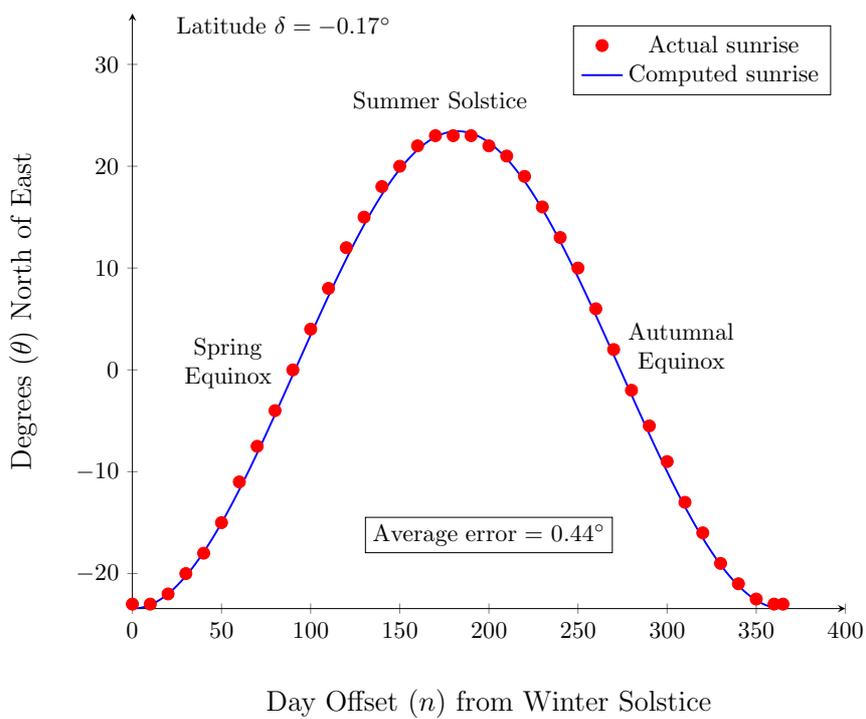
\addplot [
	red,
	only marks,
	mark size=2.5pt
] table {Graph-quito.dat};
\addlegendentry{Actual sunrise};

\addplot [
	blue,
	domain=0:364,
	samples=365,
	variable=n,
	line width=0.8pt
]
{sunrise(\latitude, n)} 
[xshift=-6pt,yshift=3pt] node [pos=0.25,black,anchor=east] {Spring \\ Equinox}
[xshift=6pt,yshift=5pt] node [pos=0.75,black,anchor=west] {Autumnal \\ Equinox}
[xshift=-7pt,yshift=-2pt] node [pos=0.5,black,anchor=south] {Summer Solstice}
;
\addlegendentry{Computed sunrise};

\path (axis cs:200,-18) node[draw,inner sep=3pt,anchor=south] {Average error = $\avgerror^\circ$};

\path (axis cs:20, \ytop) node[below right=-2pt and 0pt] {Latitude $\delta = \latitude^\circ$};

\end{axis}

\end{tikzpicture}
\end{center}
\end{figure}

\newpage

\subsection{Perpetual Day/Night Predictions}

\label{sec:perpetual-day-night}

Using the same circular orbit/constant speed model as in Appendix~\ref{sec:actual-sunrise-data}, consider the location of Jan Mayen, Norway at latitude $71\dg$ north. The argument to the $\arcsin$ goes out of range when either:
\begin{eqnarray*}
\cos \p & > & \frac{\cos \de}{\sin \al} = \frac{\cos 71\dg}{\sin \earthAxialTilt\dg} \simeq 0.81845 \simeq \cos 35\dg, \\
\mbox{or} \hspace{1em} \cos \p & < & -\frac{\cos \de}{\sin \al} \simeq -0.81845.
\end{eqnarray*}

These correspond with \p\ ranges of $[0, 35], [145, 215], [325, 360]$, which for the year 2018-2019 correspond with the dates: 25/1/2019, 17/5/2019, 27/7/2019, 15/11/2019, for entering/exiting periods of perpetual day or night. The actual dates from timeanddate.com, \cite{timeanddate:sun-calculator} are: \\

\begin{tabular}{l|llll}
Actual date & 22/1/2019 & 13/5/2019 & 1/8/2019 & 21/11/2019 \\
Estimated date & 25/1/2019 & 17/5/2019 & 27/7/2019 & 15/11/2019 \\
Error (days) & 3 & 4 & 5 & 6 \\ 
Average error & 4.5 days
\end{tabular} \\[2ex]

For Cape Adare, Antartica at latitude $71\dg$ south, the estimated dates for 2018-2019 are as for Jan Mayen and the actual dates are: \\

\begin{tabular}{l|llll}
Actual date & 31/1/2019 & 19/5/2019 & 26/7/2019 & 14/11/2019 \\
Estimated date & 25/1/2019 & 17/5/2019 & 27/7/2019 & 15/11/2019 \\
Error (days) & 6 & 2 & 1 & 1 \\ 
Average error & 2.5 days
\end{tabular} \\[2ex]

\vspace{2ex}

For Longyearbyen, Svalbard, Norway at latitude $78\dg$ north, the $\arcsin$ goes out of range when $\cos \p > 0.52267 \simeq \cos 58.5\dg$ or $\cos \p < -0.52267$, ie in \p\ ranges $[0, 58.5], [121.5, 238.5], [301.5, 360]$, which for 2018-2019 correspond with the dates: 18/2/2019, 23/4/2019, 20/8/2019, 23/10/2019. The actual dates are: \\

\begin{tabular}{l|llll}
Actual date & 16/2/2019 & 19/4/2019 & 25/8/2019 & 27/10/2019 \\
Estimated date & 18/2/2019 & 23/4/2019 & 20/8/2019 & 23/10/2019 \\
Error (days) & 2 & 4 & 5 & 4 \\ 
Average error & 3.75 days
\end{tabular} \\[2ex]

For McMurdo Station, Antartica at latitude $78\dg$ south, the estimated dates for 2018-2019 are as for Longyearbyen and the actual dates are : \\

\begin{tabular}{l|llll}
Actual date & 20/2/2019 & 25/4/2019 & 19/8/2019 & 24/10/2019 \\
Estimated date & 18/2/2019 & 23/4/2019 & 20/8/2019 & 23/10/2019 \\
Error (days) & 2 & 2 & 1 & 1 \\ 
Average error & 1.5 days
\end{tabular} \\[2ex]

The average errors overall is 3.1 days which though not as good as the sunrise predictions of \S\ref{sec:actual-sunrise-data} or the axial tilt prediction of Appendix~\ref{sec:axial-tilt-estimate} are still reasonable approximations for the simplified model.


\newgeometry{top=0.5in, bottom=0.8in, hmargin=0.5in, footnotesep=0in}

\section{\textcolor{white}{Axial Tilt Estimates}}

\label{sec:axial-tilt-estimate}

\setcounter{table}{0}
\renewcommand\thetable{\thesection.\arabic{table}}

\vspace{2ex}
{\Large \bfseries Axial Tilt Estimates} \\

\noindent The axial tilt estimates of the sunrise direction formula obtained by applying equation~(\ref{eq:sunrise-formula-axial-tilt}) to the sunrise data of Table~\ref{table:actual-sunrise}, using the same circular orbit/constant speed model as in Appendix~\ref{sec:actual-sunrise-data}, are shown in Table~\ref{table:axial-tilt-estimates} below. The estimates can become abnormally inaccurate when observed values of $\sin \te$ and $\cos \p$ are close to zero near the equinoxes ($n \simeq 90$ for spring and $n \simeq 270$ for autumn), for then small absolute changes in these two terms can result in large \% changes, causing large \% change in $\sin \al$ on lhs of (\ref{eq:sunrise-formula-axial-tilt}). The rhs of (\ref{eq:sunrise-formula-axial-tilt}) is thus unstable near the equinoxes (and at the equinoxes it has form $0/0$). The $\cos \de$ term does not cause instability so long as we are not too close to the poles. A good direction approximation can have a large \% error when we are near zero - eg approximating direction of $0.1\dg$ by $0.2\dg$ has $100\%$ error but is still a good direction approximation. Thus the particularly bad 5 estimates near the spring equinox in the table which were negative are excluded in calculating the averages --- though any other bad estimates are left in. Over the remaining 299 data points the overall average estimate for the Earth's axial tilt is $23.52\dg$, which is within $0.1\dg$ of the currently accepted value of $\earthAxialTilt\dg$. \\

\vspace{-1ex}

\begin{table}[!h]
\begin{center}
\begin{tabular}{l|l|l|l|l|l|l|l|l|l}
Day Offset&		Abu Dhabi&		Edinburgh&		Melbourne&		Milan&		Quito&		Reykjavik&		Rio&		Stanley&	Overall	\\
$n$&				$24.45\dg$&		$55.95\dg$&		$-37.81\dg$&	$45.47\dg$&	$-0.17\dg$&	$64.15\dg$&		$-22.88\dg$& $-51.69\dg$	& Average \\ \hline
0&				 22.63&			 22.89&			 24.01&			 23.09&		 23.00&		 22.64&			 23.82&		 24.00		\\
10&				 22.99&			 22.80&			 24.39&			 22.81&		 23.37&		 22.54&			 24.20&		 24.38		\\
20&				 23.16&			 22.48&			 24.01&			 22.56&		 23.45&		 22.58&			 23.46&		 23.92		\\
30&				 23.09&			 22.24&			 24.36&			 22.25&		 23.16&		 22.24&			 23.38&		 24.14		\\
40&				 22.57&			 21.93&			 24.08&			 22.57&		 23.59&		 22.20&			 23.47&		 24.05		\\
50&				 22.64&			 21.70&			 23.86&			 21.59&		 23.39&		 21.66&			 22.93&		 23.70		\\
60&				 21.68&			 20.84&			 23.52&			 20.75&		 21.86&		 21.08&			 22.90&		 23.82		\\
70&				 20.74&			 19.80&			 23.73&			 19.90&		 21.40&		 19.63&			 22.38&		 22.95		\\
80&				 19.27&			 16.19&			 23.17&			 16.61&		 21.25&		 16.03&			 19.51&		 23.11		\\
90&				-21.67&			-42.94&			-18.69&			-34.67&		  0.00&		-62.12&			  0.00&		 14.56		\\
100&			 31.92&			 33.47&			 24.40&			 34.72&		 27.70&		 35.40&			 25.36&		 25.58		\\
110&			 26.68&			 28.15&			 22.93&			 27.37&		 26.03&		 28.78&			 23.84&		 23.98		\\
120&			 25.54&			 26.21&			 23.73&			 26.36&		 25.96&		 26.88&			 23.79&		 23.79		\\
130&			 25.48&			 25.58&			 23.24&			 25.13&		 24.73&		 25.72&			 24.23&		 23.05		\\
140&			 24.73&			 24.88&			 23.44&			 24.87&		 24.54&		 25.15&			 23.77&		 23.02		\\
150&			 23.73&			 24.57&			 23.20&			 24.44&		 23.80&		 24.76&			 22.93&		 22.81		\\
160&			 23.57&			 24.36&			 22.79&			 24.36&		 23.86&		 24.45&			 22.88&		 22.88		\\
170&			 24.11&			 24.35&			 23.08&			 24.31&		 23.58&		 24.35&			 23.49&		 23.00		\\
180&			 23.54&			 24.20&			 22.54&			 24.37&		 23.02&		 24.21&			 22.94&		 22.98		\\
190&			 23.73&			 24.18&			 22.72&			 23.93&		 23.20&		 24.23&			 23.12&		 22.64		\\
200&			 23.76&			 24.26&			 22.85&			 24.25&		 23.10&		 24.34&			 23.10&		 22.71		\\
210&			 24.07&			 24.38&			 22.90&			 24.31&		 23.74&		 24.40&			 22.82&		 22.92		\\
220&			 24.11&			 24.64&			 22.73&			 24.34&		 24.05&		 24.72&			 23.23&		 22.83		\\
230&			 24.29&			 24.94&			 23.27&			 24.65&		 23.77&		 24.99&			 23.19&		 22.52		\\
240&			 25.42&			 25.54&			 23.38&			 25.91&		 24.20&		 26.01&			 23.96&		 22.73		\\
250&			 25.91&			 27.29&			 24.41&			 27.17&		 25.90&		 27.05&			 23.73&		 22.98		\\
260&			 28.24&			 29.77&			 24.24&			 29.58&		 26.47&		 29.80&			 24.25&		 24.43		\\
270&			 47.61&			 49.16&			 32.29&			 49.32&		 32.75&		 55.46&			 29.90&		 24.78		\\
280&			  8.51&			 10.48&			 18.72&			  9.84&		 18.97&		  4.06&			 17.42&		 23.75		\\
290&			 18.42&			 17.44&			 21.93&			 18.03&		 20.31&		 17.54&			 20.41&		 22.95		\\
300&			 21.22&			 20.04&			 22.10&			 19.51&		 20.99&		 19.97&			 22.61&		 23.04		\\
310&			 22.14&			 21.04&			 23.29&			 21.16&		 22.64&		 20.86&			 22.42&		 23.42		\\
320&			 21.86&			 21.58&			 23.34&			 21.57&		 22.69&		 21.54&			 23.48&		 23.61		\\
330&			 22.20&			 22.03&			 23.43&			 21.91&		 23.27&		 21.97&			 23.62&		 23.84		\\
340&			 23.04&			 22.29&			 23.67&			 22.69&		 23.22&		 22.21&			 23.34&		 23.94		\\
350&			 22.52&			 22.33&			 24.11&			 22.60&		 23.32&		 22.49&			 23.75&		 24.07		\\
360&			 22.72&			 22.54&			 24.11&			 22.54&		 23.09&		 22.51&			 23.92&		 24.09		\\
365&			 22.63&			 22.89&			 24.01&			 23.09&		 23.00&		 22.64&			 23.82&		 24.00		\\ \hline
Average\footnotemark&		23.8&			23.88&			23.57&			23.9&		23.06&		23.98&			22.67&		23.29&		23.52
\end{tabular}
\vspace{1ex}
\caption{Axial Tilt Estimates in Degrees ($\dg$)}
\label{table:axial-tilt-estimates}
\end{center}
\end{table}
\vspace{-0.5ex}
\footnotetext{excluding any negative estimates of axial tilt.}

\restoregeometry


\section{\textcolor{white}{Perl Scripts}}

\label{sec:perl-scripts}

\vspace{3ex}
{\Large \bfseries Perl Scripts} \\


\def\KeywordColor{blue}
\def\CommentColor{green!50!black}
\def\StringColor{red!90!black}
\def\IdentifierColor{green!65!black!60!blue}
\def\BackgroundColor{gray!6!white}

\noindent The scripts \ref{sec:script-sunrise} and \ref{sec:script-axial-tilt} below use the circular orbit/constant speed model for Earth with stationary days evenly spaced around the circle and the orbital angle \p\ defined as $\frac{d}{N}(360\dg)$ for day $d \in [0, N - 1]$, where $d$ is the day offset from the day of the winter solstice, and $N = 365$ is the number of days in the year. To use the more accurate mapping from day of the year to \p\ described in \cite{jenkins:sun-position} make the following changes to these scripts (the time of noon is used for each day, and each day duration is taken to be a mean solar day of 24 hours). A similar change can be made to script \ref{sec:script-analemma}.

\begin{itemize}
\item replace the line \verb+my $winter_solstice = DateTime->new(day => 21, month => 12, year => 2018);+ with \\ 
\verb+my $base_date = DateTime->new(day => 1, month => 1, year => 2013);+

\item replace the line \verb+my $day_offset = $date->delta_days($winter_solstice)->in_units('days');+ with \\
\verb#my $t = $date->delta_days($base_date)->in_units('days') + 0.5;#

\item replace the line \verb+$psi = ($day_offset / 365) * pi2;)+ with the formula of \cite{jenkins:sun-position} as a function of \verb+$t+, plus \verb+$pip2+ (ie $90\dg$)
\end{itemize}

\subsection{Calculate Sunrise Direction}

\label{sec:script-sunrise}

\begin{lstlisting}[
columns=fixed,
basewidth=0.5em,
escapechar=|,
language=Perl,
breaklines,
tabsize=4,
showstringspaces=false,
basicstyle=\ttfamily\footnotesize,
keywordstyle=\color{\KeywordColor},
commentstyle=\color{\CommentColor},
stringstyle=\color{\StringColor},
identifierstyle=\color{\IdentifierColor},
backgroundcolor=\color{\BackgroundColor},
]
# Compare actual sunrise with sunrise computed using axial tilt of 23.44 degrees. Calculate average error.
# Usage : perl calc-sunrise.pl <file 1>  ... <file N>
# Each input file line has format :-
# <Day> <Month> <Year> <Actual Sunrise> <Elevation>
# or
# <Day> <Month> <Year> <Actual Sunrise>
# where <Elevation> is the Pole Star (or southern equivalent) elevation above the horizon
# (ie the 'latitude' in the spherical model).
# When elevation is omitted from a line the last specified value is used.
# Lines containing only white space or beginning with '#' are ignored.
# Annotation lines begin with field 'annot' (case-insensitive) and are echoed to output.

use strict;
use warnings;

use 5.012_003;

use DateTime;
use DateTime::Duration;
use Math::Trig;
use Math::Trig qw(:pi);

my $AXIAL_TILT = deg2rad(23.44);

my $winter_solstice = DateTime->new(
	day        => 21,
	month      => 12,
	year       => 2018
);

print("Actual Sunrise   Computed Sunrise   Abs Error\n");
print("--------------   ----------------   ---------\n");

my $saved_elevation;
my $line_counter = 0;
my $data_counter = 0;
my $abs_total = 0;

while (<>) {
	$line_counter++;
	my @fields = split;
	my $fields_count = scalar @fields;

	# ignore any line containing only white-space
	next if (!$fields_count);

	# skip any line beginning with '#'
	next if ( substr($fields[0], 0, 1) eq '#' );

	# echo annotation line (eg a blank line can be inserted with this)
	if ( lc($fields[0]) eq 'annot') {
		shift @fields;
		printf("@fields\n");
		next;
	}

	# 4 and 5 are the only valid numbers of fields on a line
	if ($fields_count != 4 && $fields_count != 5) {
		printf(STDERR "Invalid line : line number %d\n", $line_counter);
		exit 1;
	}

	my ($day, $month, $year, $actual_sunrise, $elevation) = @fields;

	if ( !defined($elevation) ) {
		# only 4 fields were specified on the line - elevation not specified, so we use the last
		# specified elevation, exiting with an error if the latter does not exist
		if (!defined($saved_elevation)) {
			printf(STDERR "Input files do not specify elevation data on first data-containing line.\n");
			exit 1;
		}
		$elevation = $saved_elevation;
	} else {
		# 5 fields specified on the line - save off the elevation
		$saved_elevation = $elevation;
	}

	my $date = DateTime->new(
		day        => $day,
		month      => $month,
		year       => $year
	);

	# zero-based day offset from Winter Solstice
	my $day_offset = $date->delta_days($winter_solstice)->in_units('days');

	# various angles
	my ($psi, $delta, $theta);

	# day of year angle
	$psi = ($day_offset / 365) * pi2;

	# calculate Pole Star Elevation in radians
	$delta = deg2rad($elevation);

	# sine of sunrise angle
	my $sin_theta = -sin($AXIAL_TILT) * cos($psi) / cos($delta);

	# sunrise angle
	$theta = rad2deg( asin($sin_theta) );

	$data_counter++;

# here 'absolute' means absolute error as opposed to relative error, not modulus
	my $abs_error = $theta - $actual_sunrise;
	$abs_total += abs($abs_error);
	printf("%5.1f%12s%6.2f%13s%5.2f\n", $actual_sunrise, "", $theta, "", $abs_error);
}

printf("Average magnitude abs error = %.2f degrees\n", $abs_total / $data_counter);
printf("%d data points\n", $data_counter);
\end{lstlisting}

\vspace{6ex}
\noindent Sample output :
\vspace{3ex}

\begin{lstlisting}[
columns=fixed,
basewidth=0.5em,
breaklines,
tabsize=4,
showstringspaces=false,
basicstyle=\ttfamily\footnotesize,
backgroundcolor=\color{\BackgroundColor},
]
C:\>perl calc-sunrise.pl iedinburgh.txt
Actual Sunrise   Computed Sunrise   Abs Error
--------------   ----------------   ---------
....
-43.0            -44.42             -1.42
-40.0            -41.97             -1.97
-36.0            -38.16             -2.16
....
Edinburgh
Average magnitude abs error = 2.14 degrees
38 data points
\end{lstlisting}

\newpage

\subsection{Calculate Axial Tilt}

\label{sec:script-axial-tilt}

\begin{lstlisting}[
columns=fixed,
basewidth=0.5em,
escapechar=|,
language=Perl,
breaklines,
tabsize=4,
showstringspaces=false,
basicstyle=\ttfamily\footnotesize,
keywordstyle=\color{\KeywordColor},
commentstyle=\color{\CommentColor},
stringstyle=\color{\StringColor},
identifierstyle=\color{\IdentifierColor},
backgroundcolor=\color{\BackgroundColor},
]
# Calculate axial tilt from actual sunrise data.
# Usage : perl calc-axial-tilt.pl <file 1>  ... <file N>
# Each input file line has format :-
# <Day> <Month> <Year> <Actual Sunrise> <Elevation>
# or
# <Day> <Month> <Year> <Actual Sunrise>
# where <Elevation> is the Pole Star (or southern equivalent) elevation above the horizon
# (ie the 'latitude' in the spherical model).
# When elevation is omitted from a line the last specified value is used.
# Lines containing only white space or beginning with '#' are ignored.
# Annotation lines begin with field 'annot' (case-insensitive) and are echoed to output.

use strict;
use warnings;

use 5.012_003;

use DateTime;
use DateTime::Duration;
use Math::Trig;
use Math::Trig qw(:pi);

my $winter_solstice = DateTime->new(
	day        => 21,
	month      => 12,
	year       => 2018
);

print("PSI        cos(PSI)      Actual Sunrise     Axial Tilt\n");
print("------     ----------    --------------     ----------\n");

my $saved_elevation;
my $line_counter = 0;
my $data_counter = 0;
my $total = 0;

while (<>) {
	$line_counter++;
	my @fields = split;
	my $fields_count = scalar @fields;

	# ignore any line containing only white-space
	next if (!$fields_count);

	# skip any line beginning with '#'
	next if ( substr($fields[0], 0, 1) eq '#' );

	# echo annotation line (eg a blank line can be inserted with this)
	if ( lc($fields[0]) eq 'annot') {
		shift @fields;
		printf("@fields\n");
		next;
	}

	# 4 and 5 are the only valid numbers of fields on a line
	if ($fields_count != 4 && $fields_count != 5) {
		printf(STDERR "Invalid line : line number %d\n", $line_counter);
		exit 1;
	}

	my ($day, $month, $year, $actual_sunrise, $elevation) = @fields;

	if ( !defined($elevation) ) {
		# only 4 fields were specified on the line - elevation not specified, so we use the last
		# specified elevation, exiting with an error if the latter does not exist
		if (!defined($saved_elevation)) {
			printf(STDERR "Input files do not specify elevation data on first data-containing line.\n");
			exit 1;
		}
		$elevation = $saved_elevation;
	} else {
		# 5 fields specified on the line - save off the elevation
		$saved_elevation = $elevation;
	}

	my $date = DateTime->new(
		day        => $day,
		month      => $month,
		year       => $year
	);

	# zero-based day offset from Winter Solstice
	my $day_offset = $date->delta_days($winter_solstice)->in_units('days');

	# various angles
	my ($psi, $delta, $theta);

	# day of year angle
	$psi = ($day_offset / 365) * pi2;

	# calculate Pole Star elevation in radians
	$delta = deg2rad($elevation);

	# sunrise angle
	$theta = deg2rad($actual_sunrise);

	my $sin_axial_tilt = -sin($theta) * cos($delta) / cos($psi);
	my $axial_tilt = rad2deg( asin($sin_axial_tilt) );

	# exclude any estimates < 0 as abnormally inaccurate due to cos($psi) and sin($theta) becoming small,
	# thus introducing potentially large percentage errors, even though $psi and $theta are good
	# direction estimates. Inaccurate estimates on the +ve side will still be included though.
	if ($axial_tilt >= 0) {
		$total += $axial_tilt;
		$data_counter++;
	}

	printf("%6.2f%5s%5.2f%9s%5.1f%14s%7.2f\n", rad2deg($psi), "", cos($psi), "", $actual_sunrise, "", $axial_tilt);
}

printf("%d data points\n", $data_counter);
printf("Average axial tilt = %6.2f\n", $total/$data_counter);
\end{lstlisting}

\vspace{6ex}
\noindent Sample output :
\vspace{3ex}

\begin{lstlisting}[
columns=fixed,
basewidth=0.5em,
breaklines,
tabsize=4,
showstringspaces=false,
basicstyle=\ttfamily\footnotesize,
backgroundcolor=\color{\BackgroundColor},
]
C:\>perl calc-axial-tilt.pl iedinburgh.txt
PSI        cos(PSI)      Actual Sunrise     Axial Tilt
------     ----------    --------------     ----------
....
147.95     -0.85          39.0                24.57
157.81     -0.93          43.0                24.36
167.67     -0.98          46.0                24.35
....
Edinburgh
37 data points
Average axial tilt =  23.88

\end{lstlisting}

\newpage

\subsection{Calculate Analemma}

\label{sec:script-analemma}

\begin{lstlisting}[
columns=fixed,
basewidth=0.5em,
escapechar=|,
language=Perl,
breaklines,
tabsize=4,
showstringspaces=false,
basicstyle=\ttfamily\footnotesize,
keywordstyle=\color{\KeywordColor},
commentstyle=\color{\CommentColor},
stringstyle=\color{\StringColor},
identifierstyle=\color{\IdentifierColor},
backgroundcolor=\color{\BackgroundColor},
]
# Outputs a table of solar altitude and azimuth values for a given location on Earth and time of day. 
# The output can be used directly as input to TIKZ/pgfplots package to produce an analemma graph of altitude versus 
# azimuth (analemma-graph.tex). Uses a circular orbit/constant speed model to calculate approximate orbital angles 
# for days of the year.
# Usage :
# perl calc-analemma.pl <Latitude> <ClockTimeExDST> <LongitudeDiff> <DayStep> <NumberOfDays> ...
# ... <AdjustAzimuthAngleRange> <PerihelionOffset>
# <Latitude> = latitude in degrees.
# <ClockTimeExDST> = clock time in your time zone in decimal hours, excluding any DST shift. 24 hour clock.
# <LongitudeDiff> = difference in degrees between your longitude and longitude of the defining meridian for your time zone.
# The defining meridian for the time zone UTC+<n> is at longitude a multiple of <n> times 15 degrees (<n> +ve or -ve).
# Your time zone's clock (ie civil time, ex-DST) gives the mean solar time (MST) at your time zone's defining 
# meridian longitude - the MST at your longitude at <ClockTimeExDST> is the latter plus (4 mins * <LongitudeDiff>), 
# where <LongitudeDiff> is +ve for easterly, and -ve for westerly. Default value of <LongitudeDiff> is 0, in which
# case the analemma is calculated for the time zone's defining meridian longitude rather than your own longitude.
# Ref : https://web.archive.org/web/20190820154547/http://aa.usno.navy.mil/faq/docs/eqtime.php
# <DayStep> = step between days (default value = 1).
# <NumberOfDays> = length of period covered, starting from winter solstice, 365 for whole year (default value = 365).
# <AdjustAzimuthAngleRange> = if set to true (1) adjust azimuth angle range from [0, 360) to (-180, 180] (default = 0),
# use if angles congegrate around zero, to avoid discontinuity in the graph.
# <PerihelionOffset> = offset in days of perihelion from winter solstice (default 14 = 2020 value). Used for sine wave
# approximation of the eccentricity component of the Equation of Time.
# Number of command line arguments = 2-7, the first two always required, and the remaining five optional in the 
# order specified. If any optional argument is specified as '.' then it is set to the default value.

use strict;
use warnings;
use 5.012_003;
use Math::Trig;
use Math::Trig qw(:pi);

my $AXIAL_TILT = deg2rad(23.44);

if (@ARGV == 0 || $ARGV[0] eq "-h") {
	print(	"Usage:\n" .
			"perl calc-analemma.pl <Latitude> <ClockTimeExDST> <LongitudeDiff> <DayStep> <NumberOfDays> " .
			"<AdjustAzimuthAngleRange> <PerihelionOffset>\n" .
			"First 2 parameters required, parameters 3-7 optional with default values : 0, 1, 365, 0 (false), 14\n" .
			"Specifying an optional parameter as '.' selects default value.\n"
	);
	exit;
}

# number of command line arguments must be 2-7
if (@ARGV < 2 || @ARGV > 7) {
	print(STDERR "Invalid command line : 2-7 parameters required\n");
	exit 1;
}

# defaults
my ($longitude_diff, $day_step, $num_days, $adjust_azimuth_angle_range, $perihelion_offset) = (0, 1, 365, 0, 14);

if (@ARGV == 7) {
	$perihelion_offset = pop @ARGV;
	$adjust_azimuth_angle_range = pop @ARGV;
	$num_days = pop @ARGV;
	$day_step = pop @ARGV;
	$longitude_diff = pop @ARGV;
} elsif (@ARGV == 6) {
	$adjust_azimuth_angle_range = pop @ARGV;
	$num_days = pop @ARGV;
	$day_step = pop @ARGV;
	$longitude_diff = pop @ARGV;
} elsif (@ARGV == 5) {
	$num_days = pop @ARGV;
	$day_step = pop @ARGV;
	$longitude_diff = pop @ARGV;
} elsif (@ARGV == 4) {
	$day_step = pop @ARGV;
	$longitude_diff = pop @ARGV;
} elsif (@ARGV == 3) {
	$longitude_diff = pop @ARGV;
}

# if any optional argument was specified as '.' then reset it to default value
if ($perihelion_offset eq '.') {
	$perihelion_offset = 14;
}
if ($adjust_azimuth_angle_range eq '.') {
	$adjust_azimuth_angle_range = 0;
}
if ($num_days eq '.') {
	$num_days = 365;
}
if ($day_step eq '.') {
	$day_step = 1;
}
if ($longitude_diff eq '.') {
	$longitude_diff = 0;
}

my $clock_time_ex_dst = pop @ARGV;
my $latitude_degrees = pop @ARGV;

# Calculate mean solar time corresponding to clock time, in decimal hours.
my $mean_solar_time = $clock_time_ex_dst + ($longitude_diff / 15);

# latitude in radians
my $delta = deg2rad($latitude_degrees);

my($sin_delta, $cos_delta) = (sin($delta), cos($delta));


# sub sin_cos_lambda
# ==================
# Returns the sin/cos of the solar declination for a day of the year.
# Usage : sin_cos_lambda($day_offset)
# $day_offset = day offset from winter solstice (starting from 0)
sub sin_cos_lambda {
	my $day_offset = $_[0];

	# day of year angle
	my $psi = ($day_offset / 365) * pi2;

	my $sin_lambda = -sin($AXIAL_TILT) * cos($psi);
	my $cos_lambda = sqrt(1 - $sin_lambda**2); # note lambda is always in range [-90, 90]

	return ($sin_lambda, $cos_lambda);
}

# sub angle_from_sin_cos
# ======================
# Returns an angle in the range [0, 360) from its sine and cosine values.
# Usage : angle_from_sin_cos($sin, $cos)
sub angle_from_sin_cos {
	my $sin = $_[0];
	my $cos_negative = ($_[1] < 0);
	my ($angle, $q1_angle, $mod_sin);

	$mod_sin = abs($sin);
	$q1_angle = asin($mod_sin);

	if ($sin >= 0) {
		if ($cos_negative) {
			$angle = pi - $q1_angle;
		} else {
			$angle = $q1_angle;
		}
	} else {
		if ($cos_negative) {
			$angle = pi + $q1_angle;
		} else {
			$angle = pi2 - $q1_angle;
		}
	}
	return rad2deg($angle);
}

# sub solar_hour_angle
# ====================
# Return the solar hour angle in radians using equation of time.
# Usage : solar_hour_angle($day_offset)
sub solar_hour_angle {
	my $day_offset = $_[0];

	# EOT component amplitudes in decimal hours
	my ($EC_amplitude, $OB_amplitude) = (0.1277, 0.1645);

	my $EC_component = $EC_amplitude * -sin( ($day_offset - $perihelion_offset) * pi2 / 365 );
	my $OB_component = $OB_amplitude * -sin( $day_offset * 2 * pi2 / 365 );

	my $EOT_hours_adjustment = $EC_component + $OB_component;

# uncomment one of these to investigate effect on analemma of the separate EOT components
#	$EOT_hours_adjustment = $EC_component;
#	$EOT_hours_adjustment = $OB_component;
#	$EOT_hours_adjustment = 0;

	my $apparent_solar_time = $mean_solar_time + $EOT_hours_adjustment;

	return (($apparent_solar_time - 12) / 24) * pi2;
}


# the hash allows the output to be used directly as a table in TIKZ/pgfplots
print("Day      Altitude      Azimuth\n");
print("#----    --------      -------\n");

for(my $day_offset = 0; $day_offset < $num_days; $day_offset += $day_step) {
	# altitude
	my ($mu, $sin_mu, $cos_mu);

	# azimuth
	my ($phi, $sin_phi, $cos_phi);

	# solar declination
	my ($sin_lambda, $cos_lambda) = &sin_cos_lambda($day_offset);

	# solar hour angle for clock time $mean_solar_time
	my $tau = &solar_hour_angle($day_offset);
	my ($sin_tau, $cos_tau) = (sin($tau), cos($tau));

	$sin_mu = $cos_tau * $cos_delta * $cos_lambda + $sin_delta * $sin_lambda;
	$cos_mu = sqrt(1 - $sin_mu**2); # note mu is always in range [-90, 90]
	$mu = rad2deg( asin($sin_mu) );

	# $cos_mu == 0 means sun is at zenith or nadir, therefore azimuth is undefined
	if ($cos_mu != 0) {
		$sin_phi = -$sin_tau * $cos_lambda / $cos_mu; 
		$cos_phi = ($cos_delta * $sin_lambda - $cos_tau * $sin_delta * $cos_lambda) / $cos_mu;
	}
	$phi = &angle_from_sin_cos($sin_phi, $cos_phi);

	if ($adjust_azimuth_angle_range && $phi > 180) {
		$phi -= 360;
	}

	if ($cos_mu != 0) {
		printf("%3d%6s%6.2f%8s%6.2f\n", $day_offset, "", $mu, "", $phi);
	} else {
		printf("%3d%6s%6.2f%8s%s\n", $day_offset, "", $mu, "", "Undefined");
	}
}

\end{lstlisting}

\vspace{6ex}
\noindent Sample output :
\vspace{3ex}

\begin{lstlisting}[
firstline=2,
escapechar=+,
columns=fixed,
basewidth=0.5em,
breaklines,
tabsize=4,
showstringspaces=false,
basicstyle=\ttfamily\footnotesize,
backgroundcolor=\color{\BackgroundColor},
]
C:\>perl calc-analemma.pl 37.98 16 -6.27 | perl -S tee.pl +\textcolor{red}{analemma.dat}+
Day      Altitude      Azimuth
#----    --------      -------
  0       10.27        229.12
  1       10.35        229.04
....
364       10.21        229.21
\end{lstlisting}

\newpage

\noindent Analemma graph generator \verb+analemma-graph.tex+ :
\vspace{3ex}

\begin{lstlisting}[
firstline=3,
lastline=44,
columns=fixed,
basewidth=0.5em,
escapechar=|,
language={[latex]tex},
breaklines,
tabsize=4,
showstringspaces=false,
basicstyle=\ttfamily\footnotesize,
keywordstyle=\color{\KeywordColor},
commentstyle=\color{\CommentColor},
stringstyle=\color{\StringColor},
identifierstyle=\color{\IdentifierColor},
backgroundcolor=\color{\BackgroundColor},
]
\begin{verbatim}
% changes from original version : escape to Latex code to highlight 'analemma.dat' in red
\documentclass[a4paper]{article}

\usepackage[margin=0.3in]{geometry}
\usepackage{tikz}

\usepackage{pgfplots}
\pgfplotsset{width=12cm,compat=1.14}

\newcommand{\dg}{^{\circ}}

\begin{document}

\begin{tikzpicture}[
	node font=\Large
]

\definecolor{Grid Color}{HTML}{aaaaaa};
\definecolor{Analemma Color}{HTML}{ff0000}

\begin{axis}[
	clip=false,
	axis equal=true,
	xlabel={Azimuth ($\dg$)},
	xlabel style={anchor=north, below=8pt},
	ylabel={Altitude ($\dg$)},
	ylabel style={anchor=south, above=8pt},
	grid=both,
	grid style={Grid Color, line width=0.3pt, opacity=0.3},
]
\addplot [
	Analemma Color,
	only marks,
	mark size=1.1pt,
] table[x index=2, y index=1, skip first n=0]  {|\textcolor{red}{analemma.dat}|} -- cycle;

\path (axis description cs:0.5, 1.1) node[] {$<$Title$>$};

\end{axis}

\end{tikzpicture}

\end{document}
\end{verbatim}
\end{lstlisting}


\newpage

\section{\textcolor{white}{Rodrigues Rotation Formula}}

\label{sec:rodrigues-rotation-formula}

\setcounter{figure}{0}
\renewcommand\thefigure{\thesection.\arabic{figure}}

\vspace{3ex}
{\Large \bfseries Rodrigues Rotation Formula} \\

\noindent In the following EV denotes the set of all Euclidean vectors. Rotations about an axis follow the right hand rule convention. \\

{\setlength{\parindent}{0cm}
\textbf{Theorem.}
Given a unit vector $\vc{u} \in \mathrm{EV}$ then the rotation $\vc{r}'$ of vector $\vc{r} \in \mathrm{EV}$ about the axis $\vc{u}$ is given by :
\begin{equation}
\vc{r}' = (\cos \te)\,\vc{r} + (\sin \te)\,(\vc{u} \times \vc{r}) + (1 - \cos \te)\,(\vc{u} \cdot \vc{r})\,\vc{u}
\label{eq:rodrigues}
\end{equation}

\textbf{Proof.}
As shown in Figure~\ref{fig:rodrigues} express $\vc{r}$ as a sum of components parallel and perpendicular to $\vc{u}$ :
\begin{eqnarray*}
\vc{r} & = & \rPar + \rPerp, \\
\mbox{where} \hspace{2em} \rPar & = & (\vc{u} \cdot \vc{r})\,\vc{u}\:, \\
\mbox{and} \hspace{2em} \rPerp & = & \vc{r} - \rPar = \vc{r} - (\vc{u} \cdot \vc{r})\,\vc{u}\:.
\end{eqnarray*}

Since we are using the right hand rule convention the image of $\rPerp$under the rotation is given by :
\begin{eqnarray*}
\rpPerp & = & (\cos \te)\,\rPerp + (\sin \te)\,(\vc{u} \times \rPerp) \\
& = & (\cos \te)\,\rPerp + (\sin \te)\,(\vc{u} \times \vc{r})\:, \hspace{2em} \mbox{since}\ \vc{u} \times \rPar = \vc{0}\:,
\end{eqnarray*}

and clearly $\rpPar = \rPar$, thus :
\begin{eqnarray*}
\vc{r}' & = & \rpPar + \rpPerp \\
& = & (\vc{u} \cdot \vc{r})\,\vc{u} + (\cos \te)\,(\vc{r} - (\vc{u} \cdot \vc{r})\,\vc{u}) + (\sin \te)\,(\vc{u} \times \vc{r})
\end{eqnarray*}

which readily rearranges to (\ref{eq:rodrigues}). \hfill QED 
}

\vspace{1ex}

\begin{figure}[!h]
\begin{center}
\begin{tikzpicture} [
    scale=0.29,
    node font=\normalsize,
	coord axis/.style = {very thin, double arrow=3mm, opacity=0.28},
	single arrow/.style = -{Stealth[length=#1]},
	double arrow/.style = {Stealth[length=#1]}-{Stealth[length=#1]},
	rotation arrow/.style = {single arrow=0.4cm, line width=1.5pt},
	dashed line/.style = {dash pattern=on 4pt off 4pt},
	right angle/.style = {line width=1pt},
	->-/.style={
	decoration={
		markings,
		mark=at position #1 with {\arrow{Stealth[length=0.3cm]}}
	},
	postaction={decorate}
	},
	-<-/.style={
	decoration={
		markings,
		mark=at position #1 with {\arrow{Stealth[length=0.3cm,reversed]}}
	},
	postaction={decorate}
	},
]

\definecolor{Rotating Vector Color}{HTML}{358af3}
\definecolor{Rotated Vector Color}{HTML}{e20006}
\definecolor{Rotation Axis Color}{HTML}{00bb00}

\def\circleRadius{18}
\def\rotationAngle{60}
\def\pointRadius{6pt}
\def\rightAngleSize{1.4}

\def\mainXRotation{65}
\def\mainZRotation{0}

\def\coordSystemAlpha{-24}
\def\coordSystemBeta{-7}
\def\coordSystemGamma{0}

\tdplotsetmaincoords{\mainXRotation}{\mainZRotation}
\tdplotsetrotatedcoords{\coordSystemAlpha}{\coordSystemBeta}{\coordSystemGamma}

\begin{scope}[tdplot_main_coords, tdplot_rotated_coords]


\draw[right angle] (\rightAngleSize, 0) -- ++(0, \rightAngleSize) -- ++(-\rightAngleSize, 0);

\coordinate (O) at (0, 0, 0);
\path (\circleRadius, 0, 0) coordinate (P)  node[below right=2pt and 2pt] {$P$};
\path (\rotationAngle:\circleRadius) coordinate (P') node[above right=2pt and 2pt] {$P'$};
\coordinate (Q) at (0, \circleRadius, 0);
\coordinate (S) at (0, 0, -1.5*\circleRadius);
\coordinate (M) at (0, 0, 0.3*\circleRadius);
\coordinate (N) at (0, 0, \circleRadius);

\draw[->-=0.35, ->-=0.85] (O) circle[radius=\circleRadius];

\begin{scope}[Rotating Vector Color]
\fill (O) circle[radius=\pointRadius] (P) circle[radius=\pointRadius] (Q) circle[radius=\pointRadius] (S) circle[radius=\pointRadius];
\end{scope}
\fill[Rotated Vector Color] (P') circle[radius=\pointRadius];
\fill[Rotation Axis Color] (M) circle[radius=\pointRadius];

\draw[rotation arrow] (15:1.1*\circleRadius) arc [start angle=15, end angle=45, radius=1.1*\circleRadius];

\draw[->-=0.55, Rotating Vector Color] (S) -- (P) node[right=6pt, pos=0.51] {
$\vc{r}$
};

\draw[->-=0.55, Rotating Vector Color] (S) -- (O) node[left=3pt, pos=0.51] {
$\rParB$
};

\draw[->-=0.57, Rotating Vector Color] (O) -- (P) node[below=4pt, pos=0.55] {$\rPerpB$};

\draw[->-=0.65, Rotating Vector Color] (O) -- (Q) node[left=4pt, pos=0.76] {$\vc{u} \times \rPerpB$};

\draw[->-=0.65, Rotated Vector Color] (O) -- (P') node[below right=-1pt and -1pt, pos=0.62] {$\rpPerpB$};

\draw[single arrow=3mm] (0.4*\circleRadius, 0) arc[start angle=0, end angle=\rotationAngle, radius=0.4*\circleRadius] node[pos=0.4, left=8pt] {$\theta$};

\draw[Rotation Axis Color, dashed line] (O) -- (M);
\draw[single arrow=3mm, Rotation Axis Color, line width=1pt] (M) -- (N) node[left=5pt, pos=0.95] {$\vc{u}$};
\end{scope}

\draw[->-=0.55, Rotated Vector Color, dashed line] (S) -- (P') node[left=4pt, pos=0.55] {
$\vc{r}'$
};

\end{tikzpicture}
\vspace{1.2ex}
\caption{Rotating Vector $\vc{r}$ by \te\ About Axis $\vc{u}$}
\label{fig:rodrigues}
\end{center}
\end{figure}


\newpage

\section{\textcolor{white}{Latitude as Elevation of Pole Star}}

\label{sec:latitude-pole-star}

\setcounter{figure}{0}
\renewcommand\thefigure{\thesection.\arabic{figure}}

\vspace{3ex}
{\Large \bfseries Latitude as Elevation of Pole Star} \\

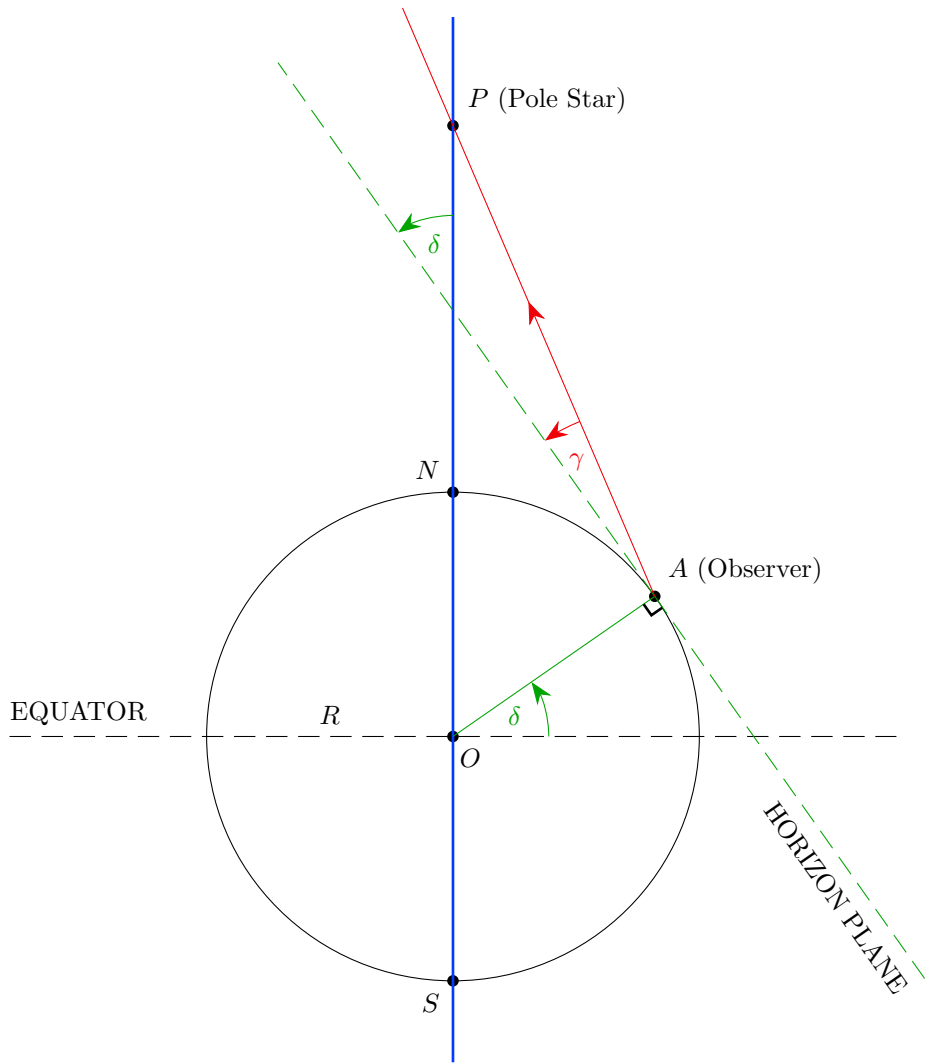
\begin{figure}[!h]
\begin{center}
\begin{tikzpicture} [
	scale=0.36,
	node font=\normalsize,
	polar axis style/.style = {Polar Axis Color, line width=1pt},
	equator/.style = {dash pattern=on 8pt off 4pt},
	horizon plane style/.style = {Horizon Plane Color, dash pattern=on 8pt off 4pt},
	sight line/.style = {Sight Line Color},
	right angle style/.style = {line width=1pt},
	single arrow/.style=-{Stealth[length=#1]},
	->-/.style={
	decoration={
		markings,
		mark=at position #1 with {\arrow{Stealth[length=0.3cm]}}
	},
	postaction={decorate}
	},
]

\definecolor{Polar Axis Color}{HTML}{003cff}
\definecolor{Sight Line Color}{HTML}{ef0006}
\definecolor{Horizon Plane Color}{HTML}{00a400}

\def\latitude{35}
\def\R{9}
\def\rightAngleSize{0.5}
\def\pointRadius{6pt}

\fill (0, 0) coordinate (O) circle[radius=\pointRadius] node[below right=1pt and -1pt] {$O$};
\fill (0, -\R) coordinate (S) circle[radius=\pointRadius] node[below left=2pt] {$S$};
\fill (0, \R) coordinate (N) circle[radius=\pointRadius] node[above left=2pt] {$N$};
\fill (0, 2.5*\R) coordinate (P) circle[radius=\pointRadius] node[above right=1pt and 2pt] {$P$ (Pole Star)};
\fill (\latitude:\R) coordinate (A) circle[radius=\pointRadius] node[above right=2pt] {$A$ (Observer)};

\draw (O) circle[radius=\R];

\draw[polar axis style, name path=polar axis] ($(S) + (0, -3)$) -- ($(P) + (0, 4)$);

\draw[equator] (-1.8*\R, 0) -- (1.8*\R, 0) node[sloped, above right=4pt and 0pt, pos=0, inner sep=0pt, black] {EQUATOR};

\path (O) -- (-\R, 0) node[above=1pt, pos=0.5] {$R$};

\draw[Horizon Plane Color] (O) -- (A);
\draw[Horizon Plane Color, single arrow=0.3cm] (3.5, 0) arc [start angle=0, delta angle=\latitude, radius=3.5] node[pos=0.36, left=6pt] {$\delta$};

\draw[right angle style] (A) ++(\latitude:-\rightAngleSize) -- ++(\latitude-90: \rightAngleSize) -- ++(\latitude:\rightAngleSize);

\draw[horizon plane style, name path=horizon plane] (O) (A) -- ([turn]90:24);
\draw[horizon plane style] (O) (A) -- ([turn]-90:18) node[sloped, below left=6pt and 0pt, pos=0.95, inner sep=0pt, black] {HORIZON PLANE};
\path [name intersections={of=horizon plane and polar axis, by={B}}];

\draw[sight line, ->-=0.5] (A) -- ($(A)!1.25!(P)$);

\def\gammaStartAngle{113}

\draw[Horizon Plane Color, single arrow=0.3cm] ($(B) + (0, 3.5)$) arc [start angle=90, delta angle=\latitude, radius=3.5] node[pos=0.32, below=3pt] {$\delta$};
\draw[Sight Line Color, single arrow=0.3cm] ($(A) + (\gammaStartAngle:7)$) arc [start angle=\gammaStartAngle, end angle=90+\latitude, radius=7] node[pos=0.08, below=8pt] {$\gamma$};

\end{tikzpicture}
\vspace{3ex}
\caption{Latitude as Elevation of Pole Star}
\label{fig:latitude-pole-star}
\end{center}
\end{figure}

\noindent In Figure~\ref{fig:latitude-pole-star}, an observer at latitude \de\ above the equator sees an angle of elevation \g\ of the Pole Star above their horizon. Because the distance $OP$ is vastly much greater than the planet radius $R$ the lines $PA$ and $PN$ are close to parallel, thus the angles of incidence \de\ and \g\ of the horizon line onto these two lines are close to equal. (And as we approach the equator the Pole Star falls to the horizon with $\g < \de$, $\de \rightarrow \sin^{-1}(\frac{R}{OP})^+, \g \rightarrow 0^+$, and $R \ll OP$). Thus \g\ always well approximates \de\ as a direction.

\vspace{5ex}

\end{appendices}


\newpage

\let\OLDthebibliography\thebibliography
\renewcommand\thebibliography[1]{
  \OLDthebibliography{#1}
  \setlength{\parskip}{0pt}
  \setlength{\itemsep}{2pt plus 0.3ex}
}

\setcounter{secnumdepth}{0}
\section{References}

All web links retrieved on 31\textsuperscript{st} August 2020.

\begingroup
\renewcommand{\section}[2]{}

\endgroup

\end{document}